\documentclass[usegraphicx,useAMS,usenatbib]{mn2e}\pdfoutput=1

\usepackage{subfigure}
\usepackage{lscape}
\usepackage{amsmath}
\usepackage{ulem}
\usepackage{footnote}
\usepackage{color}

\definecolor{darkgreen}{rgb}{0,.4,0}

\makesavenoteenv{tabular}

\title[Golden gravitational lensing systems from the Sloan Lens ACS
  Survey. II. SDSS J1430+4105]{Golden gravitational lensing systems from the Sloan Lens ACS
  Survey. II. SDSS J1430+4105: A precise inner total mass
    profile from lensing alone}

\author[Thomas Eichner, Stella Seitz, Anne Bauer]{Thomas
  Eichner$^{1,2}$\thanks{E-mail: eichner@usm.lmu.de}, Stella Seitz$^{1,2}$, Anne Bauer$^{1,3}$\\
$^{1}$ Universit\" ats-Sternwarte M\" unchen, Scheinerstr. 1, 81679
  Muenchen, Germany\\
$^{2}$ Max-Planck-Institut f\" ur extraterrestrische Physik,
  Giessenbachstra\ss e, 85748 Garching, Germany\\
$^{3}$ Institut de Ci\`encies de l'Espai, CSIC/IEEC, F. de Ci\`encies,
  Torre C5 par-2, Barcelona 08193, Spain\\
}

\begin{document}

\date{Accepted 2012 August 28. Received 2012 August 28; in original form 2012 June 28}
\maketitle
\label{firstpage}

\begin{abstract}
We study the Sloan Lens ACS survey (SLACS) strong lensing system SDSS
J1430+4105 at $z_{\rm l}=0.285$. The lensed source ($z_{\rm s}=0.575$)
of this system has a complex morphology with several subcomponents. 
Its subcomponents span a radial range from $4
\, \rm kpc$ to $10\, \rm kpc$ in the plane of the lens. Therefore we
can constrain the slope of the total projected mass profile around the
Einstein radius from lensing alone. 
We measure a density profile that is slightly but not
significantly shallower than isothermal at the Einstein radius.
We decompose the mass of the lensing galaxy into a de Vaucouleurs
component to trace the stars and an additional dark component.
The spread of multiple image components over a large radial range
also allows us to determine the amplitude of the de Vaucouleurs and dark
matter components separately. 
We get a mass to light ratio of $\frac{\rm
  M_{de Vauc}}{\rm L_{\rm B}}\approx (5.5\pm1.5)\,\frac{\rm M_{\odot}}{\rm L_{\odot,\rm B}}$
and a dark matter fraction within the Einstein radius of $\approx 20\%
\; \rm to \; 40\%$. 
Modelling the star formation history assuming
  composite stellar populations at solar metallicity to the galaxy's
  photometry yields a mass to light ratio of $\frac{\rm
  M_{\star, \rm salp}}{\rm L_{\rm B}}\approx 4.0_{-1.3}^{+0.6}\,\frac{\rm M_{\odot}}{\rm
  L_{\odot,\rm B}}$ and $\frac{\rm M_{\star, \rm chab}}{\rm L_{\rm
    B}}\approx 2.3_{-0.8}^{+0.3}\,\frac{\rm M_{\odot}}{\rm
  L_{\odot,\rm B}}$ for Salpeter and Chabrier IMFs, respectively.
Hence, the mass to light ratio derived from lensing
is more Salpeter-like, in
agreement with results for massive Coma galaxies and other nearby
massive early type galaxies. 
We examine the consequences of the galaxy group in which the lensing
galaxy is embedded, showing that it has little influence on the mass
to light ratio obtained for the de Vaucouleurs component of the lensing galaxy. 
Finally, we decompose the projected, azimuthally averaged 2D
density distribution of the de Vaucouleurs and dark matter component
of the lensing signal into spherically averaged 3D density profiles. We
can show that the 3D dark and luminous matter density within the
Einstein radius ($\rm R_{Ein} \approx 0.6 \rm R_{eff}$) of this SLACS
galaxy is similar to the values of Coma galaxies with the same velocity
dispersions.
\end{abstract}
\section{Introduction}

Early-type galaxies contain a large fraction of the total stellar mass
observed in the Universe (e.g., \citealt{baryonic_mass_dist};
\citealt{Bell_baryonic_content}).
Studying the internal structure
of early-type galaxies is crucial for understanding the baryonic
physics that plays a key role in the formation and evolution of these
objects. Several studies have shown that the stars assembled in
early-type galaxies are embedded in massive dark matter haloes
(e.g.,\citealt{slacs6}; \citealt{Lagattuta_2009}; \citealt{weijmans_NGC2974}), but the
precise amount of dark matter contained in the galaxies' inner regions is still under debate.\\
Dark matter only simulations have found indications of a universal
density profile for dark matter haloes, present also in galaxies (the so called NFW
profile; \citealt{nfw_profile}). Nevertheless, more recent and
realistic simulations that include also the physics of baryons (e.g.,
\citealt{gnedin_condensation}; \citealt{baryon_physics};
\citealt{elz01}; \citealt{ber03}; \citealt{ma04}; \citealt{blu86};
\citealt{jes02}), like radiative cooling and
supernova and black hole feedback, have suggested that the inner
profile of the dark matter component can be significantly affected by
the interactions between baryonic and dark matter. \\
The internal structure of nearby early-type galaxies has been for
decades the object of intense dynamical analyses (e.g.,
\citealt{roberto_darkmatter_ellipticals};
\citealt{ortwin_giant_ellitpcals};
\citealt{thomas_modelling_coma_galaxies}; \citealt{tho09};
\citealt{thomas_slacs_vs_coma_IMF};
\citealt{pu10}). One focus of these studies is to
  compare stellar with dynamical mass to light ratios. The dynamical studies, e.g. \cite{ortwin_giant_ellitpcals},
  \cite{thomas_slacs_vs_coma_IMF} find ratios for nearby elliptical
  galaxies of $M/L_{B}\approx$ 4 to 10. Similar values are also found
  by \cite{sauronIV}. Only in the
last few years has gravitational lensing also contributed
significantly to our understanding of the luminous and dark matter
composition of early-type galaxies
beyond the local Universe (\citealt{slacs2}; \citealt{claudio_1538};
\citealt{barnabe_0728}). Strong gravitational lensing in early-type
galaxies has also proved to be a powerful cosmological tool to probe
the geometry of the universe independently from other
diagnostics (e.g., \citealt{claudio_2008a}; \citealt{Suyu_B1608+656_I,{Suyu_B1608+656_II}}).\\
By combining strong gravitational lensing and stellar dynamics in a
sample of first 15, then 58 early-type galaxies,  \cite{slacs3} and \cite{Koopmans_etg_dynamics_2009} have found
that the average total (luminous and dark) mass density distribution
within the effective radius -- the radius of the isophote
  containing half of the total light of the galaxy -- is well represented by an isothermal
distribution ($\rho \propto r^{-2}$), although significant deviations from
this result can be observed in individual galaxies. Only rare
 systems where an extended or several distinct
sources are gravitationally lensed over an extended radial area on the lens plane can be used to
determine the total mass density profile of the
lens galaxy over a wide radial range through lensing only
(e.g. \citealt{claudio_ClJ0152, {claudio_1538}}; \citealt{fadely_q0957}).  Moreover, combining lensing total mass
measurements with photometric stellar mass estimates in these systems
offers a unique way to disentangle their luminous and dark matter components.\\
In this paper, we study the gravitational lensing system SDSS
J1430+4105 that is composed of a massive early-type galaxy acting
as a lens for an irregular background source. This
  galaxy was part of the SLACS survey\footnote{http://www.slacs.org}
  and has been studied as part of their lens sample, especially in
  \cite{slacsV}, \cite{slacs9} and \cite{slacs10}:
  \cite{slacsV} fit a singular isothermal lens model
    to the observed multiple images, while \cite{slacs9} and
    \cite{slacs10} combine the measured Einstein radii and masses with
    photometric and dynamical data. The surface brightness
distribution of the lensed source shows several peaks that extend
from 4 to 10 kpc from the lens galaxy centre. This fact provides the
opportunity to investigate the lens galaxy mass distribution on
radial ranges larger than those explored in similar
analyses of other gravitational lensing systems (e.g., \citealt{Xanthopoulos_B1030+074}, \citealt{Cohn_B1933+503},\citealt{claudio_1538}).\\
The paper is organised as follows: Section 2 gives an overview of the
observations and data used in this work and
  introduces the environment of SDSSJ 1430+4105; Section 3 describes the details of the strong lensing
models. Section 4 states the implications for the total
mass and the mass profile. In Section 5 the mass to light
  ratio for the de Vaucouleurs component is constrained, in Section 6
  the results are discussed. Appendix A contains
    further variants of strong lensing models. 
Appendix B gives further details of the
environment implementation.\\
The cosmological model adopted in this paper is parametrized by
$\Omega_{\mbox{m}}=0.3,\Omega_{\Lambda}=0.7,H_0=70 \,
\mbox{km}\, \mbox{s}^{-1}\, \mbox{Mpc}^{-1}$. In the cosmology assumed, 1 arcsec
in the lens ($z_{\rm l}=0.285$) and source
($z_{\rm s}=0.575$) plane corresponds to 4.30 and 6.55 kpc.\\

\section{Observations}

The SLACS Survey\footnote{http://www.slacs.org} aimed at finding strong
gravitational lenses among the galaxies observed in the SDSS. The lens
detection strategy is presented in \cite{Bolton_preslacs} and is based
on the examination of the SDSS galaxy spectra, taken with a 3 \arcsec
diameter fibre, to identify emission lines not associated with the
primary target galaxy but with an additional source, aligned with the
first galaxy and located at a higher redshift. The lens candidates are
then ranked in terms of their probability of being lensing systems and
are consequently observed with the HST/ACS and WFPC2.

\begin{figure}
\centering
\includegraphics[width=80mm]{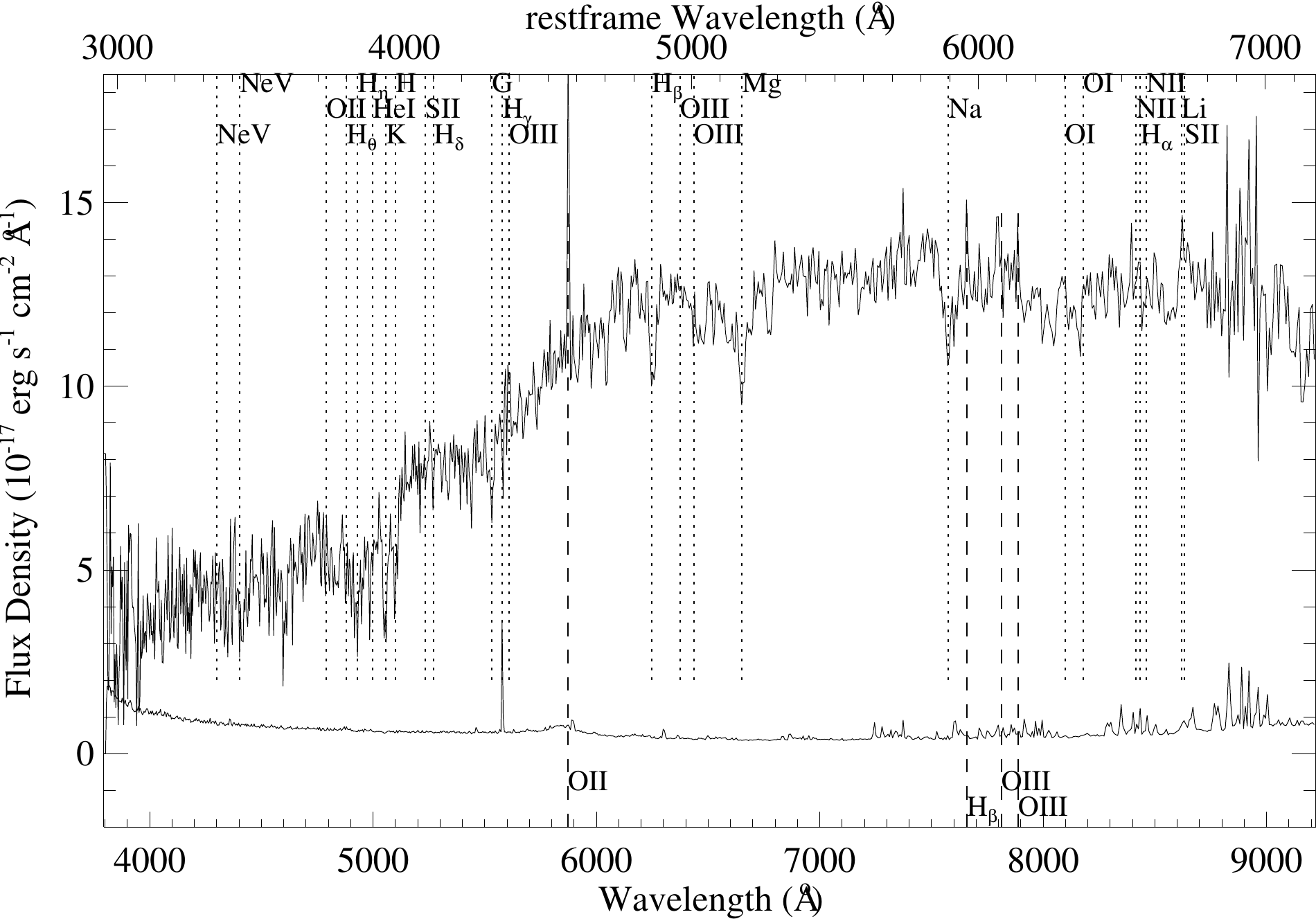}
\caption{The SDSSJ 1430+4105 spectrum as observed by the SDSS
  3 \arcsec diameter fibre. On the bottom, the
    observed wavelength is stated, while on the top, this is converted
  into the restframe wavelength of the lens ($z=0.285$). The dotted
  vertical lines give the SDSS emission / absorption line sample at
  the redshift of the lens. The dashed vertical lines give some
  prominent emission lines at the redshift of the source. Overplotted in
  the lower part of the figure is the flux uncertainty given again by
  SDSS. The spectrum shows several absorption lines
  typical of an early type galaxy at $z=0.285$ and some additional 
  emission lines at redshift $z=0.575$ (e.g. the lines at
  $5872\mbox{\AA}$, $7661\mbox{\AA}$ and $7813\mbox{\AA}$, which can be
  identified as the redshifted [OII]3728, $\rm H\beta$ and [OIII]4960
  lines respectively). Data taken from the Sloan
    digital sky survey, www.sdss.org.}
\label{observations:spectrum:SDSS}
\end{figure}

Up to now, 85 confirmed (grade-A) lenses
(\citealt{slacsI}, \citealt{slacs9}) were discovered in this way, and SDSSJ 1430+4105 is
one of these.
In Fig. \ref{observations:spectrum:SDSS} we show the SDSS spectrum,
from which lens and source redshifts of
$z_{\rm l}=0.285$ and $z_{\rm s}=0.575$ are measured, together with the lens
aperture velocity dispersion of $\sigma_{\rm SDSS}=(322 \pm 32)\, \rm
km\,\rm s^{-1}$.

\subsection{Galaxy light profile and lensing observables}
\label{sec:Galaxy light profile and lensing observables}

\begin{figure}
\centering
\includegraphics[height=70mm]{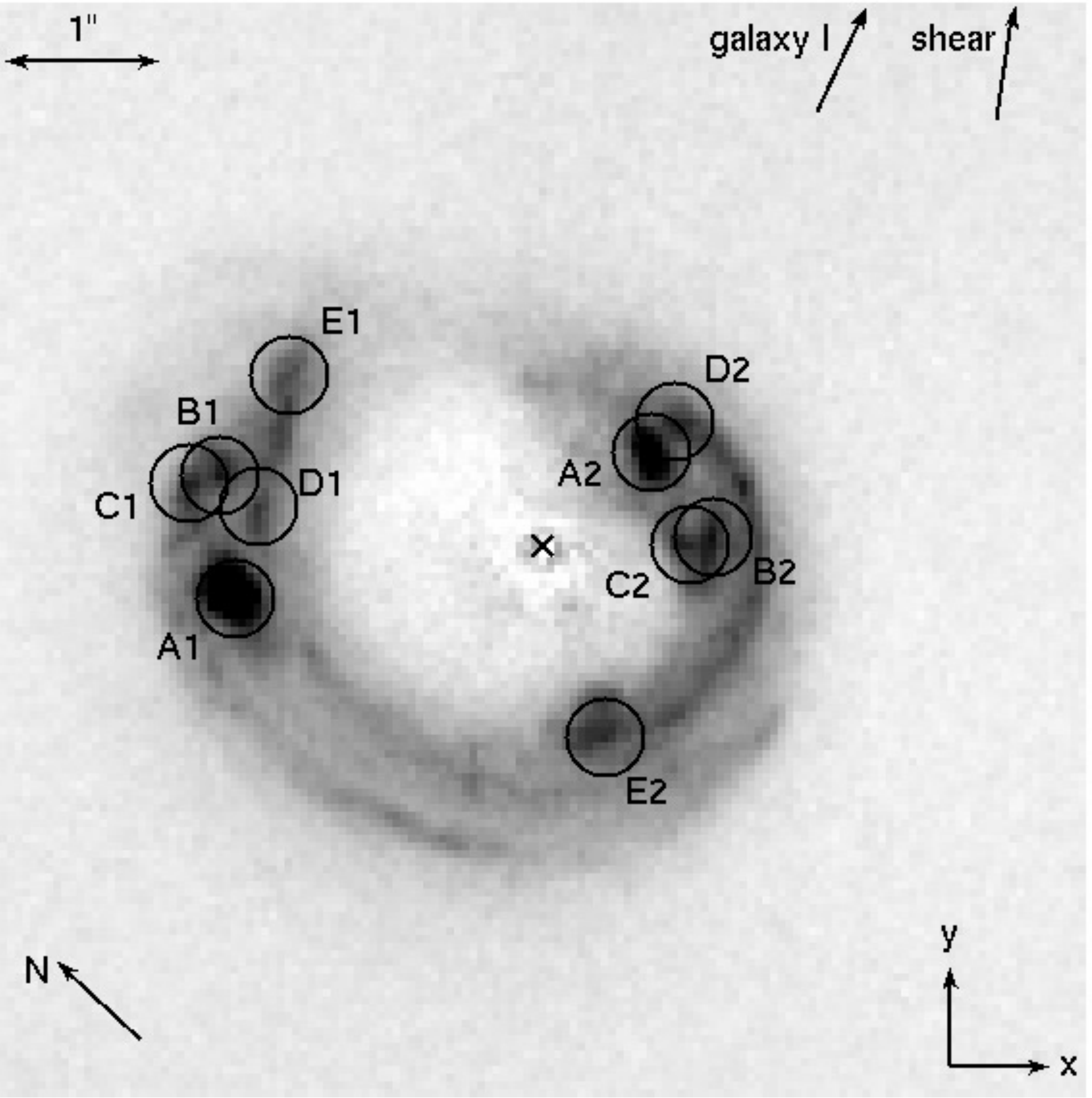}
\caption{The multiple image systems that are identified after the lens
  galaxy subtraction and used as input for the lensing analysis. The labels A1
  to E2 mark the positions used for the strong lensing analysis. The same 
  letters correspond to images coming from the same source feature, the labelling is
  done according to Table \ref{lensing:parametric:table:input_pos}. The cross marks the centre of
the subtracted lens galaxy light. Also indicated are the derived shear
direction from Sec. \ref{subsec:clumpy group} and the direction of
galaxy I. For orientation north is given as well. Angles are
measured in the local coordinate system counterclockwise as (-x) over
y if not otherwise stated. The image is rotated relative to the WCS J2000 by $47.21^{\circ}$.}
\label{lensing:parametric:image1}
\end{figure}

The basic photometric and spectroscopic properties of SDSSJ 1430+4105, taken from \cite{slacsV}, are stated in Table
\ref{observations:table:slacs_values}. For these, \cite{slacsV} fitted a
de Vaucouleurs \citep{deVacouleurs_profile} profile with elliptical isophotes to the galaxy's surface
brightness distribution. They obtained an effective radius of
$\theta_{\rm eff} = 2.55 \arcsec
=10.96\, \rm kpc$, a minor to major axis ratio of
$q_{\rm L}=0.79$, and a major axis angle of $\Theta_{\rm q,L}=-12.8^{\circ}$.
The angles are transformed to the adopted local
reference frame shown in Fig. \ref{lensing:parametric:image1}, and measured
counterclockwise with the y-axis equals to $0^{\circ}$.\\
We retrieve the public HST images from the Hubble Space Telescope archive
at ESO\footnote{http://archive.eso.org/archive/hst/}. Three
filters were available for this system: HST/WFPC2 F606W with a total
integration time of 4400s (ua1l4501m, ua1l4502m, ua1l4503m,
ua1l4504m), HST/ACS F814W with a total integration time of 2128s
(j9op36010) and HST/WFC3 F160W with a total integration time of
2497s. For the lensing
analysis we use the ACS F814W filter observations, since the PSF of
the ACS camera is smaller than the one of the WFPC2 and WFC3. First, we subtract the 
 lensing galaxy's light contribution with GALFIT (\citealt{galfit}) by using a
de Vaucouleurs profile, with the parameters of Table \ref{observations:table:slacs_values}.
Then, in order to refine the lens galaxy subtraction and especially
remove the residuals still present in the central region, an additional Sersic profile (\citealt{sersic_profile})
with index $1.2$ is subsequently subtracted.\\
Fig. \ref{lensing:parametric:image1} shows the final galaxy subtracted
image. The lensed source has a complex surface brightness
distribution, with 5 surface brightness maxima which are imaged 2 times
each. We mark and label the $5\times2$ multiple image positions, identified as the
brightest pixels, in Fig. \ref{lensing:parametric:image1}.  Their
coordinates are reported in Table
\ref{lensing:parametric:table:input_pos} together with  approximate
error estimates.
We assume in the following that all subcomponents A-E are at the same
redshift and not unlikely line-of-sight projections at different
redshifts. 
The distances of the multiple images from the centre of the lens galaxy light
distribution span a range from $0.93\arcsec$ to
$2.32\arcsec$.
In the rest of the paper, if not otherwise stated, we adopt the
coordinate system introduced in Fig. \ref{lensing:parametric:image1} which is
rotated relative to the WCS J2000 (world coordinate system) by $47.21^{\circ}$.

\begin{table*}
\centering
\begin{minipage}{100mm}
\centering
\caption{Photometric and spectroscopic quantities of the lens system}
\begin{tabular}{cccccccc}
\hline
RA & Dec & $z_{\rm l}$ & $z_{\rm s}$ & $q_{\rm L}$ & $\Theta_{\rm q,L}$
& $\vartheta_{\rm eff}$ & $\sigma_{\rm SDSS}$\\
(J2000)& (J2000) & & & $(\frac{b}{a})$ & $(^{\circ})$ &(\arcsec)
&$(\rm km\,\rm s^{-1})$\\ 
\hline
14:30:04.10 & +41:05:57.1 & 0.285 & 0.575  & 0.79 &
-12.8\footnotemark[1] & 2.55 & $322 \pm 32$ \\
\hline
\end{tabular}
Given are the position of the galaxy(RA, Dec), the redshifts
  of galaxy and source ($z_{\rm l}$ $z_{\rm s}$), the axis ratio($q_{\rm L}$), the
  orientation ($\Theta_{\rm q,L}$), the effective radius
  ($\vartheta_{\rm eff}$) of the lens' light distribution and the velocity dispersion $\sigma_{\rm SDSS}$. 
Values are taken from \cite{slacsV}\\
\footnotemark[1]{This angle is equivalent to $-59.3^{\circ}$ in the WCS coordinate
  system, defined as (-E) over N.}
\label{observations:table:slacs_values}
\end{minipage}
\end{table*}

\begin{table}
\centering
\caption{Observed positions of the multiple image systems}
\begin{tabular}{cccccc}
\hline
ID &$\Theta_1$\footnotemark[1]&$\Theta_2$\footnotemark[1]&$z_{\rm
      s}$&$\delta_{\Theta}$&d\footnotemark[1]\\
&$(\arcsec)$&$(\arcsec)$& &$(\arcsec)$&$(\arcsec)$\\
\hline
A1 & -1.99 & -0.32 & 0.575 & 0.05 & 2.02\\
A2 & 0.69 & 0.62 & 0.575 & 0.05 & 0.93\\
B1 & -2.08 & 0.47 & 0.575 & 0.05 & 2.13\\
B2 & 1.08 & 0.08 & 0.575 & 0.05 & 1.08\\
C1 & -2.28 & 0.42 & 0.575 & 0.05 & 2.32\\
C2 & 0.93 & 0.03 & 0.575 & 0.05 & 0.93\\
D1 & -1.84 & 0.27 & 0.575 & 0.05 & 1.86\\
D2 & 0.84 & 0.80 & 0.575 & 0.05 & 1.35\\
E1 & 0.39 & -1.21 & 0.575 & 0.05 & 1.27\\
E2 & -1.64 & 1.11 & 0.575 & 0.05 & 1.98\\
\hline
\label{lensing:parametric:table:input_pos}
\end{tabular}\\
\footnotemark[1]{relative to the centre of the galaxy light distribution}
\end{table}

\subsection{Observed environment}
\label{sec:observed environment}

\begin{figure}
\centering
\includegraphics[height=75mm]{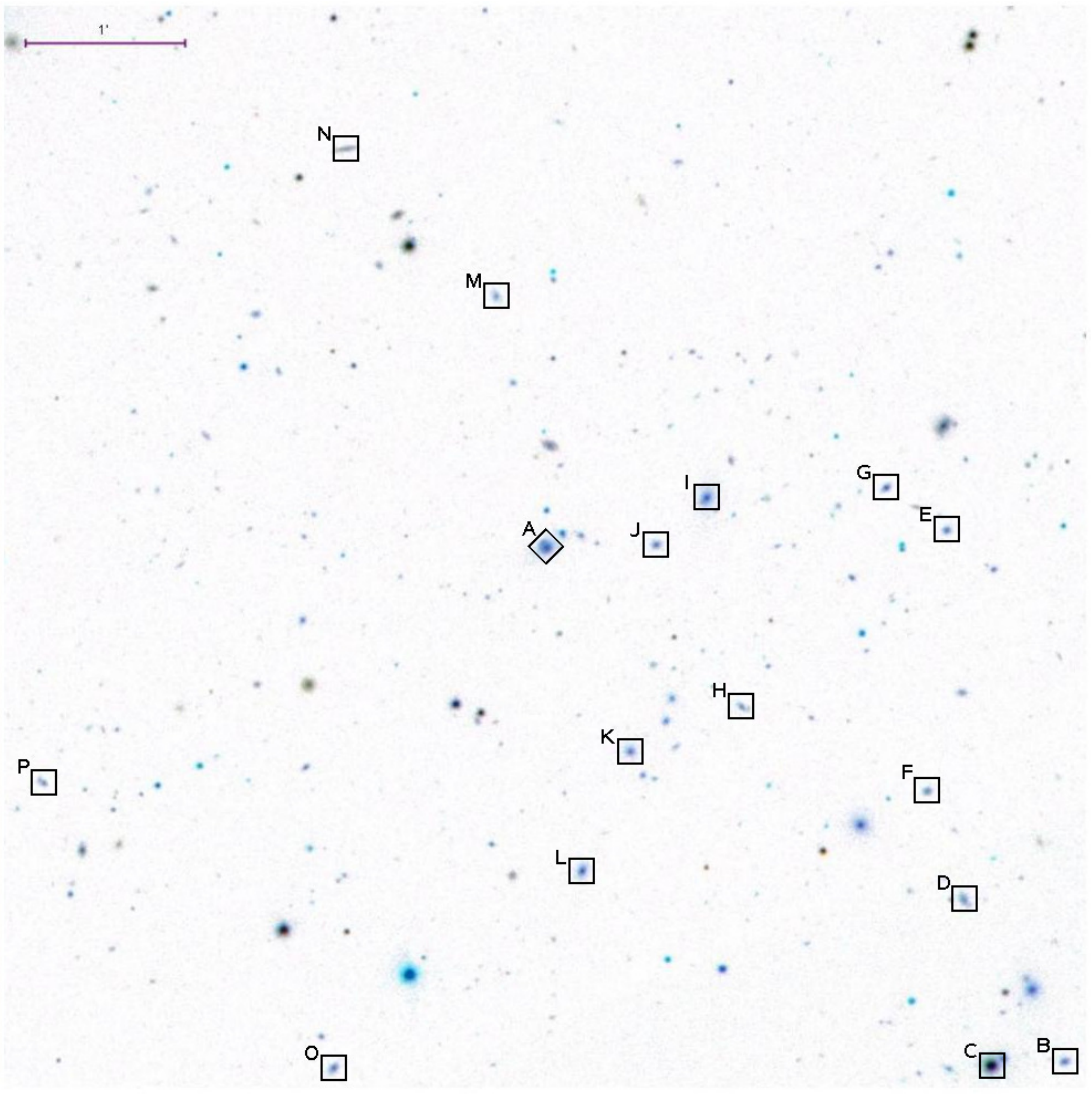}
\caption{The observed environment of J1430+4105, taken from
  SDSS. All galaxies with a photometric redshift consistent with the
  spectroscopic redshift of
  SDSSJ 1430+4105 (labelled with A and marked with a diamond) within 1
  standard deviation are marked by squares. The galaxy marked as I is
  the BCG for the group identified by \protect\cite{maxBCG}. In the image,
  north is up and east
  is left. This image is rotated relative to Fig.
  \ref{lensing:parametric:image1} by $47.21^{\circ}$. The size is
  $7.5\arcmin\,\times\,7.5\arcmin$.  For the properties of the galaxies,
  see Table \ref{table:observations:environment:galaxies}.}
\label{observation:environment:image1}
\end{figure}

SDSSJ 1430+4105 is not an isolated galaxy. It coincides in redshift
and location with a galaxy group at $z=0.287$, 
listed in the
maxBCG cluster catalogue, \citep{maxBCG}. 
Therefore, we should consider the
light deflection by the lens' environment when we model this lens. We
show the environment of SDSSJ 1430+4105
(labelled as A) in
Fig. \ref{observation:environment:image1}. The galaxy
labelled as I (and called GI in the following) was proposed to be the
brightest cluster galaxy (BCG) of this group found in \cite{maxBCG}.
The photometric redshift of the group GI is estimated
to be $z=0.287$ with a typical redshift error in the
maxBCG catalogue of 0.01. Within this error the photometric redshift
of the group is identical to the spectroscopic redshift (0.28496) of
the lensing galaxy. The group consists of 12 members within
the $\rm R_{200}$
of this group, $\rm N_{(\rm gal,200)}=12$. At this
richness level, the maxBCG cluster is
typically more than 90\% pure and complete, based on tests with mock catalogues.
We now estimate the group members based on astrometric and photometric data from 
SDSS DR7 (\citealt{sdss7}).  We consider each galaxy within $10\arcmin$
from the main lens J1430+4105. 
We allow for galaxies which have at least one of the photometric redshift
estimates (template based (Template-z) (\citealt{sdss_dr5}) or neural network based (CC2 z
and D1 z) (\citealt{sdss_photo_z_d1_cc2})) consistent within one
standard deviation with the spectroscopic redshift value of SDSS
J1430+4105 and the photometric redshift of the group GI. The neighbours which pass 
these requirements are listed in Table
\ref{table:observations:environment:galaxies}.
This table shows that for our definition the galaxy A is a group member from both
its photometric and spectroscopic redshift. The magnitudes of the galaxies A and I
in the ugriz filters are 20.43, 19.02, 17.74, 17.12, 16.87 and
22.54, 19.44, 17.92, 17.35, 17.00 for A and I respectively, and thus A
is formally the brightest galaxy ('BCG') of this group. 
Since the group
membership and group redshift estimate of the maxBCG catalogue \citep{maxBCG} is
mostly based on the g-r colour, the contaminated g-r colour of A due to
lensing has likely led to A not being considered as a group member and therefore as the
BCG. Using the same definition for the $r_{200}$ as stated in
\citet{maxBCG} we find 11 group members from Table \ref{table:observations:environment:galaxies}.
\\
From \cite{Johnston_2007}, who tested the maxBCG cluster finder on
simulated groups and clusters of galaxies, we derive the probability that a group of
the richness given in the maxBCG catalogue $\rm N_{(\rm gal,200)}=12$ is centred on the correct
  BCG in the maxBCG cluster catalogue to
be $\rm p_c(\rm N_{200}=12)=0.63$. Therefore both 
A or I could be the true mass centre of the group GI.
Finally, the \cite{maxBCG} group catalogues could also contain false
positive detections. \cite{song_finder_test}
have shown that the false detection rate of such groups can be as
large as 40\%. 
This motivates why we will consider lens models
with and without an external group
  contribution. We will show
  that including a galaxy group centred on galaxy I has only minor
  influence on the lens parameters.

\begin{table*}
\centering
\begin{minipage}{190mm}
\caption{Observed environment of SDSSJ 1430+4105}
\begin{tabular}{l|llcccccccccc}
\hline
Obj & RA &Dec &
  $\rm mag_{\rm i}$ &$\sigma_{\rm i}$ & Template z & Error & NN CC2 z &
Error &  NN D1 z & Error &
Spectro-z & Error \\ 
& (J2000) & (J2000) & & & & & & & & & \\
\hline

A & 217.51704 & 41.0992 & 17.216 & 0.006 & 0.228 & 0.033 &
0.231 & 0.027 & 0.234 & 0.026 & 0.28496 & 2.13E-4  \\
B & 217.44551 & 41.0456 & 18.729 & 0.018 & 0.202 & 0.015 &
0.216 & 0.073 & 0.260 & 0.067 & &  \\
C & 217.45406 & 41.0460 & 18.495 & 0.020 & 0.262 & 0.019 &
0.287 & 0.045 & 0.316 & 0.048 & &  \\
D & 217.45941 & 41.0625 & 18.657 & 0.019 & 0.179 & 0.032 &
0.312 & 0.055 &  0.309 & 0.064 & & \\
E & 217.46178 & 41.1010 & 18.870 & 0.016 & 0.266 & 0.019 &
0.287 & 0.055 & 0.343 & 0.073 & & \\
F & 217.46445 & 41.0738 & 19.134 & 0.020 & 0.286 & 0.082 &
0.309 & 0.096 & 0.344 & 0.073 & &   \\
G & 217.47010 & 41.1054 & 18.708 & 0.013 & 0.224 & 0.011 &
0.228 & 0.073 & 0.267 & 0.063 & &  \\
H & 217.49009 & 41.0825 & 19.356 & 0.024 & 0.200 & 0.024 &
0.259 & 0.070 & 0.339 & 0.060 & &  \\
I & 217.49493 & 41.1044 & 17.348 & 0.008 & 0.273 & 0.019 &
0.272 & 0.020  & 0.272 & 0.016 & & \\
J & 217.50190 & 41.0995 & 18.697 & 0.018 & 0.251 & 0.021 &
0.293 & 0.062 &  0.326 & 0.063 & &  \\
K & 217.50545 & 41.0779 & 18.283 & 0.015 & 0.263 & 0.020 &
0.267 & 0.037 &  0.293 & 0.032 \\
L & 217.51212 & 41.0655 & 18.008 & 0.0100 & 0.268 & 0.015 &
0.273 & 0.022 &  0.279 & 0.023 & &  \\
M & 217.52402 & 41.1253 & 19.358 & 0.026 & 0.298 & 0.084 &
0.451 & 0.042 & 0.424 & 0.042 & &  \\
N & 217.54462 & 41.1407 & 18.947 & 0.026 & 0.275 & 0.063 &
0.284 & 0.057 & 0.296 & 0.048 & &  \\
O & 217.54641 & 41.0450 & 18.500 & 0.016 & 0.169 & 0.056 &
0.210 & 0.043 &  0.236 & 0.057 & &   \\
P & 217.58642 & 41.0746 & 19.373 & 0.027 & 0.371 & 0.088 &
0.401 & 0.059 & 0.403 & 0.064 & &   \\
\hline
\end{tabular}\\
Properties of the galaxies considered as part of the
  environment of J1430+4105. Given in the rows are the object name (Obj),
  the position (RA, Dec), its i band magnitude $\rm mag_{\rm i}$
  and its error $\sigma_{\rm i}$, and its various photometric
  redshift estimates together with its errors: first the template
  method and its error, summarised in \cite{sdss_dr5}, followed by the
neural network based methods CC2 and its error and D1 and its error (\citealt{sdss_photo_z_d1_cc2}).
\label{table:observations:environment:galaxies}
\end{minipage}
\end{table*}


\section{Strong gravitational lensing}
\label{sec:strong lensing}

In this section we model the  lens mass
  distribution with the public {\sc gravlens} (\cite{gravlens}) code
  (Sec. \ref{sec:parametric models:point like}) assuming
point sources and with the {\sc lensview} (\cite{lensview}) code (Sec. \ref{sec:parametric
  models:extended}) using the 2--dimensional surface brightness distribution 
   of the same system. Both approaches give consistent results. We give a
description of the influence of the environment on the lens model of
SDSSJ 1430+4105 in Sec. \ref{sec:lens modelling of the environment}.


\subsection{Parametric modelling using {\sc gravlens}}
\label{sec:parametric models:point like}

{\sc gravlens}\footnote{see http://redfive.rutgers.edu/\textasciitilde keeton/gravlens} is a public code
that uses parametric lens models to reconstruct the properties of an
observed lensing system. 
The lens-modelling we implement here is similar to
the one of \cite{claudio_1538} where the
reader can find more details.
In this subsection, we use peaks in the
surface brightness distribution of the lensed images as point-like position constraints for
the lens model (see Table \ref{lensing:parametric:table:input_pos} and
Fig. \ref{lensing:parametric:image1}). Since the complex surface brightness distribution of the lensed galaxy
makes it difficult to associate reliably a flux measurement to each
multiple image, we neglect flux constraints. 
In {\sc gravlens} the convergence $\kappa$ for a singular
isothermal ellipsoid (SIE) or an ellipsoidal powerlaw (PL) is
parametrized as

\begin{equation}
\kappa(\Theta_1,\Theta_2)=
\frac{\rm
  b^{\beta-1}}{2(1-\epsilon)^{\frac{\beta-1}{2}}\left(\frac{\Theta_{\rm c}^2}{1-\epsilon}+\Theta_1^2+\frac{\Theta_2^2}{\rm
    q^2}\right)^{\frac{\beta-1}{2}}}
\label{lensing:parametric:kappa}
\end{equation}
with 
$$
\epsilon=\frac{1-q^2}{1+q^2} \quad,
$$

where $\rm b$ is the lensing strength, $\beta$ denotes the steepness of the
density profile ($\beta=2$ in the case of an isothermal profile),
$\Theta_1$ and $\Theta_2$ are the coordinates on the plane of the sky
relative to the centre of mass of the lens, $\Theta_{\rm c}$ is the
core radius and
$\rm q$ is the axis ratio of the isocontours of the convergence ($\rm q=1$ for a circular
mass model). In the special case of a
circular lens without core radius, $\rm b$ equals the Einstein radius $\Theta_{\rm Ein}$ of the
lens defined as $\overline{\kappa}({\Theta \le \Theta_{\rm Ein}})=1$.\\
Further we use a de Vaucouleurs Model \citep{deVacouleurs_profile} parametrized as
\begin{equation}
\rm I(\rm R) =\rm I_{\rm e} \rm e^{-7.67 \left[ (\frac{\rm R}{\rm
      R_{\rm eff}})^{1/4} - 1 \right]}\quad,
\end{equation}
with $\rm R_{\rm eff}$ being the effective radius (the radius which
contains half the light) and $\rm I_{\rm e}$ the surface density at this radius.
In {\sc gravlens} this is implemented as 
\begin{equation}
\kappa=\rm b_{\rm deV}\rm e^{-7.67 \left[ \frac{(\Theta_{1}^2+\Theta_{2}^2/q^2)^{1/2}}{\Theta_{\rm eff}}\right]^{1/4}}\quad.
\label{deVauc:massmodel}
\end{equation}

In this parametrisation, $\rm b_{\rm deV}$ is the
  value of the central convergence. The Einstein radius, however,
  depends also on $\Theta_{\rm eff}$ and $q$.
Also a Navarro, Frenk and White (NFW) profile \citep{nfw_profile} is used,
defined as 
\begin{equation}
\rho(\rm r)=\frac{\delta_{\rm c}\rho_{\rm c}}{\rm r / \rm r_{\rm
    s}(1+r/\rm r_{s})^2}\quad,
\end{equation}
with $\rho_{\rm c}$ denoting the critical density of the universe at
the redshift of the lens, and
$\rm r_{\rm s}$ and $\delta_{\rm c}$ are characteristic properties of
the individual halo. For an overview of its lensing properties, see
\cite{brainerd_wright_nfw}.\\

The relation for the LOS projected surface mass density
$\Sigma$ of the lens and lensing convergence $\kappa$ is
$$\kappa=\frac{\Sigma}{\Sigma_{\rm crit}} \quad \mbox {with}\quad
\Sigma_{\rm crit}^{-1}=\frac{4 \pi \rm G}{\rm c^2}\frac{D_{\rm d}D_{\rm
    ds}}{D_{\rm s}}\quad ,$$
where $\rm D_{\rm d}$, $\rm D_{\rm s}$ and $\rm D_{\rm ds}$ are the
angular diameter distances from the observer to the lens, the source
and from the lens to the source, respectively.
The goodness of a model is judged by the $\chi^2$:
\begin{equation}
\chi^2_{\rm lens}=\sum_{\rm i}\frac{\parallel\boldsymbol{\Theta}_{\rm
    i}-\boldsymbol{\Theta}_{0,\rm i}\parallel^2}{\delta_{\Theta_{\rm i}}^2}\quad ,
\label{chi2}
\end{equation}

where $\boldsymbol{\Theta}_{\rm i}$ denote the
  model-predicted positions of the $\rm i$-th images,
$\boldsymbol{\Theta}_{0,\rm i}$ is its observed position, and
  $\delta_{\Theta_{\rm i}}$ its observed positional
uncertainty.

Any priors described in the text are added to this $\chi^2$ in {\sc Gravlens}
via
$$\chi^2_{\rm tot}=\chi^2_{\rm lens}+\chi^2_{\rm prior}$$
with
$$\chi^2_{\rm prior}=\frac{(\rm p - \rm p_{\rm prior})^2}{\sigma_{\rm
    prior}^2}\quad ,$$

where p is the used parameter value, $\rm p_{\rm prior}$ its prior
and $\sigma_{\rm prior}$ its $1\sigma$ error.
Results give best-fitting parameters and their $1\sigma$ errors. The
likelihood of a parameter set is given by $L\propto \rm e^{-\chi^2_{\rm tot}/2}$. In almost
all cases, the best-fitting values are within the 68\% error interval of
the marginalised distributions.

The values of the parameters for the minimum $\chi^2$ models are given in Tables
\ref{lensing:parametric:table:best_fit_sie} and \ref{lensing:parametric:table:best_fit_devauc}. There, we give the model
number, type, the best-fitting parameters of the model together with the
resultant $\chi^2$, the number of degrees of freedom (d.o.f.) and the
reduced $\chi_{\rm red}^2=\frac{\chi^2}{\rm d.o.f.}$ of each
model. Also, the $1\sigma$ error intervals are given.
These error estimates of the parameters are carried out using Monte
Carlo Markov Chains (MCMC) with several thousand steps each. For each
model, 10 chains are calculated with different starting
points. Convergence is reached by comparing the
  variance of the point distribution of
each of this chains with its combined distribution,
see \cite{fadely_q0957}, \cite{gelman95}. From the final
chains, the 2nd half of each chain is combined to the final MCMC point
distribution. The acceptance rate typically
  lies between $0.25$ and $0.3$. We explore potential parameter correlations from these and derive 68\%
confidence intervals\footnote{all given errors in this section are
  the 68\% confidence values of the marginalised distributions, unless
otherwise stated} on the parameters by exclusion of the lowest and highest 16 \%
of the MCMC points' distribution; the
central value is given by the median value of the MCMC points'
distribution, since there are only small deviations between the
median and the average values of the 68\% and 90\% error intervals.\\
We describe the most important different models
  without environment in the following: To check for
    the basic properties of the system, we model the lens as one
    component SIE (Model I) and PL (Model II) model. To derive the de
    Vaucouleurs masses in this lens, we combine a de Vaucouleurs
    component with a dark matter halo model (Model III) and show that
    this result is not significantly affected by also taking the
    environment into account (Models IV, V).

\begin{figure}
\centering
\subfigure[]{\includegraphics[scale=.2]{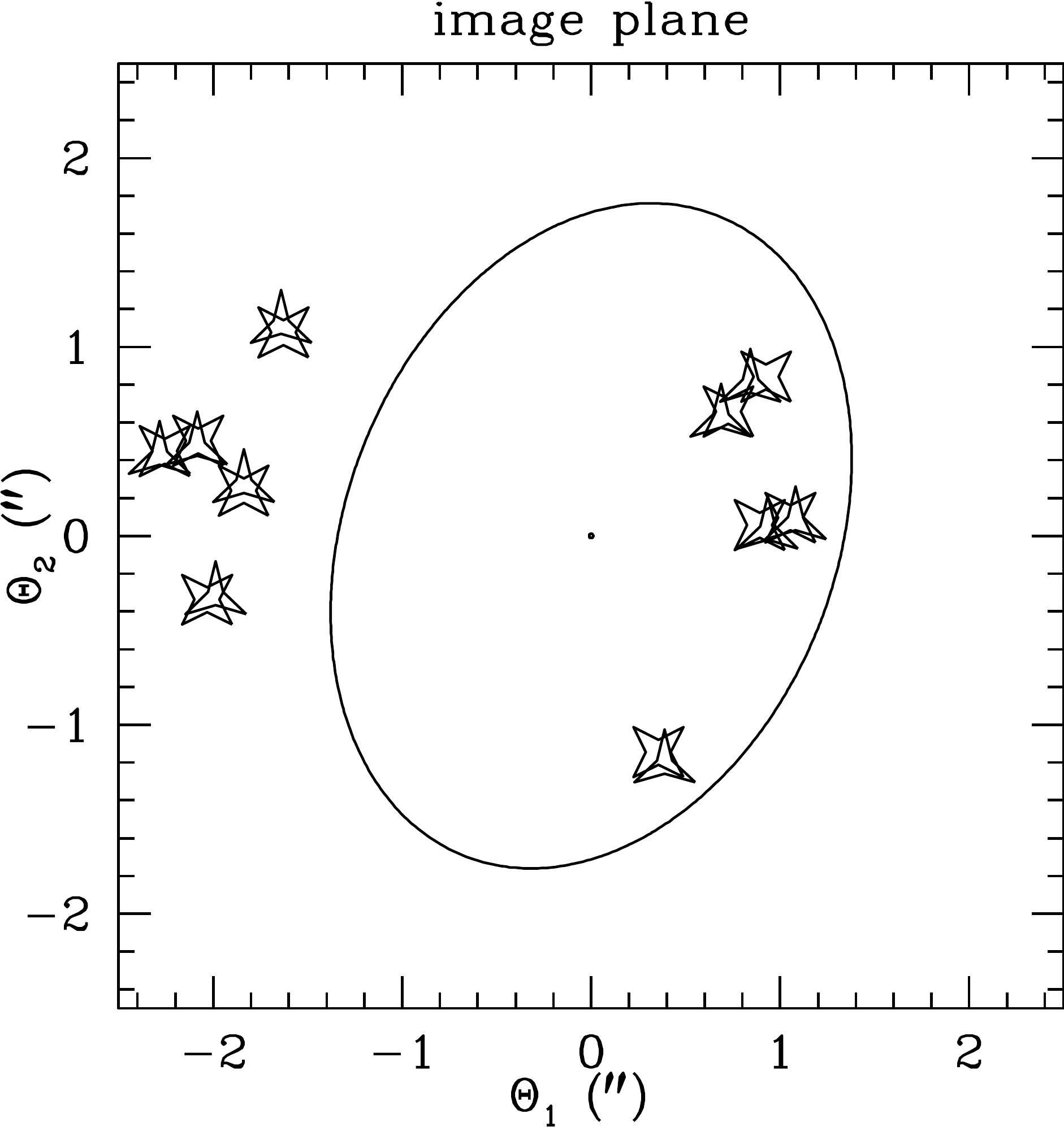}\label{bestfit_NSIE_ns_image_plane}}
\subfigure[]{\includegraphics[scale=.2]{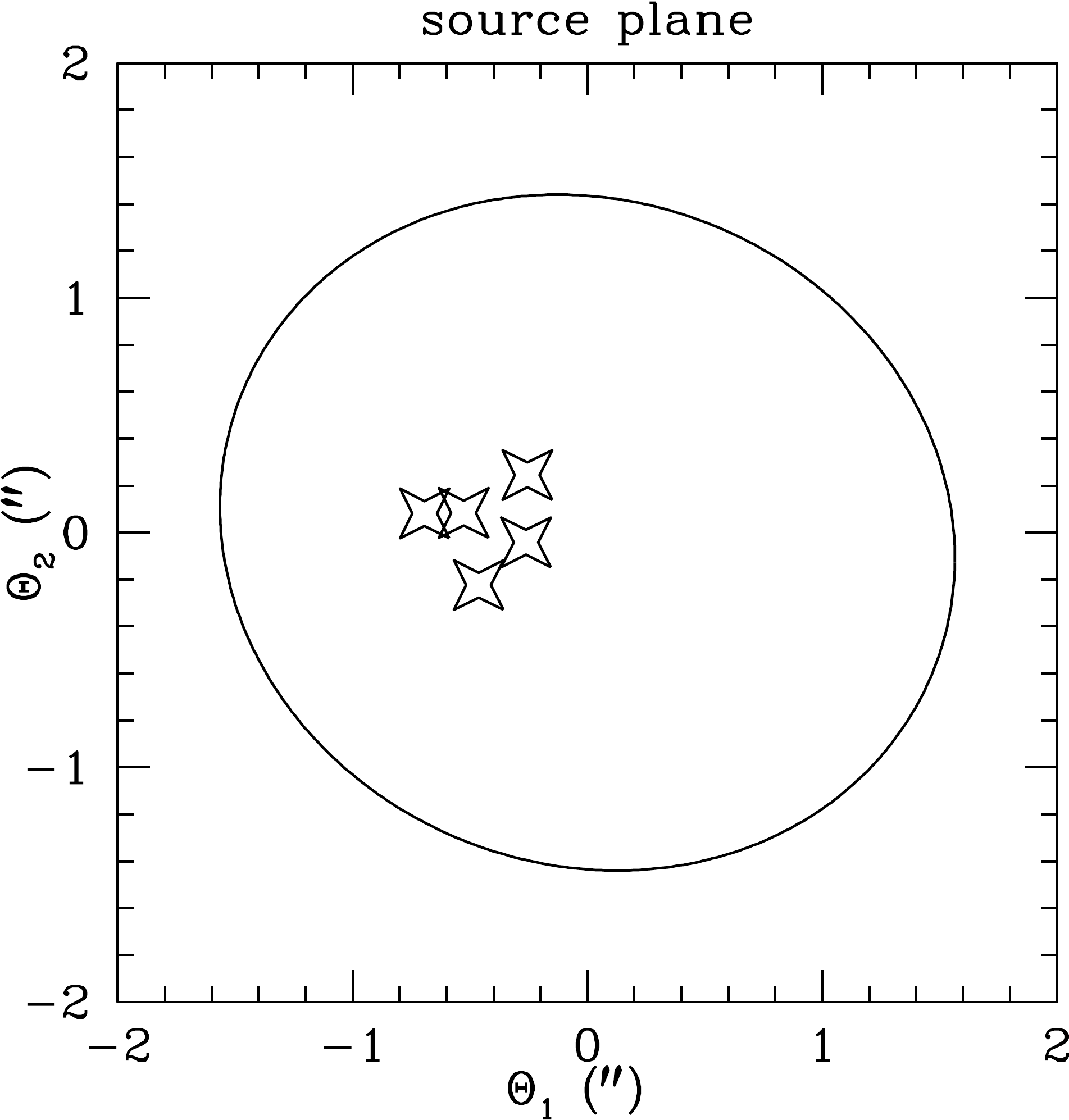}\label{bestfit_NSIE_ns_source_plane}}
\caption{The best-fitting model derived with {\sc gravlens} for the SIE case (Model I) (see Table \ref{lensing:parametric:table:best_fit_sie}), both the image plane and
  the source plane are plotted. On the image plane triangles
mark the input positions, while crosses mark the best-fitting model
positions. On the source plane, the predicted source positions are
plotted. Further, the respective
critical lines (caustics) on the image (source) plane are plotted.}
\label{best_fit_SIE_ns}
\end{figure}

\begin{figure}
\centering
\includegraphics[scale=0.4]{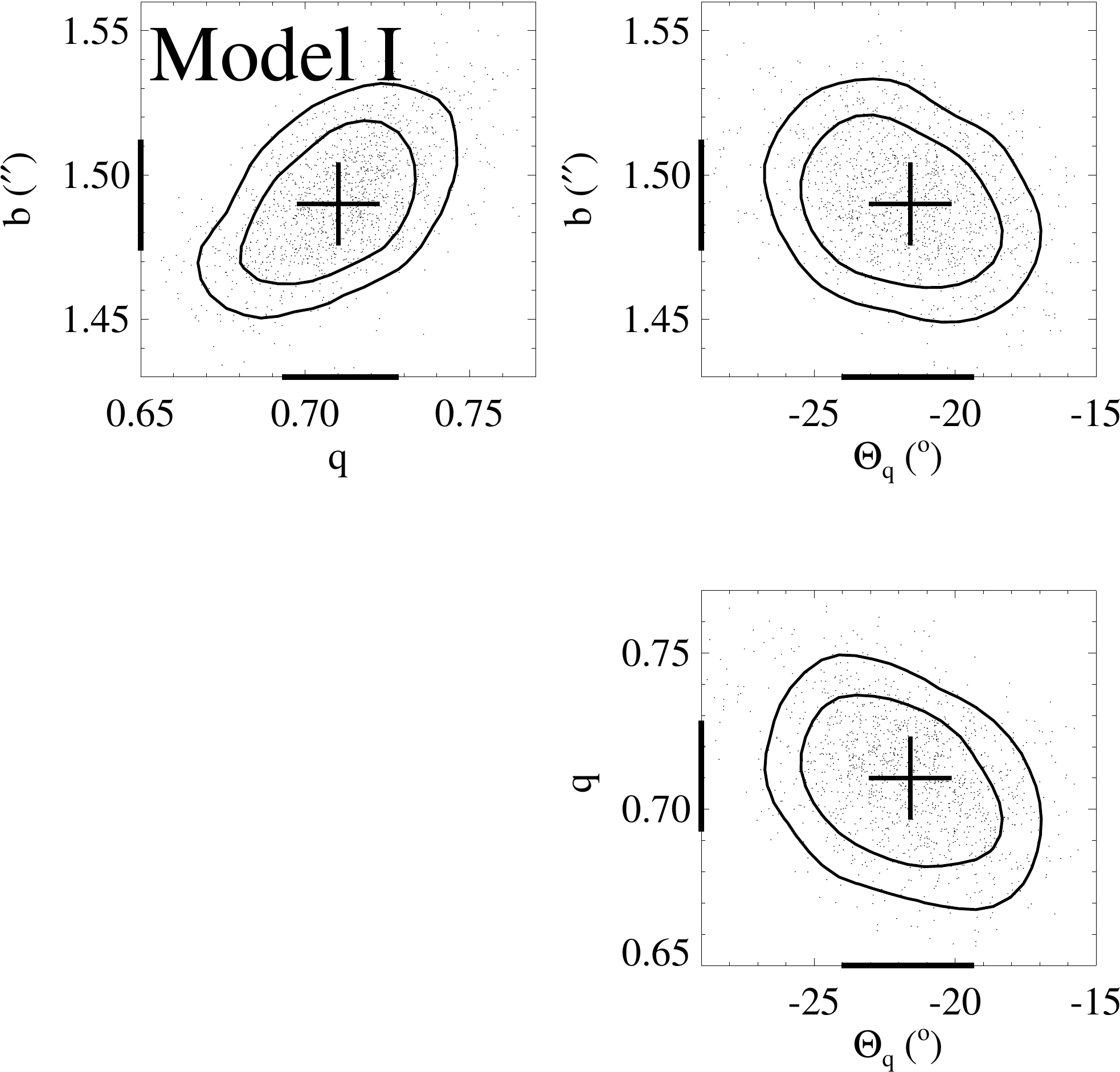}
\caption{Error estimates of the MCMC for the SIE case
  (Model I),
  plotted are the individual points of the MCMC together with the 68
  and 90 \% confidence regions of the distribution, indicated by the
  density contours. The crosses mark
  the minimum-$\chi^2$ value from Table
  \ref{lensing:parametric:table:best_fit_sie}. The bars on the axes mark
  the respective 68 \% marginalised error intervals.}
\label{lensing:parametric:SIE_nsdegeneracy}
\end{figure}

\paragraph*{Model I}:
The lens is modelled as a SIE (Eq. \ref{lensing:parametric:kappa} with
$\beta=2$); the environment of the lens is ignored. The free
parameters of this model are the lensing strength $\rm b$, the axis
ratio $\rm q$, and its position angle $\Theta_{\rm q}$. The
best-fitting model is shown in Fig. \ref{best_fit_SIE_ns}. The results
of the MCMC are
shown in Fig. \ref{lensing:parametric:SIE_nsdegeneracy}. The density
contours describe the probability density for the parameter values, whereas the best-fitting
model is marked with a cross. The reason for the apparent correlation
of $\rm q$ and $\rm b$ in Fig. \ref{lensing:parametric:SIE_nsdegeneracy} lies in the definition of $\kappa$ in Eq. \ref{lensing:parametric:kappa}. 
The marginalised 68\% confidence errors are: $\rm b=(1.49^{+0.02}_{-0.02})\arcsec$,
$\rm q=(0.71^{+0.02}_{-0.02})$ and $\Theta_{\rm q}=(-21.8^{+2.5}_{-2.3})^\circ$.
These values are in very good ($\approx 1\sigma$) agreement with the values derived by
\cite{slacsV} using a similar parameterisation for the lens total mass distribution. 

\begin{figure}
\centering
\includegraphics[scale=0.4]{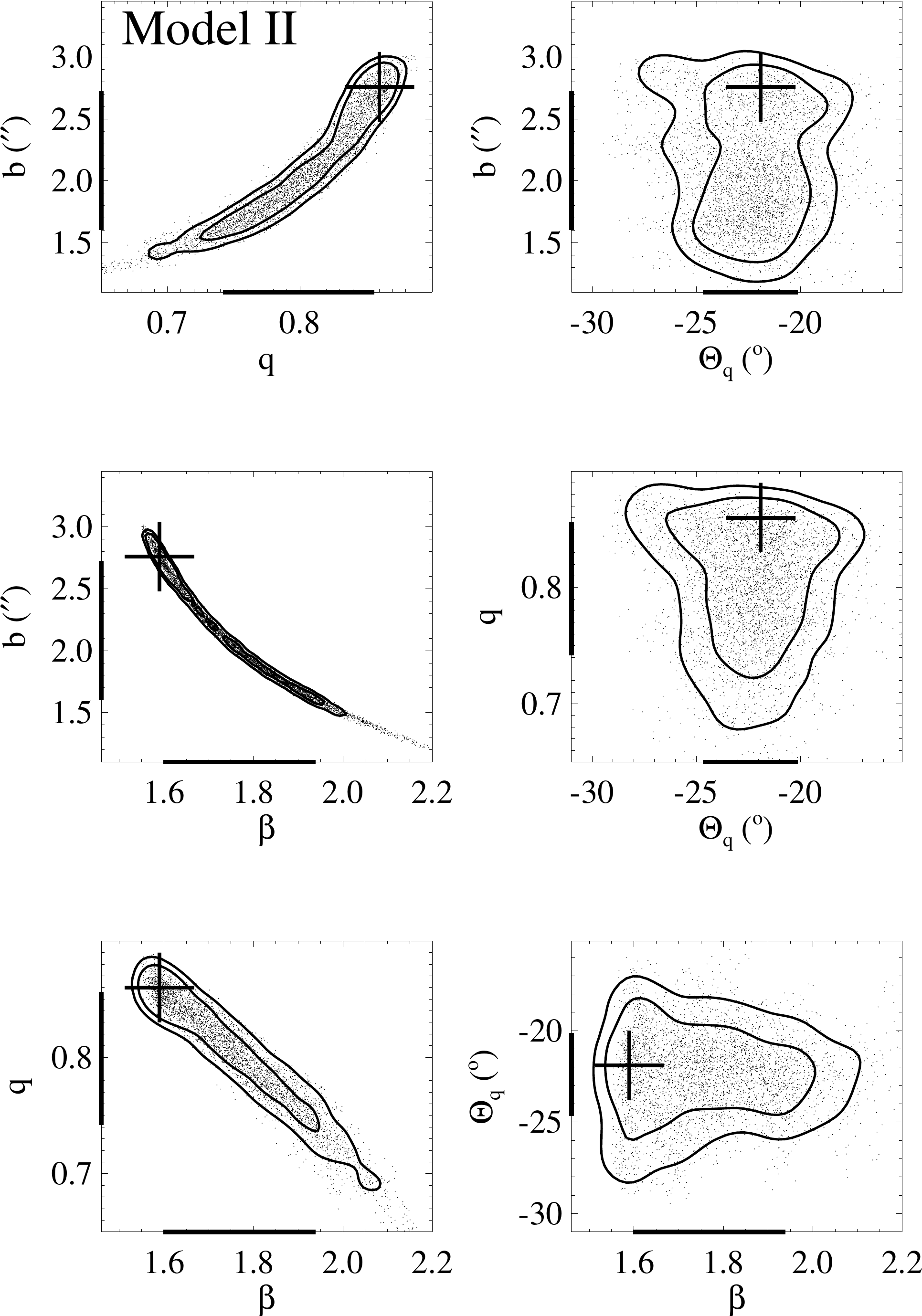}
\caption{Error estimates of the MCMC for the PL case
  (Model II),
  plotted are the individual points of the MCMC together with the 68
  and 90 \% confidence regions of the distribution. The crosses mark
  the minimum-$\chi^2$ value from Table
  \ref{lensing:parametric:table:best_fit_sie}. The bars on the axes mark
  the respective 68 \% marginalised error intervals. }
\label{lensing:parametric:PL_nsdegeneracy}
\end{figure}

\paragraph*{Model II}:
Model II follows a power law (PL) (Eq. \ref{lensing:parametric:kappa} with
arbitrary $\beta$ within the limits $[1,2.7]$), and thus has one more
free parameter relative to Model I. The values for the parameter
distributions are shown in Fig. \ref{lensing:parametric:PL_nsdegeneracy}.
The marginalised distributions change to $\rm
b=(2.12_{-0.52}^{+0.60})\arcsec$, $\rm q=(0.81_{-0.07}^{+0.04})$,
$\Theta_{\rm q}=(-22.2_{-2.5}^{+2.1})^\circ$ and
$\beta=(1.73_{-0.13}^{+0.21})$. We observe again (see Fig. \ref{lensing:parametric:PL_nsdegeneracy}) that the parameters $b$, $q$ and
$\beta$ are correlated with each other. This is entailed by
the definition of the convergence $\kappa$ in
Eq. \ref{lensing:parametric:kappa}.
The steepness parameter $\beta$  is constrained to a value shallower than
isothermal on a $1.3 \sigma$ level. The orientation $\Theta_{\rm q}$ stays at the same angle as in
the SIE case, while its axis ratio moves towards rounder solutions, now being comparable to
the axis ratio of the light distribution. 

\begin{figure*}
\centering
\includegraphics[scale=0.8]{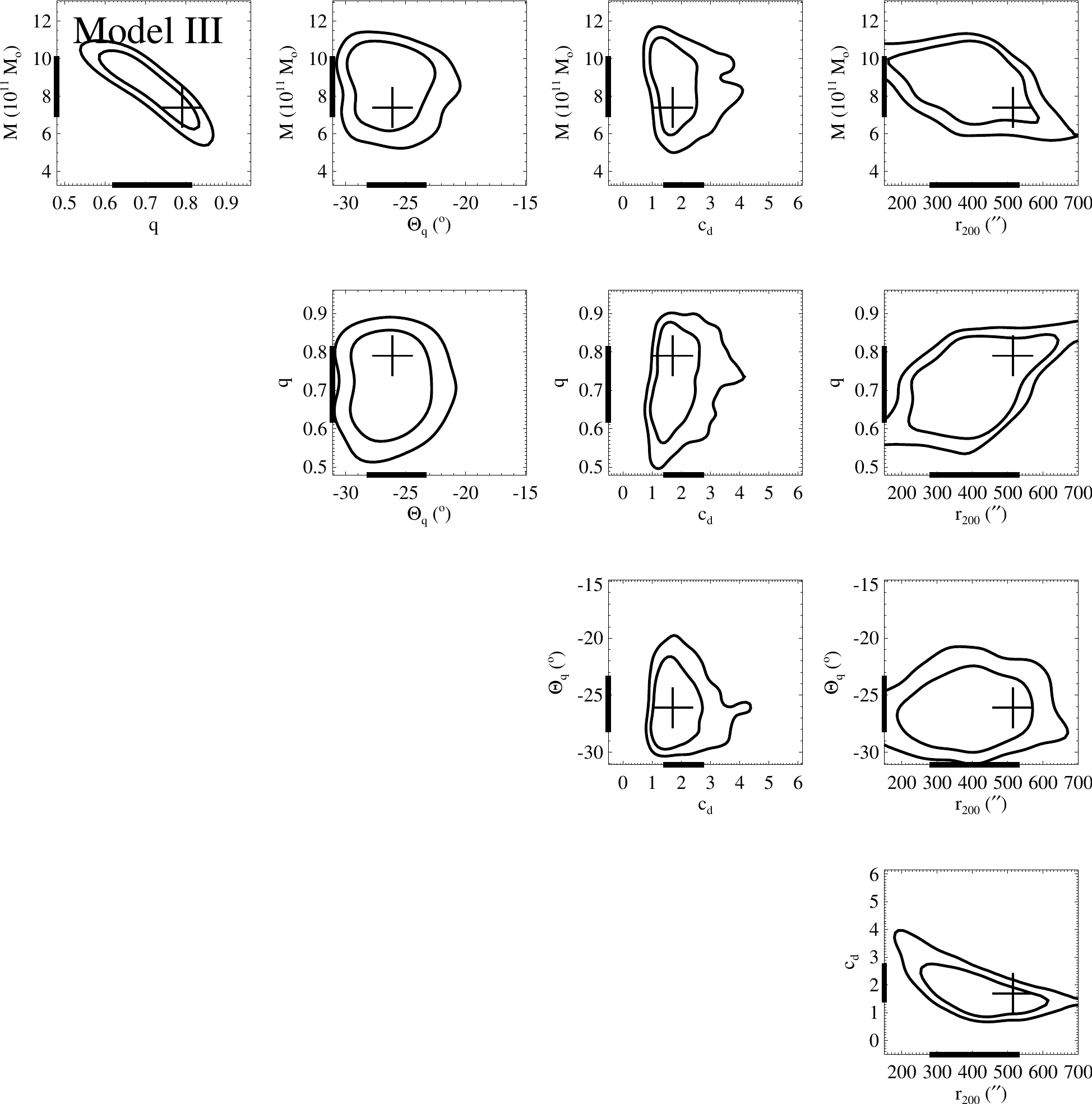}
\caption{Error estimates of the MCMC for the NFW+de Vaucouleurs model
  (Model III),
  plotted are  the 68
  and 90 \% confidence regions of the distribution. The crosses mark
  the minimum-$\chi^2$ value from Table
  \ref{lensing:parametric:table:best_fit_devauc}. The bars on the axes mark
  the respective 68 \% marginalised error
  intervals. The individual points of the MCMC are
    omitted for clarity.}
\label{lensing:parametric:gravlens:NFWdeVauc_ns}
\end{figure*}

\paragraph*{Model III}:
In the following, we split the mass distribution into different
components. We use a de Vaucouleurs like component as traced by the
stellar component and add dark matter with different profiles if
needed. Since the de Vaucouleurs component for galaxy A alone does
  not provide a good model, see Appendix
    \ref{sec:parametric models:additions}, we add a dark matter
component centred at galaxy A. We add an elliptical NFW-like
component to the de Vaucouleurs profile. 
This composition resembles the common picture of galaxy mass distribution.
For the dark matter halo, we impose a prior on the axis ratio based on the
\cite{slacsVII} work of $\rm q_{dark,prior}=(0.79\pm0.12)$. Also we set
the limit of the scale radius to values $<500\arcsec$, approximately 10
times the value we find from Sec \ref{subsec:smooth group mass
  distribution centred at galaxy A} for the scale radius.
The total mass of the de Vaucouleurs component is $\rm
M_{\rm deV}=(8.8_{-1.9}^{+1.3})\times 10^{11} M_{\odot}$ while the
parameters of the dark matter halo are, see
Fig. \ref{lensing:parametric:gravlens:NFWdeVauc_ns}:
$\rm q_{\rm d}=(0.72_{-0.1}^{+0.1})$, $\Theta_{\rm
  q}=(-26.0_{-2.3}^{+2.7})^\circ$, $\rm c_{\rm d}=(1.8_{-0.4}^{+1.0})$ and $\rm
  r_{200}=(406_{-129}^{+128})\arcsec$. We note that there is some degeneracy
between the concentration c and $\rm r_{200}$. Further we have no
constraints on $\rm r_{200}$ from the data, since we do not have observables
  outside $2.32\arcsec$. Using a NSIE-like dark
    matter component yields similar results, as described in Appendix \ref{sec:parametric
models:additions}, Model IIIb.

\begin{table*}
\centering
\begin{minipage}{290mm}
\caption{minimum-$\chi^2$ values and parameter estimates derived with
  {\sc gravlens} for the
  isothermal and powerlaw models}
\begin{tabular}{cccccccccc}
\hline
& & $b$& $q$ & $\Theta_{\rm q}$ & $\beta$  &
    $\chi^2$ &d.o.f.& $\frac{\chi^2}{\rm d.o.f}$\\
& &($\arcsec$)& & ($^{\circ}$)&  & & \\
\hline
Model I &SIE  & 1.49 & 0.71 & -21.6 & 2.00\footnotemark[1]   & 
11.5 & 7 & 1.6 \\
 & & 1.47 -- 1.51 & 0.69 -- 0.73 & -24.1 -- -19.3 &  & & & \\
Model II &PL & 2.76 & 0.86 & -21.9 &  1.59    & 10.1 &
6 & 1.7 \\
 & & 1.60 -- 2.72 & 0.74 -- 0.85 & -24.7 -- -20.1 & 1.60 -- 1.94  & &
& \\
\hline
\label{lensing:parametric:table:best_fit_sie}
\end{tabular}
\end{minipage}
\footnotemark[1]{fixed value}
\end{table*}
\begin{table*}
\centering
\begin{minipage}{290mm}
\caption{minimum-$\chi^2$ values and parameter estimates derived with
  {\sc gravlens} for the two component
   de Vaucouleurs plus dark matter models}
\begin{tabular}{ccccccccccc}
\hline
& & $\rm M$ & $q_{\rm d}$ & $\Theta_{\rm q,d}$ & $\rm
c_{\rm d}$ & $\rm r_{200}$  & $\rm b_{\rm group}$
& $\chi^2$ & d.o.f. & $\frac{\chi^2}{\rm d.o.f}$\\
& &$\left (\rm 10^{11}M_{\odot}\right )$ & & ($^{\circ}$) &($\arcsec$) & & & \\
\hline
Model III & deVauc+NFW & 7.4 & 0.79 & -26.1 & 1.7 & 514& & 7.9&
5& 1.6\\
& & 6.9 -- 10.1 & 0.62 -- 0.82 & -28.3 -- -23.3 & 1.4 -- 2.8 & 277 -- 534 &
& & &\\
Model IV & deVauc+GI & 13.5&  &  & & & 8.0  & 18.9 & 8
&2.4\\
& & 13.3 -- 13.7 & & & & & 7.3 -- 8.7 & & \\
Model V & deVauc+NFW+GI & 9.5 & 0.79 & -24.6 & 2.2 & 280 & 3.7& 7.8&
4& 2.0\\
& & 8.7 -- 11.8 & 0.68 -- 0.90 & -27.1 -- -12.8 & 1.0 -- 2.6 & 168 -- 463 &
3.2 -- 6.5 &\\
\hline
\label{lensing:parametric:table:best_fit_devauc}
\end{tabular}
\end{minipage}
\end{table*}

\subsection{Lens modelling of the environment}
\label{sec:lens modelling of the environment}

As mentioned before, this galaxy is not an
  isolated field galaxy, hence we investigate the possible impact on
  the derived lens parameters by taking the environment into account.
In the following, we centre a smooth group
  contribution at galaxy I and calculate its convergence and shear at
  the position of SDSSJ 1430+4105. Further modelling of the group
  contribution by summing up the contributions of the individual
  members (``clumpy group''), and by centreing it at galaxy A itself
  is discussed in Appendix \ref{sec:environment:clumpy group nfw}.

\subsubsection*{Smooth group mass distribution centred at galaxy I}
\label{subsec:smooth group mass distribution centred at galaxy I}

According to \cite{rozo}, we can transform the group richness into a group mass of
$\rm M_{500}=(0.72\pm0.29)\times10^{14}\rm M_{\odot}$
within $1\sigma$. This mass can
be converted into a velocity dispersion of
$\sigma_{\rm group}=519 \pm 107 \rm km \, \rm s^{-1}$, using the critical
density of the universe \[500\rho_{\rm c}(\rm z)=\frac{3 \rm
  M_{500}}{4 \pi \rm r_{500}^3}\]
 and the singular isothermal sphere equation:
\[\rm M_{500}=\rm M(\rm r_{500})=:\frac{2 \sigma^2_{\rm group}\rm r_{500}}{\rm G}.\]
There $\rho_{\rm c}(\rm z)$ denotes the critical density of the
universe at redshift z and $\sigma_{\rm group}$ the velocity dispersion of the group.
Subsequently this gives an Einstein radius of $\Theta_{\rm E}=3.6 \pm
1.5\arcsec$, using again a SIS assumption (see Sec.
\ref{sec:parametric models:point like} for details). 
\color{black}
This results in a
convergence of and shear of
\begin{equation}
\begin{aligned}
\kappa_{\rm A}^{\rm SIS\, group} = 0.029\quad,\\
\gamma_{\rm A}^{\rm SIS\, group} = 0.029\quad \,
\label{smooth_group_I_SIS}
\end{aligned}
\end{equation}
at galaxy A if galaxy I is assumed to be the group centre.\\
Alternatively, we model the smooth group as a 'typical' richness 12
galaxy group NFW (\cite{nfw_profile}) halo with concentration $\rm c=4.22$ and
$\rm r_{200}=848\, \rm kpc$ from
\cite{Johnston_2007}.
 We obtain a convergence and shear of
\begin{equation}
\begin{aligned}
\kappa_{\rm A}^{\rm NFW\, group} = 0.025\quad, \\
\gamma_{\rm A}^{\rm NFW\, group} = 0.026\quad.
\label{smooth_group_I_NFW}
\end{aligned}
\end{equation}
Further, we note that the angle of A towards I is $-26^{\circ}$,
therefore forming an angle of $16^{\circ}$  with the external shear
value derived in Sec. \ref{subsec:clumpy group}.
We examine the HST and SDSS frames which cover
  galaxy I and its vincinity for group--scale multiple images to further
  constrain the group mass distribution but do not find any sign for strong lensing.

\begin{figure}
\centering
\includegraphics[scale=0.2]{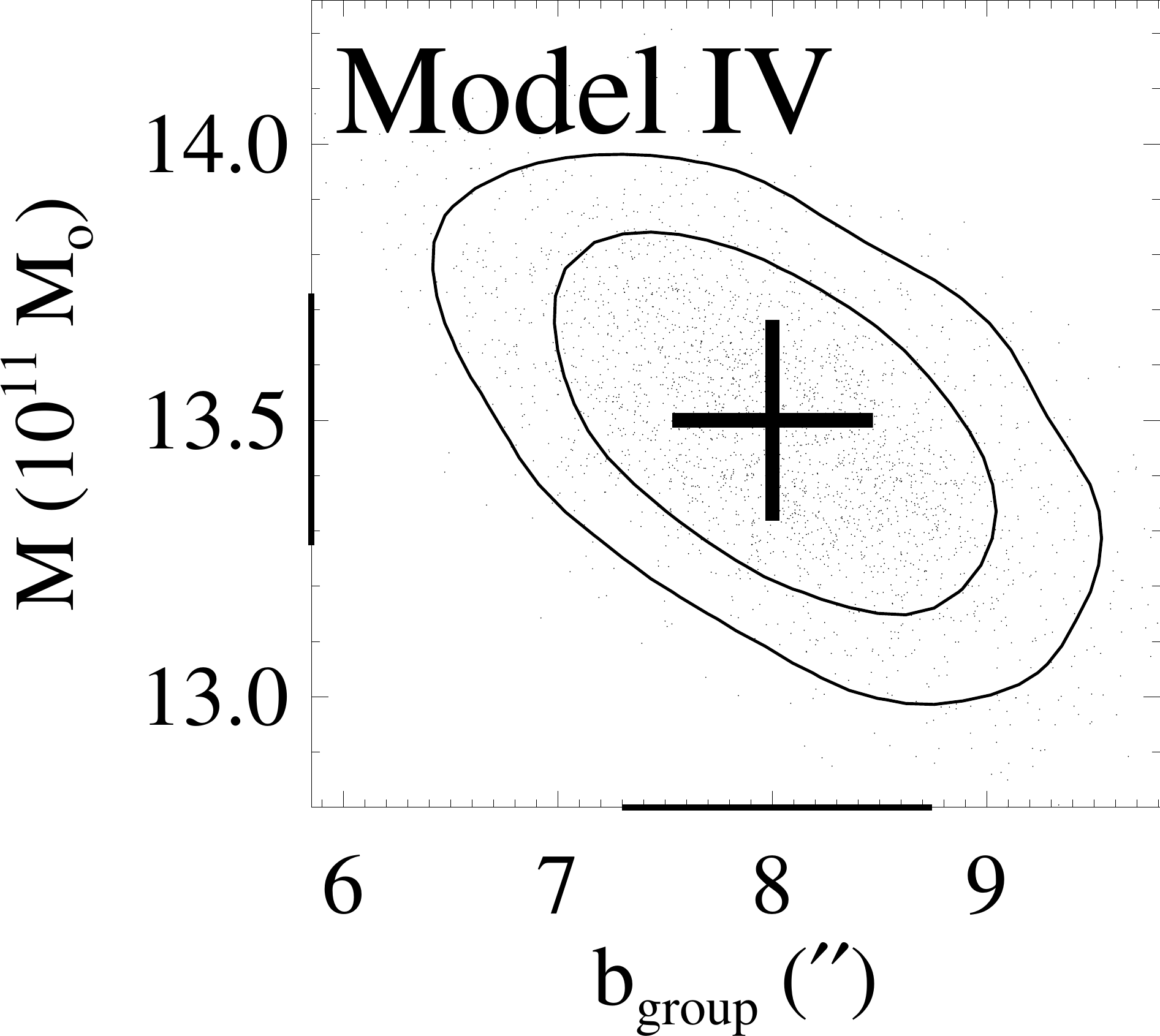}
\caption{Error estimates of the MCMC for the de Vaucouleurs+GI model
  (Model IV),
  plotted are the individual points of the MCMC together with the 68
  and 90 \% confidence regions of the distribution, indicated by the
  density contours. The crosses mark
  the minimum-$\chi^2$ value from Table
  \ref{lensing:parametric:table:best_fit_devauc}. The bars on the axes mark
  the respective 68 \% marginalised error intervals.}
\label{lensing:parametric:deVauc_ws}
\end{figure}

\paragraph*{Model IV}:
From Section \ref{sec:observed environment} we expect that there is some
environment dark matter present in this
galaxy. We check whether using this group
  dark matter contribution with a de Vaucouleurs component for galaxy
  A is sufficient to explain the observations, even though modelling this
system with a pure de Vaucouleurs component fails, see Appendix \ref{sec:parametric models:additions}.
Therefore, in this model, we combine the de Vaucouleurs profile with a
group halo centred at galaxy I. To account for the environment, we include the galaxy group explicitly as a SIS
profile centred at galaxy I in
Table \ref{table:observations:environment:galaxies}.
We use a prior on the group Einstein radius of $\rm b_{\rm
  group,prior}=(3.6 \pm 1.5) \arcsec$. The de Vaucouleurs component has shape
parameters as stated in Table
\ref{observations:table:slacs_values}. The group acts almost as a mass sheet.
We get a
$\chi^2=18.9$ for the best-fitting model. We get parameter estimates of
$\rm M_{deV}=(13.5_{-0.2}^{+0.2})\times 10^{11}M_{\odot}$ and $\rm b_{\rm
  group}=(8.0_{-0.7}^{+0.7})\arcsec$  as can be seen in Fig. \ref{lensing:parametric:deVauc_ws}.
Besides being a worse fit than most of the other models, this model
also needs a much more massive group present than what is likely from
the observations. Therefore, dark matter that is distributed almost
uniformly within $\Theta_{\rm E}$ of the galaxy does not provide a
good model for the system.

\begin{figure*}
\centering
\includegraphics[scale=0.8]{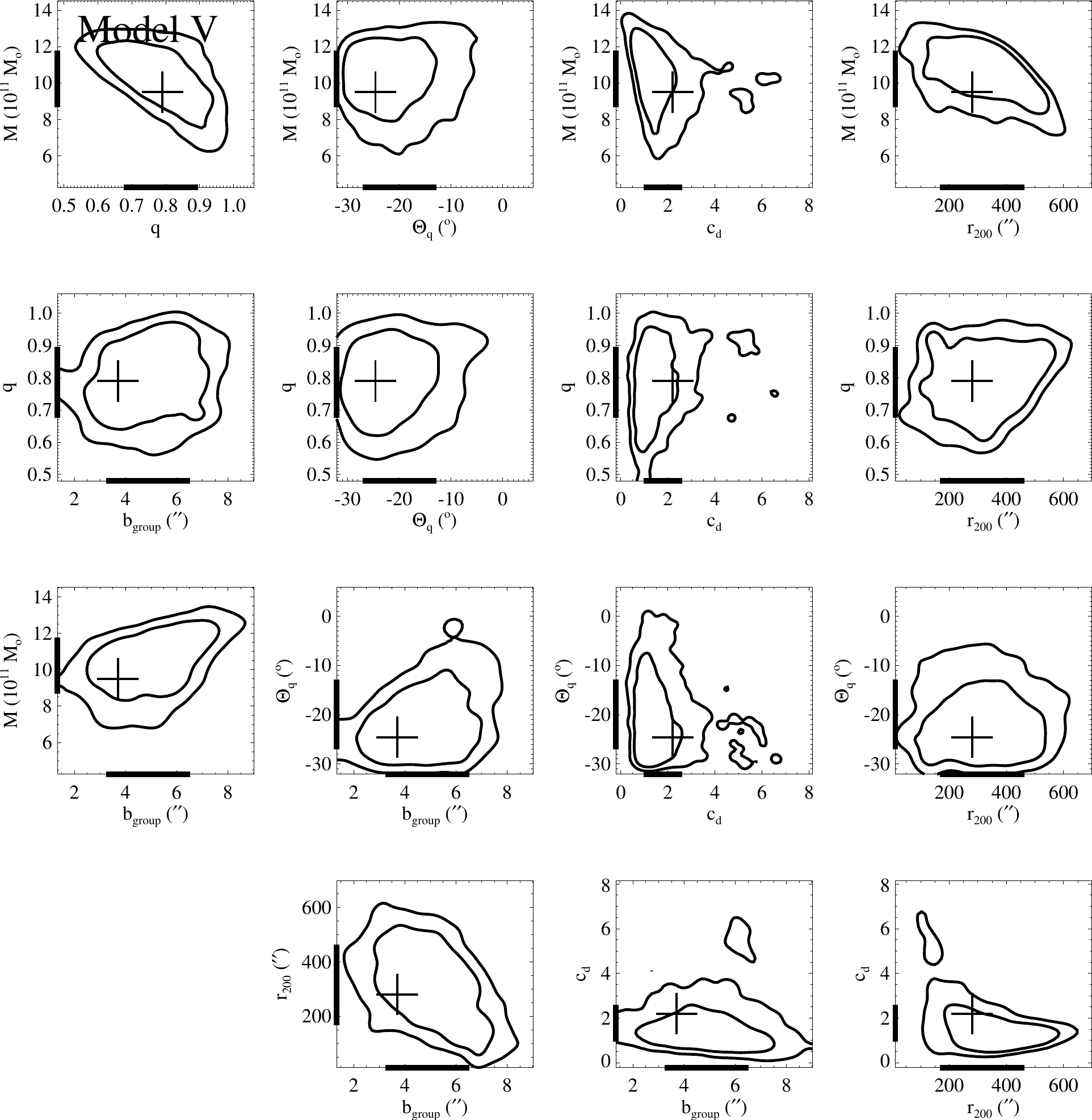}
\caption{Error estimates of the MCMC for the de Vaucouleurs+NFW model
  (Model V), plotted are the 68
  and 90 \% confidence regions of the distribution. The crosses mark
  the minimum-$\chi^2$ value from Table
  \ref{lensing:parametric:table:best_fit_devauc}. The bars on the axes mark
  the respective 68 \% marginalised error
  intervals. The individual points of the MCMC are
    omitted for clarity.}
\label{lensing:parametric:gravlens:NFWdeVauc_ws}
\end{figure*}

\paragraph*{Model V}:
This model adds environmental effects to Model III.
Therefore we add the group GI explicitly as above,
yielding 3 components: the group GI, the dark
matter associated with the galaxy as an elliptical NFW profile and a stellar
component modelled as a de Vaucouleurs profile. We use the same
constraints as for Model III. We get the following parameters, see also
Fig. \ref{lensing:parametric:gravlens:NFWdeVauc_ws}: $\rm
  M_{\rm deV}=(10.4_{-1.7}^{+1.4})\times 10^{11} M_{\odot} $, $\rm q_{\rm d}=(0.79_{-0.11}^{+0.11})$, $\Theta_{\rm
    q}=(-21.6_{-5.4}^{+8.8})^\circ$, $\rm c_{\rm
    d}=(1.4_{-0.4}^{+1.2})$, $\rm
  r_{200}=(321_{-153}^{+141})\arcsec$, and for the galaxy group $\rm b_{\rm
    group}=(4.9_{-1.7}^{+1.6})\arcsec$.
We note that these parameter estimates do not significantly change
compared to Model III, therefore the inclusion of group GI has only a small influence
on the estimated galaxy parameters; the $\rm M_{deV}$ for the de Vaucouleurs component is slightly
  increased. Again, we are not able to
constrain the concentration c or $r_{200}$ of the dark matter
halo. Models Va and Vb in Appendix \ref{sec:parametric
    models:additions} employ a NSIE-like galaxy dark matter halo
(Model Va) and an external shear contribution
  instead of a explicit group contribution (Model Vb) and again give
results very similar to Model V regarding the
  parameters for the lensing galaxy.

\subsection{Full surface brightness distribution using {\sc lensview}}
\label{sec:parametric models:extended}

We also use {\sc lensview} (\cite{lensview}) to derive models and
mass estimates for SDSSJ 1430+4105 and to reproduce the
full surface distribution of the lensed galaxy and its unlensed
source. {\sc Lensview} is a publicly
available program that fits parametric lens models to image data and
uses the best-fitting lens model to reconstruct the source and image.
The code uses the image data, a corresponding noise map, and an image
mask to minimise $\chi^2-\lambda S$, where $\chi^2$ is the chi square
difference between the reconstructed image and the data, $S$ is the
entropy in the source, and $\lambda$ is internally adjusted such that
$\chi^2$ approaches its target value. If the data are well fit by the
model, the entropy term serves to smooth the source. Because the flux
of each unmasked data pixel is used in the fit, {\sc lensview} is
well-suited to systems with extended flux like SDSS J1430+4105.
The profile used here is defined, following \cite{Barkana_SPEMD},  as

\begin{equation}
\kappa(\Theta_1,\Theta_2)=\frac{b^{\prime}}{2}\left(\frac{3-\beta}{q}\right)^{\frac{\beta-1}{2}}(\Theta_1^2+\frac{\Theta_2^2}{q^2})^{\frac{1-\beta}{2}}\, ,
\label{lensing:parametric:lensview:PL_kappa}
\end{equation}

where $b^{\prime}$ gives the Einstein radius, $q$ the axis ratio and $\beta$
again the power law exponent of the profile. We note that the
normalisation of the profiles is different from Eq. \ref{lensing:parametric:kappa}, resulting in different
values for the Einstein radius in both approaches.

The minimum-$\chi^2$ results are stated in Table
\ref{lensing:lensview:best_fit_results}. 
The SIE best-fitting parameter values derived here agree with those found
in Sec. \ref{sec:parametric models:point like}, when directly
compared to Model I in Table \ref{lensing:parametric:table:best_fit_sie}. For the PL
model, we see a consistency of the different models from {\sc gravlens} and
{\sc lensview} within the stated errors for $q$,
$\Theta_{\mbox{q}}$ and $\beta$. Since, as mentioned before, the
normalisation of the convergence profiles is different, the $b$ / $b^{\prime}$ values
do not compare directly to each other.
For the models including the environment, the direct comparison of the
SIE+ES model with
Model Ia shows again a consistency within the errors derived in Appendix \ref{sec:parametric models:additions} for the lens
parameters. However, the external shear angle shows a discrepancy, the
angle is offset relative to the expected value derived in Sec. \ref{sec:observed
  environment}. Since the external shear contributes no mass, this will
not have a significant effect on the mass estimates in Sec.
\ref{Sec:mass}. The same is true for the PL+ES case, the comparison
with Model IIa gives an agreement within the given errors in all
parameters besides $\Theta_{\gamma}$.

\begin{table}
\centering
\caption{minimum-$\chi^2$ values derived with {\sc lensview}}
\begin{tabular}{ccccccccc}
\hline
&$b$ & $\mbox{q}$ & $\Theta_{\mbox{q}}$ & $\gamma$\footnotemark[1]
&$\Theta_{\gamma}$ & $\beta$ & $\chi^2_{\mbox{red}}$\\
&($\arcsec$)& & ($^{\circ}$)& & \\
\hline
SIE  & 1.49 & 0.69 & -19.5 & 0\footnotemark[2]& 0\footnotemark[2] & 2.00\footnotemark[2] & 1.4
\\
SIE+ES & 1.45 & 0.80 & -23.0 &
0.046 & -105 & 2.00\footnotemark[2] & 1.02 \\
PL  & 1.53 & 0.77 & -20.2 & 0\footnotemark[2] & 0\footnotemark[2] & 1.83 & 1.02
 \\
PL+ES & 1.50 & 0.85 & -22 &
0.047 & -106 & 1.89 & 0.99 \\
\hline
\end{tabular}\\
\label{lensing:lensview:best_fit_results}
\footnotemark[1]{The external shear at the position of the galaxy A}\\
\footnotemark[2]{fixed value}
\end{table}

\begin{figure}
\subfigure[]{\includegraphics[height=4cm]{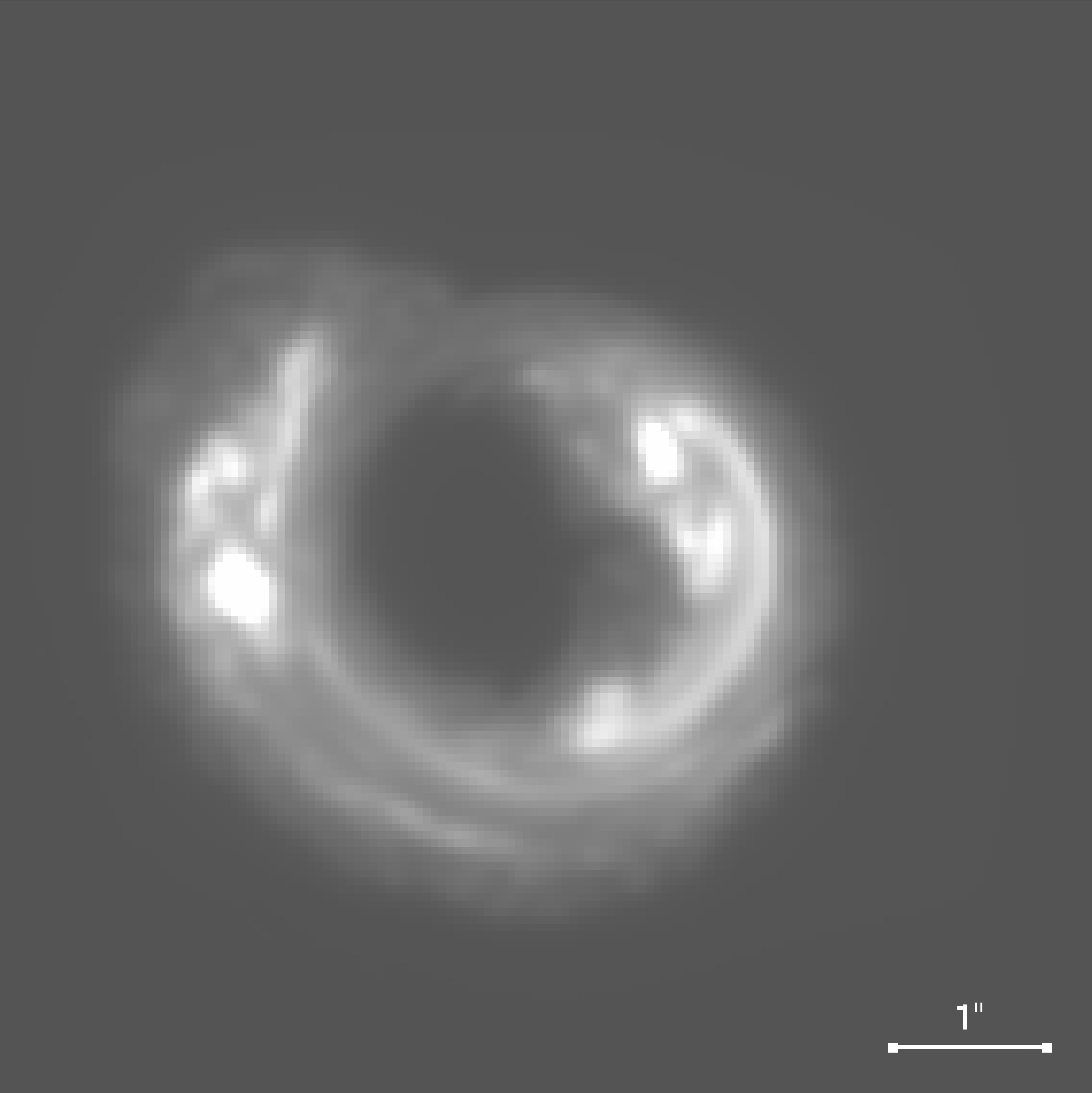}\label{lensing:lensview_best_image}}
\subfigure[]{\includegraphics[height=4cm]{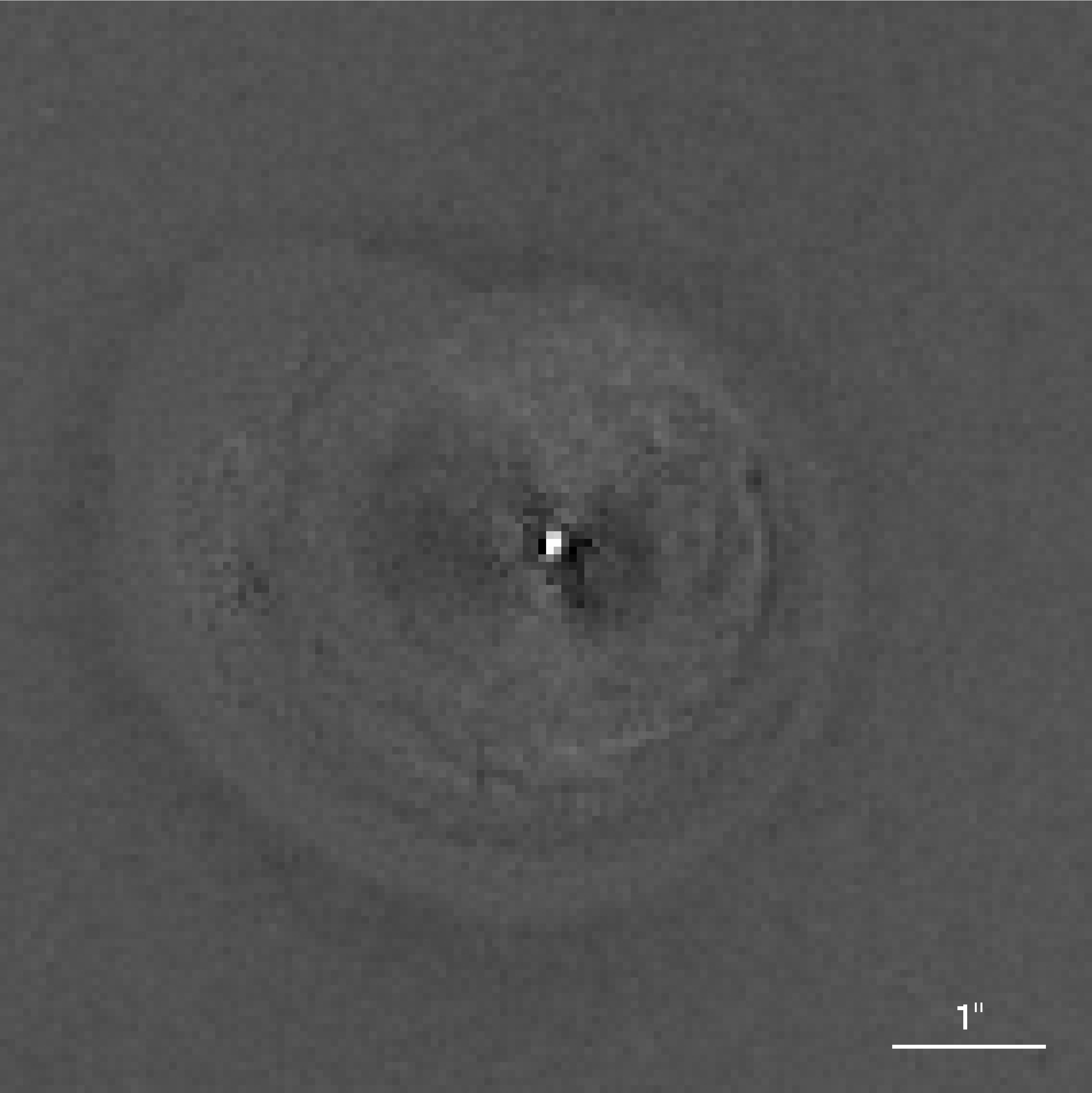}\label{lensing:lensview_best_resid}}
\subfigure[]{\includegraphics[height=4cm]{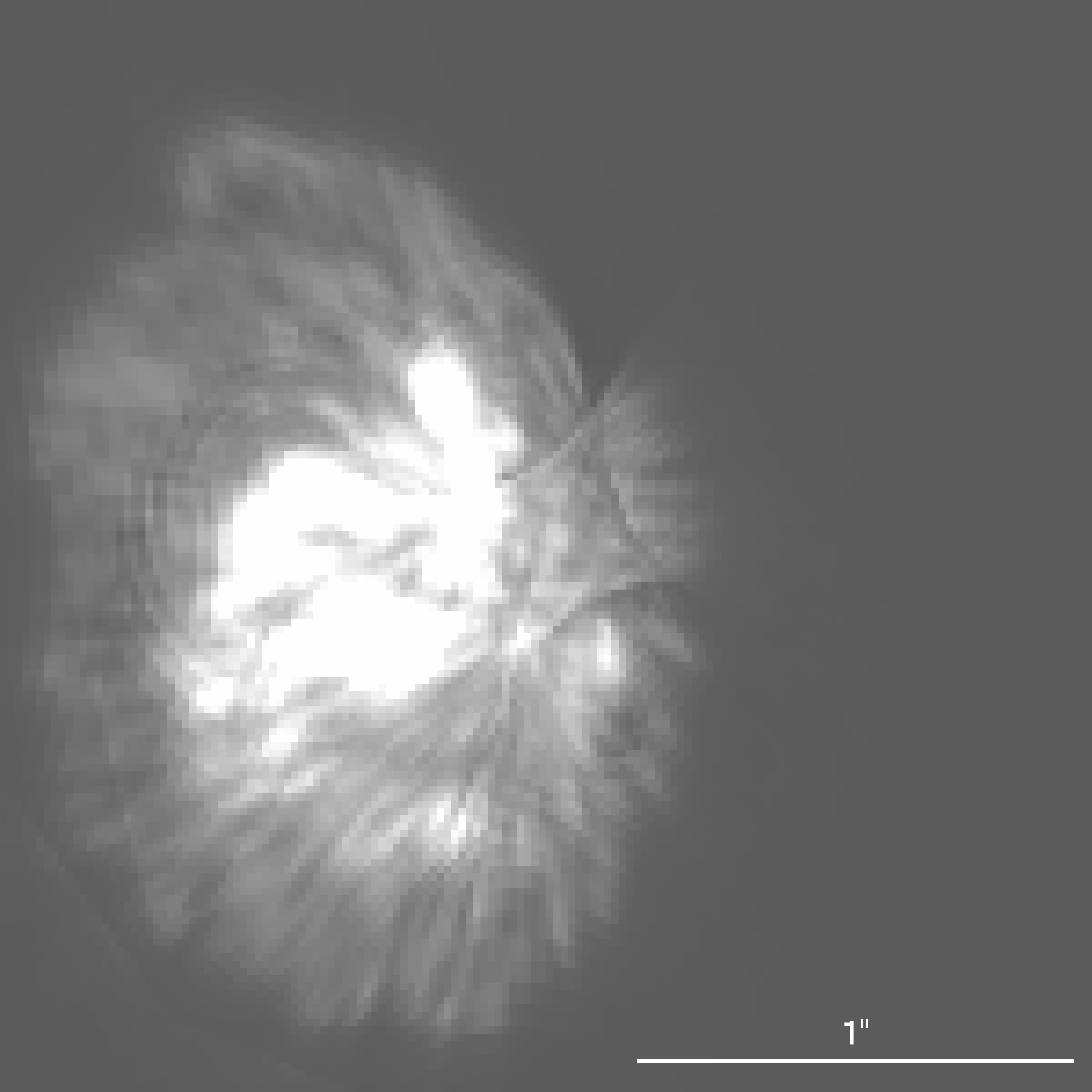}\label{lensing:lensview_best_src}}
\caption{best-fitting model for the SIE case
  \ref{lensing:lensview_best_image}, its residuals
  \ref{lensing:lensview_best_resid} and the corresponding source
  \ref{lensing:lensview_best_src}. For the images
  \ref{lensing:lensview_best_image} and
  \ref{lensing:lensview_best_resid}, the pixel size is the same as in
  Fig. \ref{lensing:parametric:image1};  the source has a 3 times
  smaller pixel size. The same flux scaling has been used on all
  images.}
\label{lensing:lensview_best_fit_sie}
\end{figure}


\subsection{Tests on the strong lensing assumptions}
\label{subsubsection:strong lensing model tests}
For all but the single, pure de Vaucouleurs  model, the centre of the light and mass
distribution do not necessarily have to coincide. To address possible
systematic effects of the assumption that the mass model profiles are
centred on the centre of light, we use Model I with a free centre with priors on the centre
position of uncertainty
$\sigma_{\Theta_{1,\rm l}}=\sigma_{\Theta_{2,\rm l}}=0.2\arcsec$.
In this case, we find: $\rm b=(1.51^{+0.02}_{-0.02})\arcsec$,
$\rm q=(0.69^{+0.06}_{-0.04})$ and $\Theta_{\rm
  q}=(-26.1^{+2.3}_{-2.4})^\circ$. The lens total mass centre moves to
$\Theta_{1,\rm l}=(-0.02^{+0.05}_{-0.07})\arcsec$ and $\Theta_{2,\rm l}=(-0.11^{+0.04}_{-0.04})\arcsec$. The
lensing strength b and the axis ratio q agree within the errors with
the values derived for Model I, but the uncertainty on q
increases. While the lens position $\Theta_{1,\rm l}$ is
still consistent with 0, there is a dependency of $\Theta_{2,\rm l}$ and
$\Theta_{\rm q}$. This has no effect on the mass profiles $\rm M(<R)$,
so we conclude that we can fix the
lens total mass centre to its centre of light without inducing
systematic effects on the derived masses.\\
In the previous sections we restricted the smooth
  mass model of the group GI to be spherically symmetric.
We alter the shape of GI to an elliptical isothermal profile (SIE) and
study the changes on the best-fitting SIE lens
models. This model is analogous to Model Ia, but
  combines a SIE lens with a SIE instead of a SIS galaxy group centred
  on galaxy I. Therefore, we create
2000 random representations of GI as a SIE.
For each of these representations, the axis ratio is randomly chosen between
$\rm q_{\rm GI}=[0.6,1]$ with a random orientation. The centre of this
SIE group model is taken from a gaussian distribution, centred on galaxy I, with
a width of $\sigma_{\rm x,GI}=\sigma_{\rm y,GI}=1.5\arcsec$.
We recalculate the best-fitting parameters for each of
  these SIE+GI(SIE) models and evaluate the scatter of the best-fitting
  parameters to estimate the systematic uncertainties emerging from
  the assumption of a simple SIS group halo.   
This results in  68\% c.l. distributions of the best-fitting parameters
for this modified Model Ia: $\rm b_{\rm lens}=[1.44,1.46] \arcsec$,
$\rm q_{\rm lens}=[0.79,0.82]$, $\Theta_{\rm q}=[-18.4,-16.2]^{\circ}$ and
$\rm b_{\rm group}=[4.27,4.42]\arcsec$. These intervals are small
compared to the statistical uncertainties for Model Ia derived from
the MCMC sampling. We conclude that the details of the group model
representation do not change the results for the lensing galaxy
significantly, therefore including the most simple SIS model for GI is
sufficient.


\section{Results for the galaxy mass profile}
\label{Sec:mass}
The total masses $\rm M(<R)$ within a cylinder of radius R
  and its derivatives  obtained from the SL analysis for Models I to
  V are shown in Figs. \ref{lensing:parametric:gravlens:masses}
  and \ref{lensing:parametric:gravlens:masses_devauc}. We have
  calculated these values within several concentric apertures with radii ([0.46 0.92 1.26
1.59 1.95 2.32 2.78] $\arcsec$), chosen 
to lie in the radial regions covered by the lensed images plus 
extrapolations towards smaller / larger radii. 
For the Models I to V of
Secs. \ref{sec:parametric models:point like} and
\ref{sec:lens modelling of the environment}, the
masses are estimated by randomly taking 1000 MCMC points and
creating convergence maps for each one of these 1000 models. The 68 \%
(90 \%) errors are estimated by
taking the central 680 (900) models at each radius.


\subsection{Mass profiles for the single component isothermal and powerlaw models}
\label{subsec:masses:isothermal}

\begin{figure}
\centering
\includegraphics[scale=0.42]{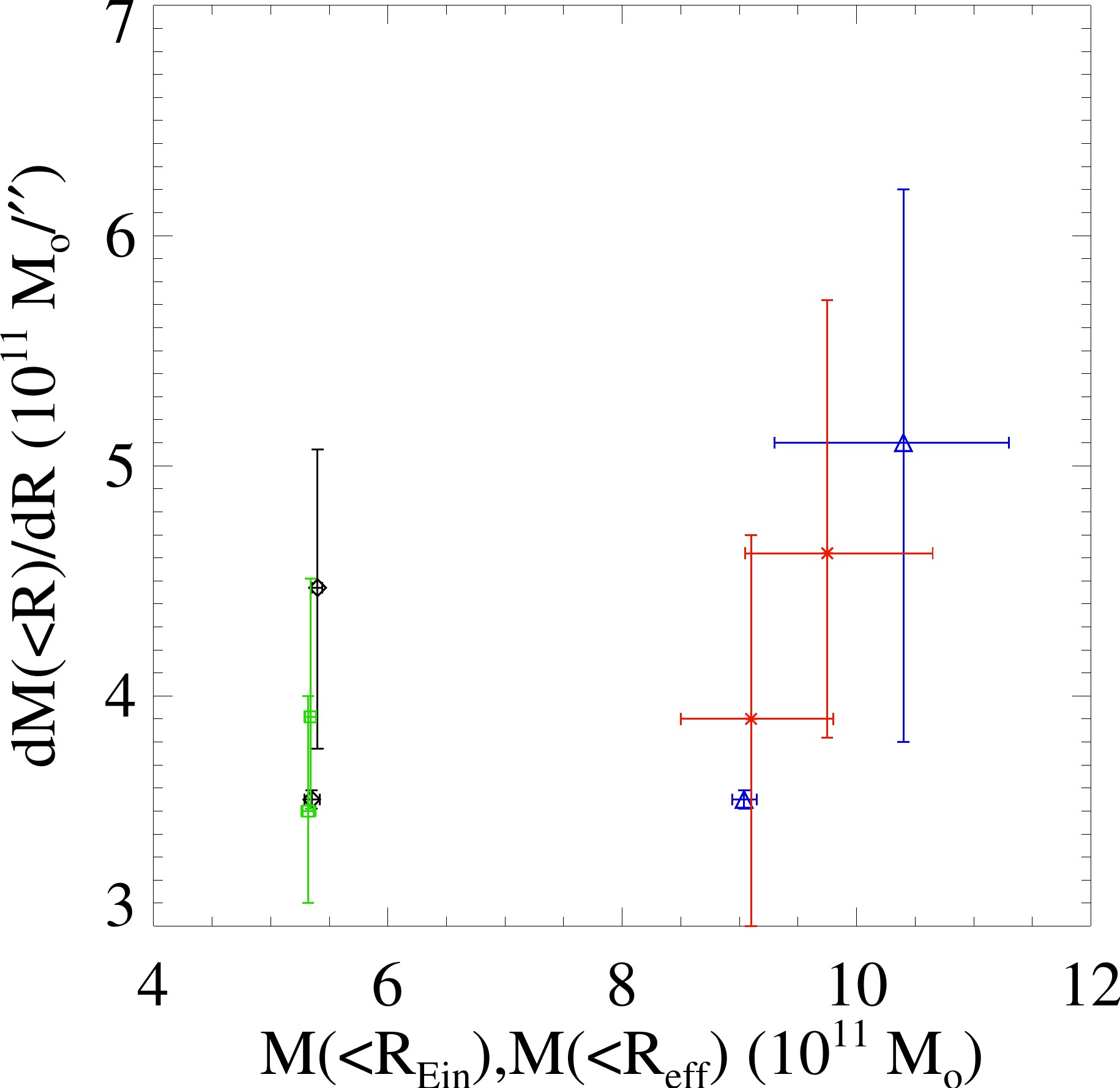}
\caption{The masses and radial derivatives for the Models stated in
  Table \ref{mass:mass_dmdr_kappa(Rein)_sie} and Table
  \ref{mass:mass_dmdr_kappa(Rein)_devauc}. Plotted are the mean
  values together with its 1 $\sigma$ error bars. For Models I and II, the black asterixes mark
  the values at the Einstein radius (fixed to $1.51 \arcsec$), while
  the blue triangles mark the extrapolations to the effective radius of the
  galaxy. For the de Vaucouleurs like models II and V, the green
  squares are the values at the Einstein radius, the red crosses are
  the ones at the effective radius. While the total mass within the Einstein radius is tightly
  constrained independent of the model used, the mass within the
  effective radius depends on the mass model used.} 
\label{sec:mass:mass_derivative_rein_reff}
\end{figure}

\begin{figure*}
\centering
\subfigure[]{\includegraphics[scale=.55]{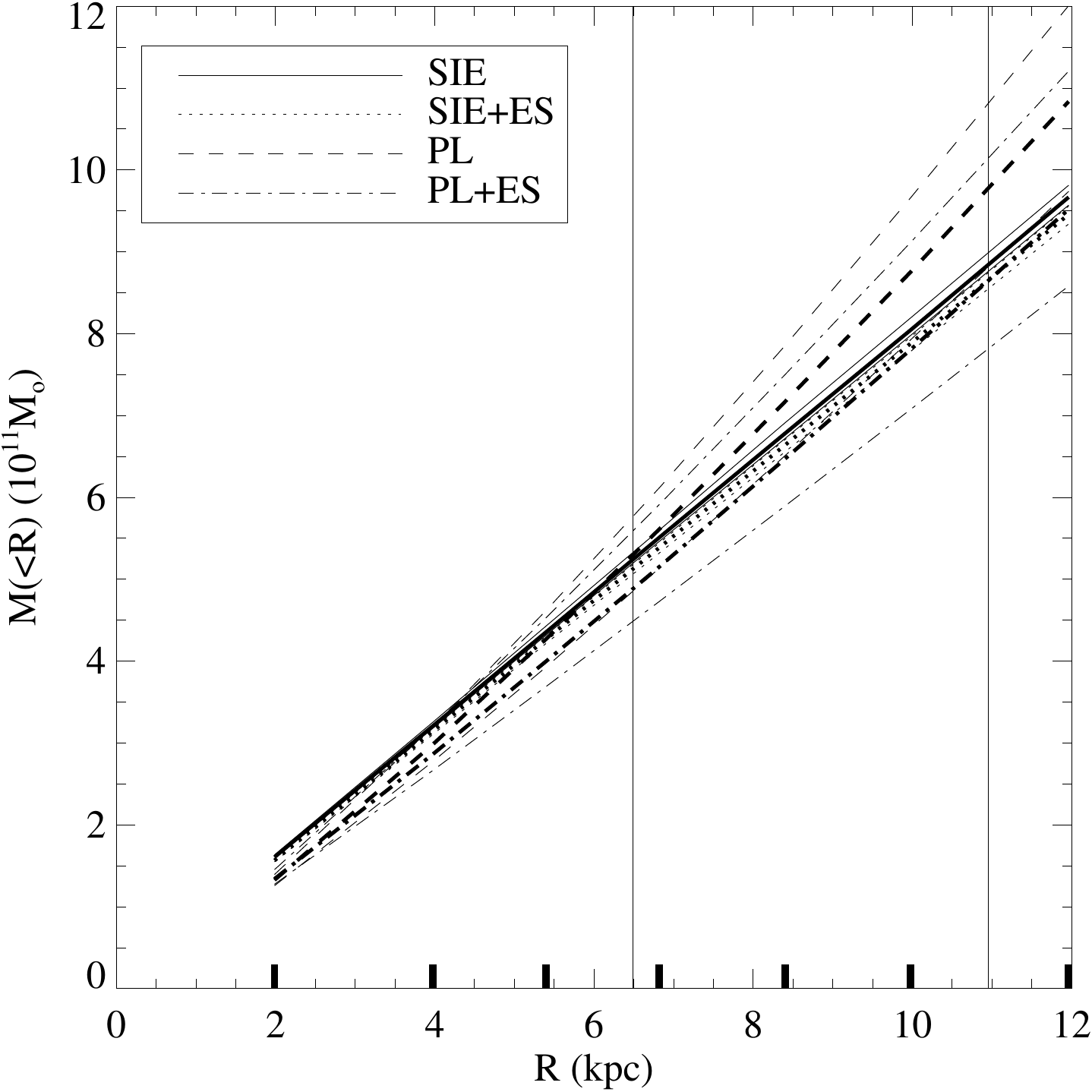}\label{lensing:parametric:SIEPLmass_ns}}
\subfigure[]{\includegraphics[scale=.55]{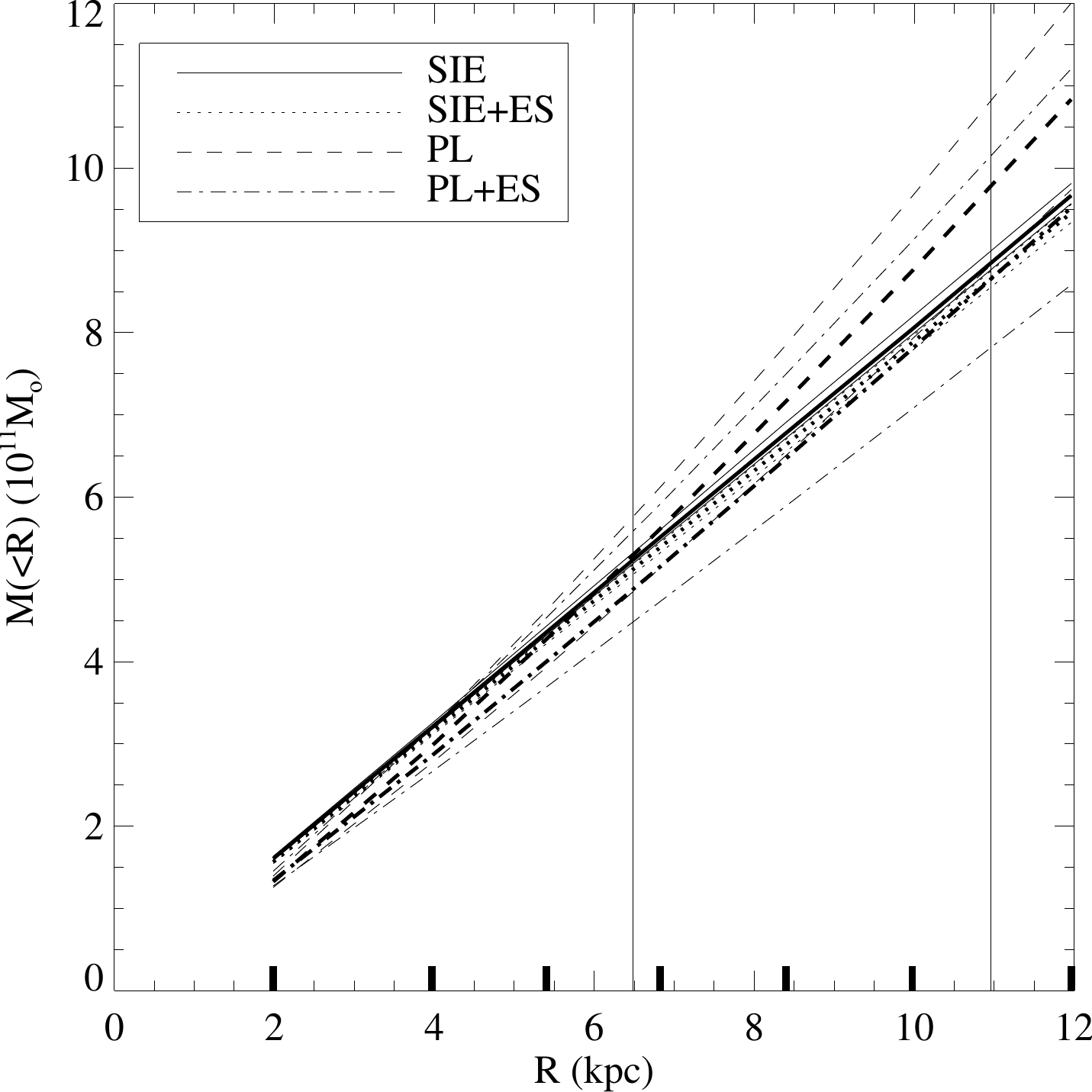}\label{lensing:parametric:lensview_masses}}
\caption{The mass within radius R, $\rm M(<R)$, 
from the models I and II, analysed in Sec.
  \ref{sec:strong lensing} is plotted in Fig. \ref{lensing:parametric:SIEPLmass_ns}. For each model, 1000 random entries from
  the MCMC are used to calculate the errors. Each time, the
  means and 90 \% errors are shown, with the respective means in bold lines. The
  bars at the bottom mark the radii of the apertures used to
  calculate the enclosed projected masses. The masses are in units of
  $10^{11}\rm M_{\odot}$, the radii are stated in kpc in the lens
  plane. 
 The  mass estimates for the Models I and II are shown as solid (I)
 and dashed (II) lines. Vertical lines indicate the Einstein and
 effective radius in both plots. 
Fig. \ref{lensing:parametric:lensview_masses} shows the same as Fig. \ref{lensing:parametric:SIEPLmass_ns} but
  for the {\sc lensview} derived masses. The
    errors are estimated by an increase of the reduced $\chi^2$ of the
    extended model by 1. As can be seen, the masses agree
  with each other in terms of derived masses within the errors.}
\label{lensing:parametric:gravlens:masses}
\end{figure*}

First we focus on the masses derived for Models I and II in Table \ref{lensing:parametric:table:best_fit_sie}.
The Einstein radii are defined as the radii within which the mean
convergence equals 1. For this, we calculate the mean convergence
around the Einstein radius in $0.03 \arcsec$ distance bins. The
results of this calculation are stated in Table \ref{mass:mass_dmdr_kappa(Rein)_sie}. Since all
models agree on an Einstein radius of $\Theta_{\rm
  E}=1.51\arcsec\widehat{=}\,6.48\,\rm kpc=R_{\rm E}$, we
adopt this value as ``the'' Einstein radius of this lens with an
uncertainty of $0.03\arcsec\widehat{=}\,0.13 \rm kpc$. We get a mean Einstein mass of $(5.37\pm
0.06)\times 10^{11}\rm M_{\odot}$ for the Models I and II
with a fixed Einstein radius of $\Theta_{\rm
  E}=1.51\arcsec$. This values are
in good agreement with the ones stated by \cite{slacs9} for this
system also based on strong lensing.\\ 
We also extrapolate the models to the effective radius $\rm r_{\rm
  eff}=2.55\arcsec\widehat{=}\,10.96 \, \rm kpc$ of the galaxy,
and calculate the mass and its derivative. We find an enclosed mass between
$\rm M_{tot,enc}=8.9\times 10^{11}\rm M_{\odot}$ and $\rm M_{tot,enc}=11.3\times 10^{11}\rm M_{\odot}$
on a $1 \sigma$ level, depending on the model used. 
The azimuthally averaged results of the included masses 
are plotted in Fig. \ref{lensing:parametric:gravlens:masses}. 
For an SIE model, the mass included within radius $r$ grows linearly with the radius,
so the derivative of it is expected to be independent of the
radius. This is the case for the singular isothermal model (Model I) in Table
\ref{lensing:parametric:table:best_fit_sie}. If we allow the steepness
to vary (Model II) the mass profile tends to be steeper at the Einstein and effective
radius.
For the radial mass derivative at the Einstein radius, we
calculate values between
$\rm \frac{d M_{tot,enc}}{d R}=0.8 \times 10^{11}\rm M_{\odot}kpc^{-1}$ and
$\rm \frac{d M_{tot,enc}}{d R}=1.2 \times 10^{11}\rm M_{\odot}kpc^{-1}$ . The extrapolation to the effective
radius ranges from $\rm \frac{d M_{tot,enc}}{d R}=0.8 \times 10^{11}\rm
M_{\odot}kpc^{-1}$ for Model I to $\rm \frac{d
  M_{tot,enc}}{d R}=1.6 \times 10^{11}\rm M_{\odot}kpc^{-1}$ for
Model II. This values are plotted in Fig. \ref{sec:mass:mass_derivative_rein_reff}. Here and in Table
  \ref{lensing:parametric:table:best_fit_sie} we state the 68 \%
  c.l. errors. 

\begin{table*}
\centering
\begin{minipage}{190mm}
\caption{
The masses and mass derivatives at the Einstein radius, the globally
adopted Einstein radius and the effective radius for the different
models. Given are the projected, enclosed masses and its derivatives}
\begin{tabular}{lcccccccc}
\hline
Model & $\Theta_{\rm Ein}$ & $\rm M_{\rm Ein}$ & $\frac{d\rm M}{d\Theta}(\Theta_{\rm Ein}) $
& $\kappa_{\rm Ein}$ & $\rm M(<1.51\arcsec)$ &
$\frac{d\rm M}{d\Theta}(\Theta=1.51\arcsec)$ & $\rm M(<2.55\arcsec)$ & $\frac{d\rm M}{d\Theta}(\Theta=2.55\arcsec)$\\
&($\arcsec$) & $10^{11}M_{\odot}$ & $10^{11}M_{\odot}/\arcsec$ &
& $10^{11}M_{\odot}$ & $10^{11}M_{\odot}/\arcsec$ & $10^{11}M_{\odot}$
& $10^{11}M_{\odot}/\arcsec$ \\
\hline
Model I	 &   $1.51\pm0.03$&	   $5.35_{-0.06}^{+0.07}$	&
$3.55\pm0.04$& $0.497_{-0.005}^{+0.006}$ & $5.35_{-0.06}^{+0.07}$
& $3.55\pm0.04$& $9.04_{-0.10}^{+0.11}$&$3.55\pm 0.04$	\\	
Model II &   $1.54\pm0.03$& $5.54_{-0.05}^{+0.06}$&
$4.5^{+0.6}_{-0.8}$	& $0.68_{-0.10}^{+0.08}$ & $5.40_{-0.05}^{+0.04}$&
$4.5^{+0.6}_{-0.7}$& $10.4_{-1.1}^{+0.9}$& $5.1_{-1.3}^{+1.1}$ \\
\hline
\label{mass:mass_dmdr_kappa(Rein)_sie}
\end{tabular}\\
\footnotemark[1]{galaxy part only, mass contribution of GI is ignored}\\
\end{minipage}
\end{table*}

\subsection{Mass profiles for the de Vaucouleurs plus dark matter halo
models}
\label{subsec:masses:devaucoulers}
\begin{figure*}
\centering
\subfigure[]{\includegraphics[scale=.55]{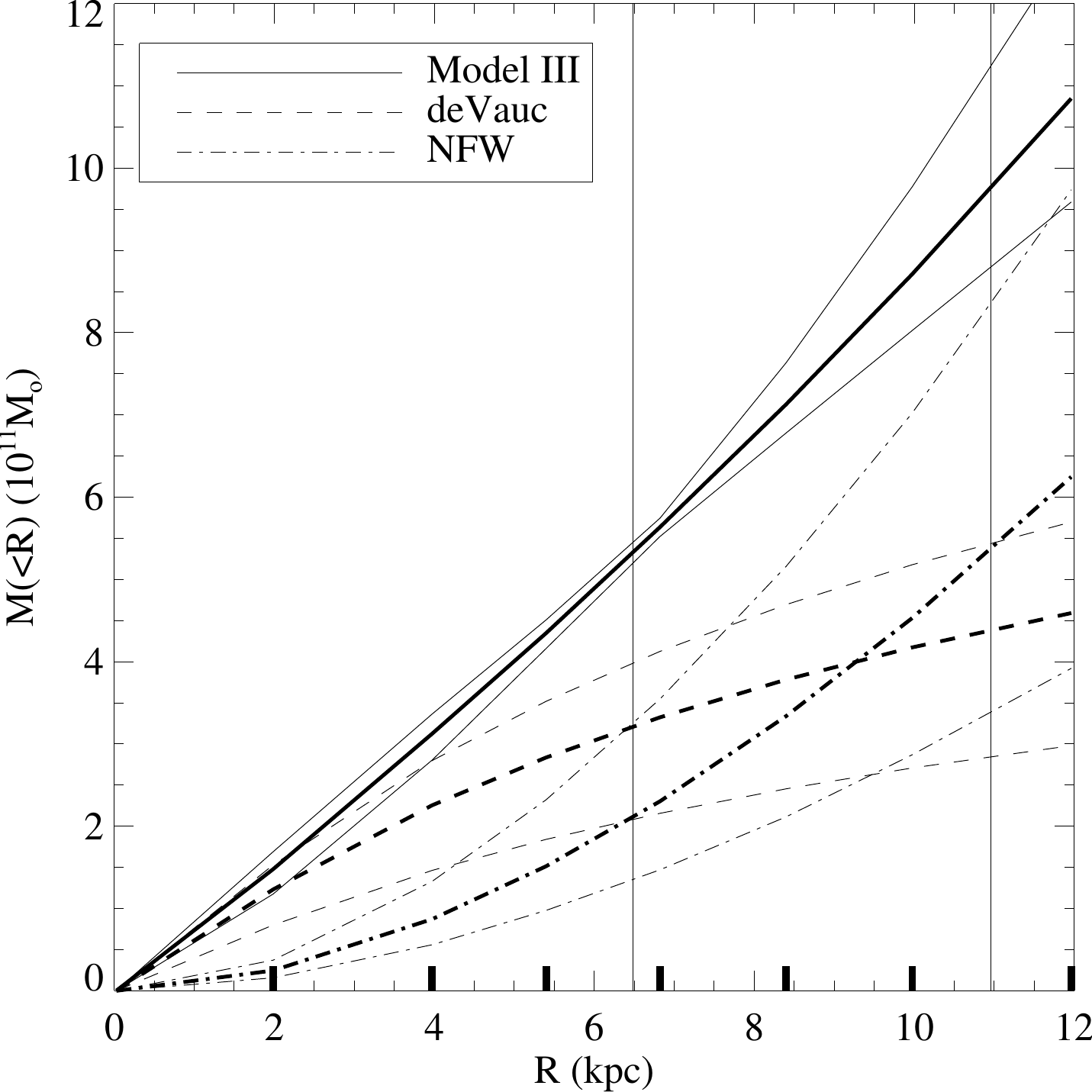}\label{lensing:parametric:mass_light_decomp_masses}}
\subfigure[]{\includegraphics[scale=.55]{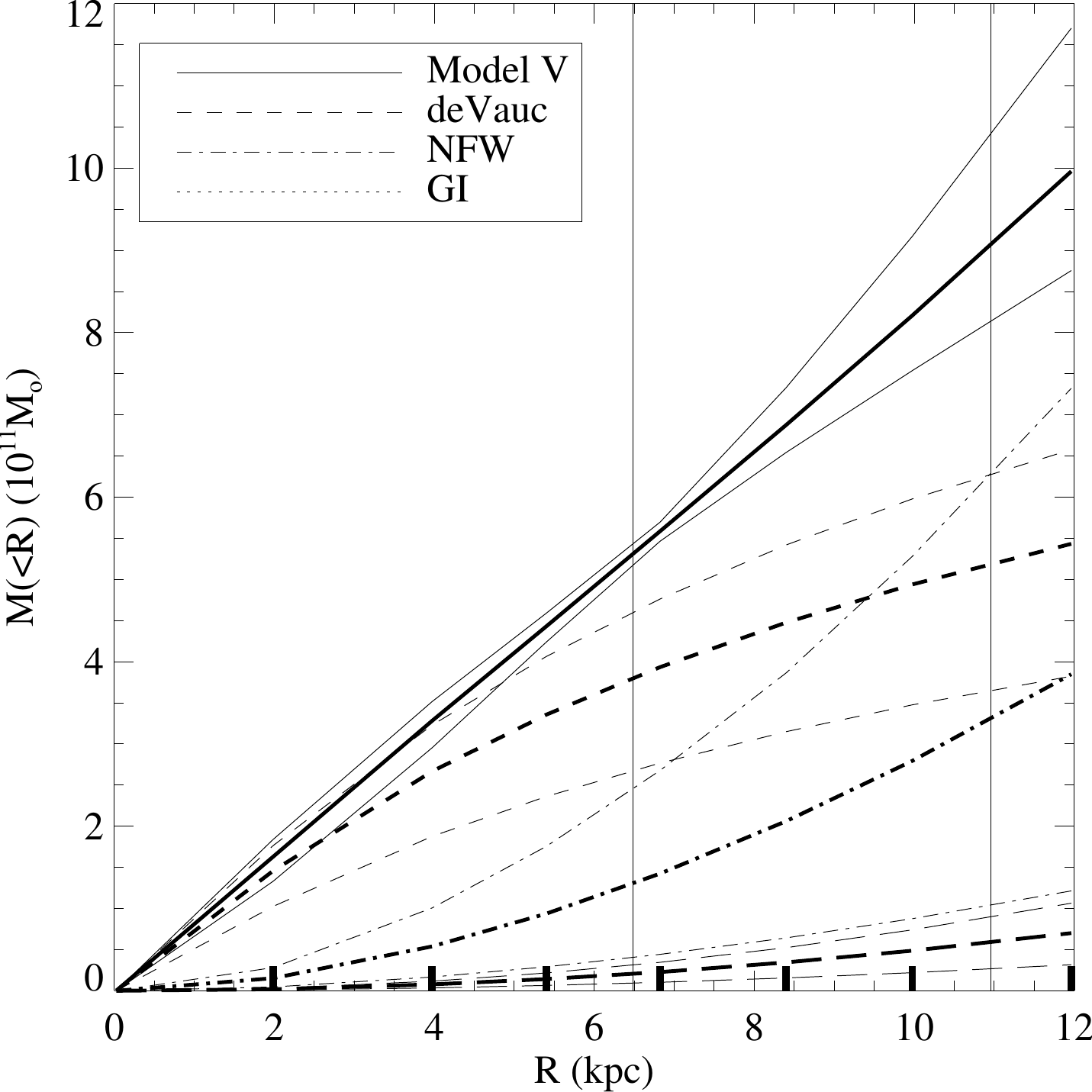}\label{lensing:parametric:mass_light_GI_decomp_masses}}
\subfigure[]{\includegraphics[scale=.55]{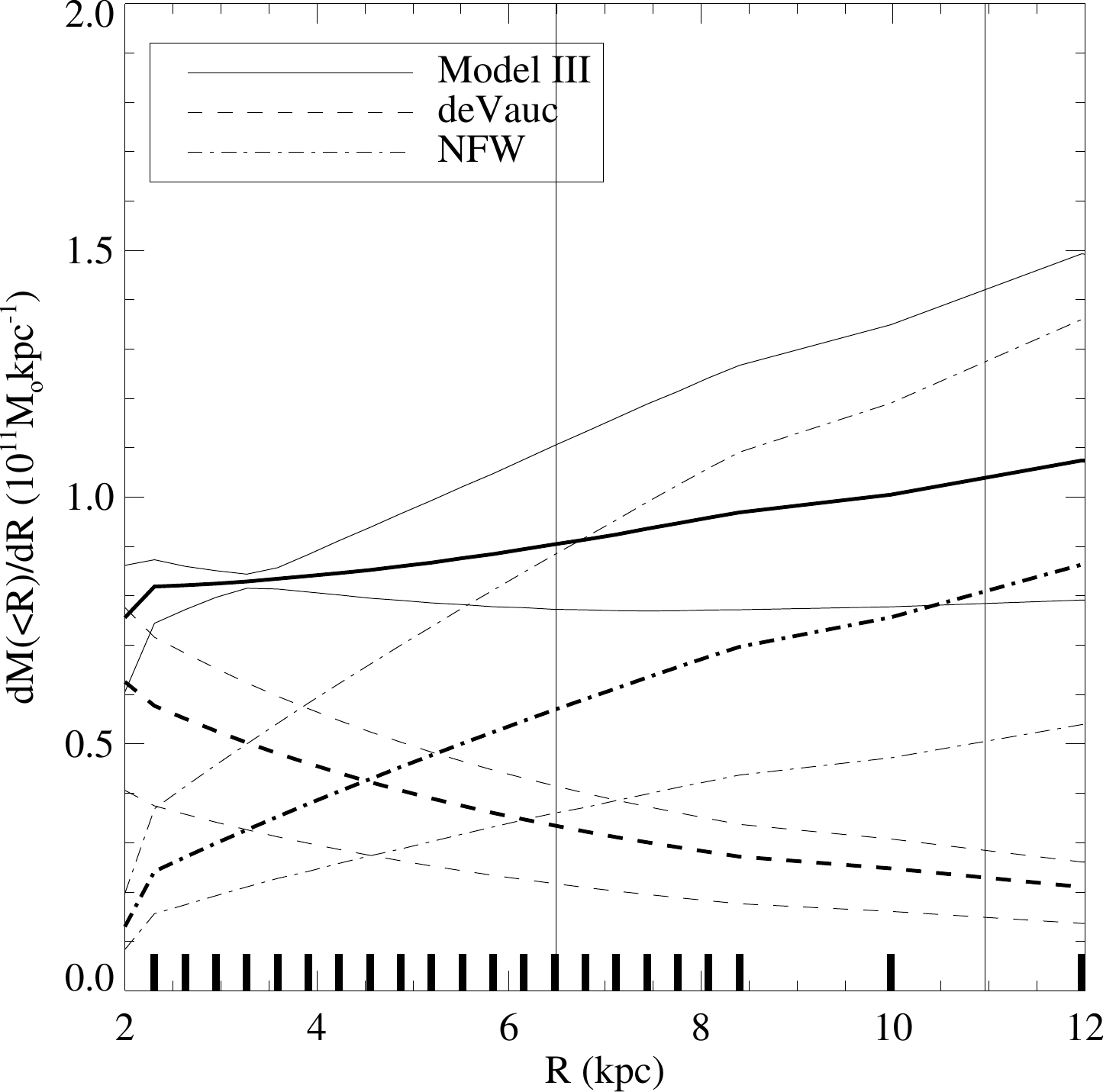}\label{lensing:parametric:mass_light_decomp_masses_deriv}}
\subfigure[]{\includegraphics[scale=.55]{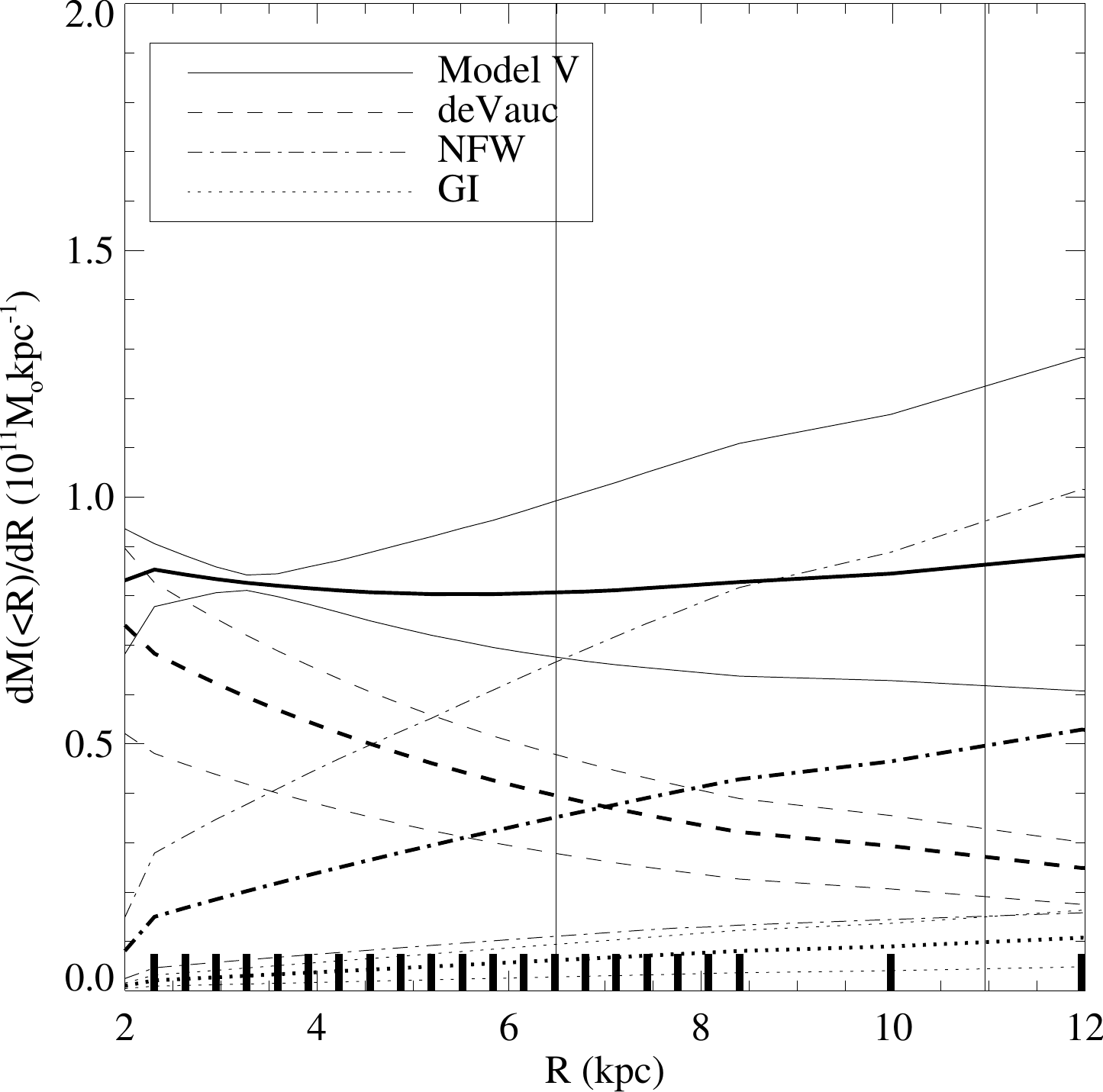}\label{lensing:parametric:mass_light_GI_decomp_masses_deriv}}
\caption{The mass within radius R, $\rm M(<R)$ from the Models III and V analysed in Sec.
  \ref{sec:strong lensing}. For each model, 1000 random entries from
  the MCMC chains are used to calculate the errors. All errors plotted are the 90 \% error
  intervals with the respective means in bold symbols. The
  bars at the bottom mark the radii of the apertures used to
  calculate the enclosed projected masses or its derivatives. The masses are in units of
  $10^{11}M_{\odot}$, the radii are stated in kpc in the lens
  plane. In \ref{lensing:parametric:mass_light_decomp_masses}, the mass estimates
  of Model III for the de Vaucouleurs (dashed), NFW (dash-dotted) and
  its sum (solid) are plotted. While the sum of these two is fairly well
  constrained, the errors on the individual parts are bigger. In \ref{lensing:parametric:mass_light_GI_decomp_masses},
the same mass estimates are plotted for Model V, together with
its 90 \% error intervals, splitted in de Vaucouleurs (dashed), NFW
(dash-dotted) and GI (dotted) parts and its total sum (solid). The radial mass derivatives are plotted in
\ref{lensing:parametric:mass_light_decomp_masses_deriv} for Model III
and \ref{lensing:parametric:mass_light_GI_decomp_masses_deriv} for Model V,
keeping the line coding. Plotted is the change in enclosed mass with
radius. This can also be interpreted as the mass in
  a thin ring with width $\rm dR$ at radius $\rm R$, $\rm M(R)$. Again, vertical lines indicate the Einstein and
 effective radius in both plots.}
\label{lensing:parametric:gravlens:masses_devauc}
\end{figure*}

From the single component lens analyses in
  Sec. \ref{subsec:masses:isothermal}, we conclude that the total projected
mass density profile is isothermal or slightly shallower than isothermal. The de Vaucouleurs
mass density drops faster with radius than the isothermal
profile. Therefore, we expect the pure de Vaucouleurs
profile to be a poor description of this lens' mass profile (as seen in Model
IV  in Sec. \ref{sec:lens modelling of the environment} and Model IVa in
Appendix \ref{sec:parametric models:additions}) and expect that we
need to add some component which follows a shallower
than isothermal density profile. In principle a constant mass sheet,
like a nearby group halo would approximately provide, could do this,
given that it provides enough mass at the position of the lens. In Model
IV, we see that this model is a poor fit to the
data. Therefore a (dark) matter component centred at the position of
the lens is necessary. We model this in Models III,
IIIa and IIIb in Sec. \ref{sec:parametric models:point like} and
Appendix \ref{sec:parametric models:additions}
with different types of dark matter halo
profiles. A SIE-like dark matter component (Model IIIa)
suppresses the de Vaucouleurs part of the matter profile,
effectively yielding a single component
model.
Because the light distribution is well-fit by a de
  Vaucouleurs profile, we require a non-zero de Vaucouleurs component
  for this 2 component fit, hence we do not consider Model IIIa in the
  following. The Models IIIb, Va and Vb in Appendix
  \ref{sec:parametric models:additions} give similar results to Models
III and V in Sec. \ref{sec:parametric models:point like}.
Therefore, in the following, we mostly consider Models III and V,
which model the lens using an NFW profile for the dark matter
component.\\
Besides the stars and the dark matter, an elliptical
  galaxy or a galaxy group also contains some amount of gas. Since we do not model this
  component individually, this gas needs to be incorporated in either
the dark matter or the de Vaucouleurs component, effectively limiting
the accuracy of our mass estimates to the gas mass fraction in
elliptical galaxies and groups of galaxies. \cite{ATLAS3D4} get typical
molecular gas masses of early-type galaxies in the $\rm ATLAS^{3D}$
project of $\rm M(H_{2})\leq 10^{9}M_{\odot}$, less than 1\% of the
total galaxy masses derived here. The hot gas component in an
elliptical galaxy or a group of galaxies can contribute up to 10\% of
the total mass in the centre of the galaxy or group of
galaxies, see e.g. \cite{sanderson03}. Hence the uncertainty of our
mass estimates due to the neglected gas is $\approx 10\%$.\\
We again adopt $\Theta_{\rm E}=1.51\arcsec$ for the Einstein radius.
First we focus on the masses within this radius, see Table
\ref{mass:mass_dmdr_kappa(Rein)_devauc}. For the total masses within
the Einstein radius of the
models III and V, we measure $\rm M_{\rm tot,Ein}=(5.33 \pm 0.04)\times 10^{11} \rm M_{\odot}$
The radial mass derivative is $\rm \frac{dM_{tot,enc}}{dR}=0.86^{+0.09}_{-0.07} \times 10^{11}\rm
  M_{\odot}\rm kpc^{-1}$ . The extrapolations to the effective radius give
$M_{\rm tot,eff}=9.5^{+0.6}_{-0.5}\times 10^{11} \rm M_{\odot}$ for the mass and 
$\rm \frac{dM_{tot,enc}}{dR}=1.00_{-0.14}^{+0.16} \times 10^{11}\rm
  M_{\odot}\rm kpc^{-1}$ for its derivative. This values are plotted
  in Fig. \ref{sec:mass:mass_derivative_rein_reff}. As can be seen,
  the enclosed masses and its derivatives at the Einstein radius and
  the effective radius agree with each other throughout Models I, II,
  III and V.  We state
    the de Vaucouleurs mass within the Einstein and effective radius
    of Model III as Component IIIA in Table
    \ref{mass:mass_dmdr_kappa(Rein)_devauc}. We get a mass of $\rm
    M_{deV,Ein}=(3.2_{-0.7}^{+0.5})\times 10^{11} \rm M_{\odot}$,
    meaning that $\rm \frac{M_{deV,Ein}}{M_{deV,tot}}\approx 35\%$ of
    the total de Vaucouleurs mass is concentrated within the Einstein
    radius for this lens. For Model V, we get
      similar values for the de Vaucouleurs component, see Component
      VA in Table \ref{mass:mass_dmdr_kappa(Rein)_devauc}.
In Fig. \ref{lensing:parametric:gravlens:masses_devauc} and Fig.
  \ref{lensing:parametric:gravlens:masses_devauc+GI}
  in Appendix \ref{sec:parametric models:additions},
  the projected, enclosed lens masses and their
  derivatives are plotted versus radius for the different 2 component
  strong lensing models. The measurements are done using circular
  apertures, so all of these values are azimuthally averaged. As one can
  see, including an explicit group halo at GI (Model V) has
  only a minor influence on the mass estimates and their
  derivatives. The total masses agree very well with the one component
  estimates in Fig. \ref{lensing:parametric:gravlens:masses}.
Also, all models agree very well on the total masses and
their radial derivatives, tending to give a shallower than isothermal
mass profile in the centre. For the Models III and V in
Fig. \ref{lensing:parametric:gravlens:masses_devauc}, the dark matter
haloes modelled as NFW-haloes agree very well with each other, meaning
  that the environment has only minor influence on the mass
  estimates. This is also true for
the de Vaucouleurs component. We note that the uncertainties on the
individual components are larger than the uncertainties on the total masses
and derivatives, giving a well-constrained total mass.

\begin{table*}
\centering
\begin{minipage}{200mm}
\caption{Same as Table \ref{mass:mass_dmdr_kappa(Rein)_sie} for the 2
component models.
}
\begin{tabular}{lcccccccc}
\hline
Model & $\Theta_{\rm Ein}$ & $\rm M_{\rm Ein}$ & $\frac{d\rm M}{d\Theta}(\Theta_{\rm Ein}) $
& $\kappa_{\rm Ein}$ & $\rm M(<1.51\arcsec)$ &
$\frac{d\rm M}{d\Theta}(\Theta=1.51\arcsec)$ & $\rm M(<2.55\arcsec)$ & $\frac{d\rm M}{d\Theta}(\Theta=2.55\arcsec)$\\
&($\arcsec$) & $10^{11}M_{\odot}$ & $10^{11}M_{\odot}/\arcsec$ &
& $10^{11}M_{\odot}$ & $10^{11}M_{\odot}/\arcsec$ & $10^{11}M_{\odot}$
& $10^{11}M_{\odot}/\arcsec$ \\
\hline
Model III& $1.51\pm0.03$  &$5.34\pm0.05$   &$3.9_{-0.4}^{+0.6}$
&$0.55_{-0.06}^{+0.08}$  & $5.34\pm0.05$
&$3.9_{-0.4}^{+0.6}$ &$9.8_{-0.7}^{+0.9}$ &$4.6_{-0.8}^{+1.1}$ 	\\	
Component IIIA\footnotemark[1]&   &	   &
&  &$3.2_{-0.7}^{+0.5}$ 
&$1.4_{-0.3}^{+0.2}$ &$4.4_{-1.0}^{+0.7}$ &$0.90_{-0.20}^{+0.15}$ 	\\	
Model V& $1.48\pm0.03$  &$5.21\pm0.06$   &$3.5\pm0.4$
&$0.50\pm0.06$  & $5.32\pm0.06$
&$3.5_{-0.4}^{+0.5}$ &$9.1_{-0.6}^{+0.7}$ &$3.9_{-0.9}^{+0.8}$ 	\\	
Component VA\footnotemark[1]&   &	   &
&  &$3.8_{-0.6}^{+0.5}$ 
&$1.6_{-0.3}^{+0.2}$ &$5.2_{-0.8}^{+0.7}$ &$1.06_{-0.17}^{+0.15}$ 	\\	
Component VB\footnotemark[2]& $1.45\pm0.03$  &	$4.91_{-0.08}^{+0.09}$   &
$3.2_{-0.4}^{+0.5}$& $0.47_{-0.06}^{+0.07}$ &$5.11_{-0.09}^{+0.10}$ 
&$3.2\pm0.5$ &$8.5_{-0.7}^{+0.8}$ &$3.3_{-0.8}^{+1.1}$ 	\\	
\hline
\label{mass:mass_dmdr_kappa(Rein)_devauc}
\end{tabular}\\
\footnotemark[1]{de Vacouleurs like part only, dark matter contribution is ignored}\\
\footnotemark[2]{galaxy part only, mass contribution of GI is ignored}\\
\end{minipage}
\end{table*}

\subsection{3d spherical reconstruction}

\begin{figure}
\centering
\includegraphics[height=80mm]{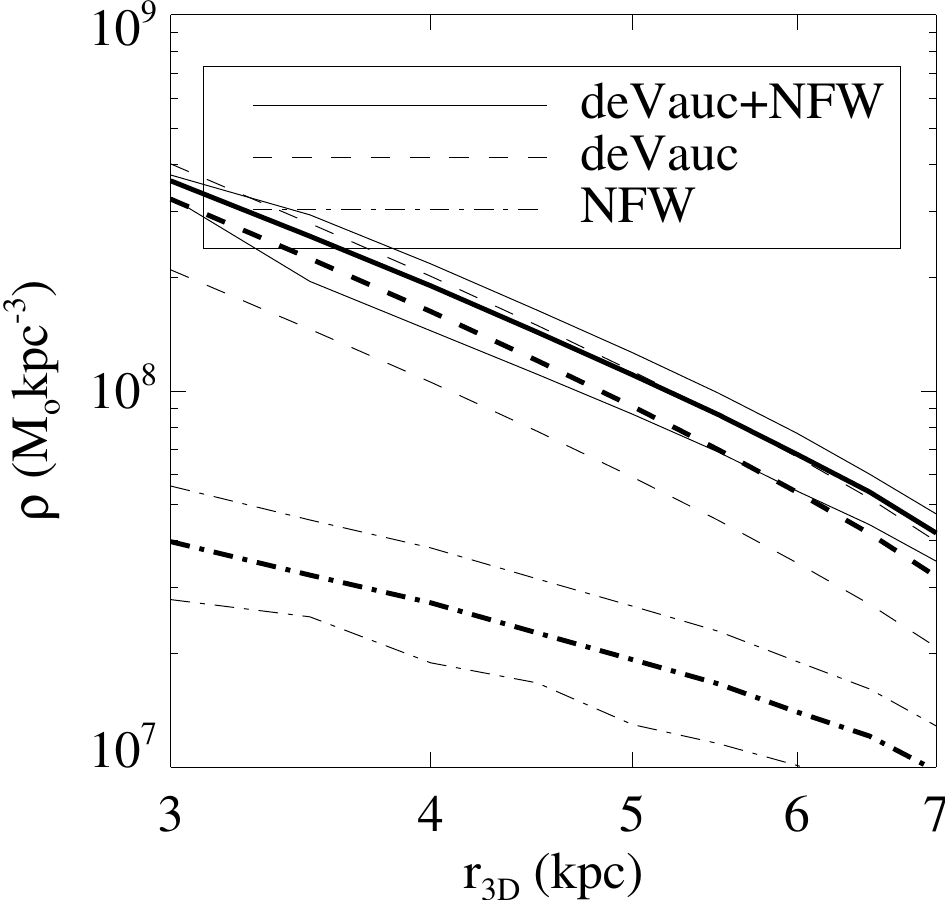}
\caption{The 3d densities for the stellar (dashed) and the dark (dash-dotted) component, as
  well as the total matter (solid) For this, the data of Model III is
  used. Plotted are the median (thick) and
  the 90 \% c.l. intervals. Spherical symmetry is assumed for this
  reconstruction.}
\label{mass_to_light:abel_transform}
\end{figure}

Further, we also reconstruct the 3d matter densities from the
2d data for Model III. For this, we employ the inverse Abel-transform:
\begin{equation}
\rho(r)=-\frac{1}{\pi}\int_r^{\infty}\frac{d\Sigma}{dR}\frac{dR}{\sqrt{R^2-r^2}}\quad ,
\label{sec:masses:abel transform}
\end{equation}
transforming a 2d circular density function $\Sigma$ into a 3d sperical
density function $\rho$. Since this only transforms
 circular to spherical profiles and vice versa, we start from the mass
measurements within a cylinder in Fig \ref{lensing:parametric:mass_light_decomp_masses}
for the azimuthally averaged profile. In Eq. \ref{sec:masses:abel
  transform} the integration extends to infinity, which is not possible
  due to our limited range of reliable data. To estimate the radial
  range at which we can use Eq. \ref{sec:masses:abel transform} only
  integrating up to our last data bin we test it on a SIS toy
  model. For a SIS toy model, we know both the spherical and the
  projected circular density. We then consider this radial range reliable where
  the deviation of the reconstructed 3d density from the analytical SIS
  density does not exceed $2\times 10^{7}\rm M_{\odot}kpc^{-3}$. 
From this comparison, we 
conclude that this inverse Abel transformation is only reliable up to
$\approx$ 6.5 kpc with a systematic error smaller than 30\%, given our
limited radial range of data.  The reconstructed 3d
profile is shown in Fig. \ref{mass_to_light:abel_transform}. The
errors plotted are only statistical, not
taking any systematic effects into account. The dark
matter accounts for only a minor fraction of the
  total mass in the 3d centre of the galaxy. 
We now turn to Fig. 7 in \cite{thomas_slacs_vs_coma_IMF}. In the lower part of
Fig. 7 they have displayed the ratios of the mean dark matter density
and mean total density within the Einstein radius of Coma galaxies as
a function of their velocity dispersion. (For the definition of the
synthetic Einstein radius for Coma galaxies, see
\citealt{thomas_slacs_vs_coma_IMF}). To see whether there are
structural differences for the Coma and the higher redshift SLACS
sample one would like to enter the corresponding deprojected values for SLACS
galaxies in these figures as well. These were not available until now
because the dark to total matter fractions were only calculated
for the line of sight projected densities within the Einstein radii
(i.e. cylindrical averages) by gravitational
  lensing. The corresponding projected values are shown for
SLACS (and Coma) galaxies in the upper part of Fig. 7 of
\cite{thomas_slacs_vs_coma_IMF}. The projected and
  deprojected values differ, since the projection along the line of
  sight mixes scales: Parts of the matter that has a large physical distance
  from the centre of the galaxy but lie on the line of sight are taken
into account when calculating the projected dark matter fractions.
Due to the
monotonic increase of the dark to total matter density ratios as a
function of radius, the projected ratios displayed in the upper part of Fig. 7
of \cite{thomas_slacs_vs_coma_IMF} are upper limits to the
central, 3--dimensional density ratios at the Einstein radius.
With the analysis described in this work we are
able to measure the 3--dimensional densities of the (spherically
averaged) dark matter and de Vaucouleurs components of the lensing galaxy
separately from gravitational lensing alone due to
  the large radial coverage of multiple images in the image plane by
  one source. Since the source is only one background object, we do
  not need to take the systematic uncertainties into account that
  arise in systems with multiple image systems from sources at
  different redshifts like e.g \cite{slacs6}. At
the Einstein radius we obtain (using
Fig. \ref{mass_to_light:abel_transform},
displaying Model III (de Vaucouleurs+NFW)) a dark to total
density ratio of 22 \% for the dark to
total density.
Doing the same for the Model V where SDSSJ 1430+4105 (consisting of
de Vaucouleurs and dark matter component)  is embedded in a DM halo
centred on galaxy I we find that the ratio of dark to total matter
density at the Einstein radius is about 14\%.
Since the dark matter fraction increases towards the outskirts,
these ratios of densities at the Einstein radius are upper limits for
the mean dark matter to total matter
density ratios of galaxy SDSSJ 1430+4105 within the same Einstein radius.
On a (90\% c.l.) basis, the density ratios
  at the Einstein radius are larger than
  15\% (Model III) and 5 \% (Model V) for the dark to total matter
  density.

\section{Mass to light ratios for the de Vaucouleurs component and
  dark to total mass ratio}
\label{subsec:mass:light_to_mass_ratios}

\begin{figure}
\centering
\includegraphics[height=80mm]{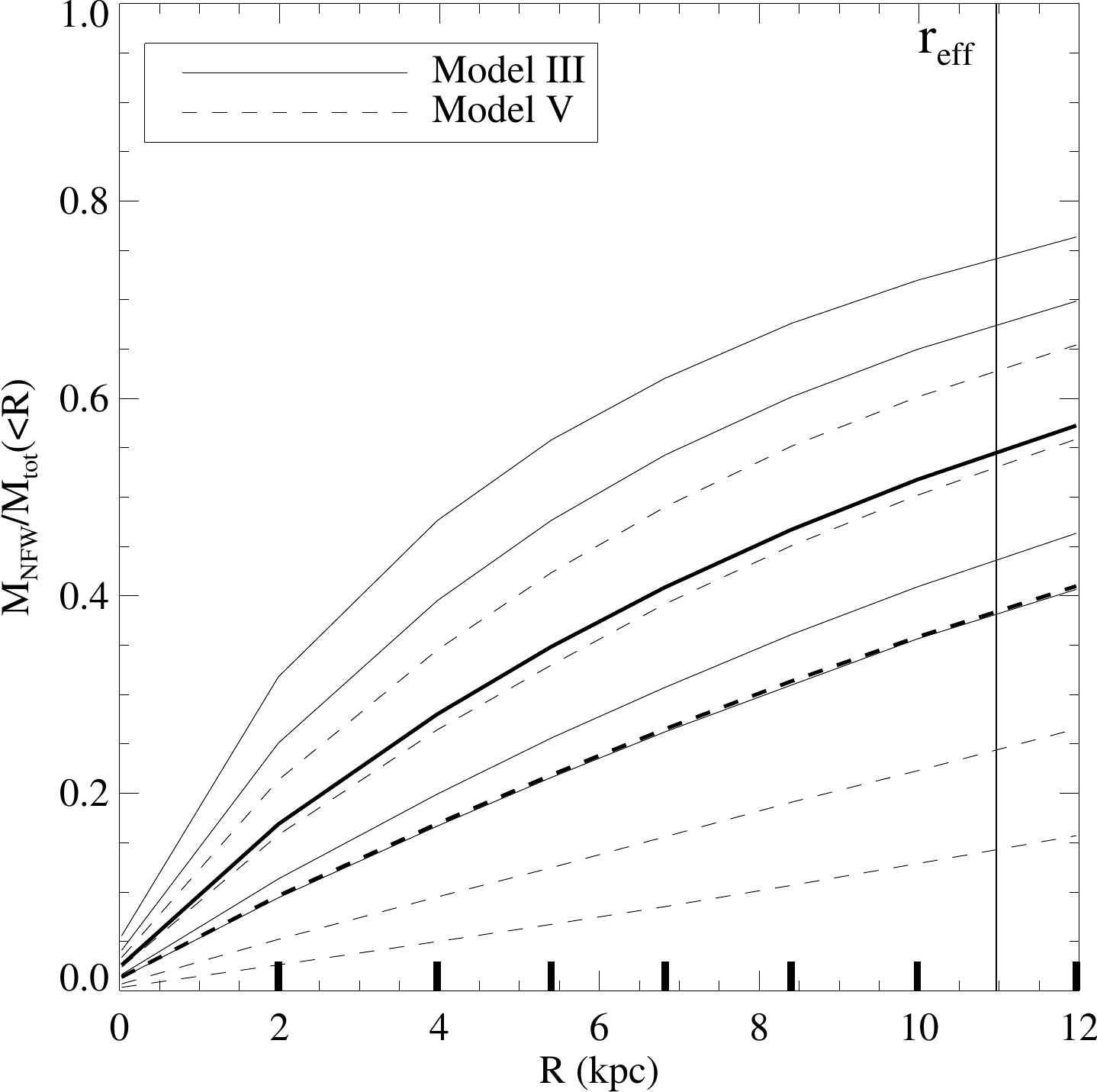}
\caption{Dark over total enclosed mass ratio for this galaxy. Shown are
  Models III (solid) and V (dashed). For Model V the group contribution
is ignored, meaning that we only consider the dark matter associated
with the galaxy itself as dark matter and only the sum of de Vaucouleurs and dark
matter as the total matter. The vertical line indicates the effective radius of the
galaxy. As one can see, both models predict dark matter
contributions in the centre of this galaxy, with changing amounts
depending on the modelling details. Plotted are the 68 \% and 90 \%
errors, respectively.}
\label{mass:devauc:darktototal}
\end{figure}

\begin{figure*}
\centering
\includegraphics[width=170mm]{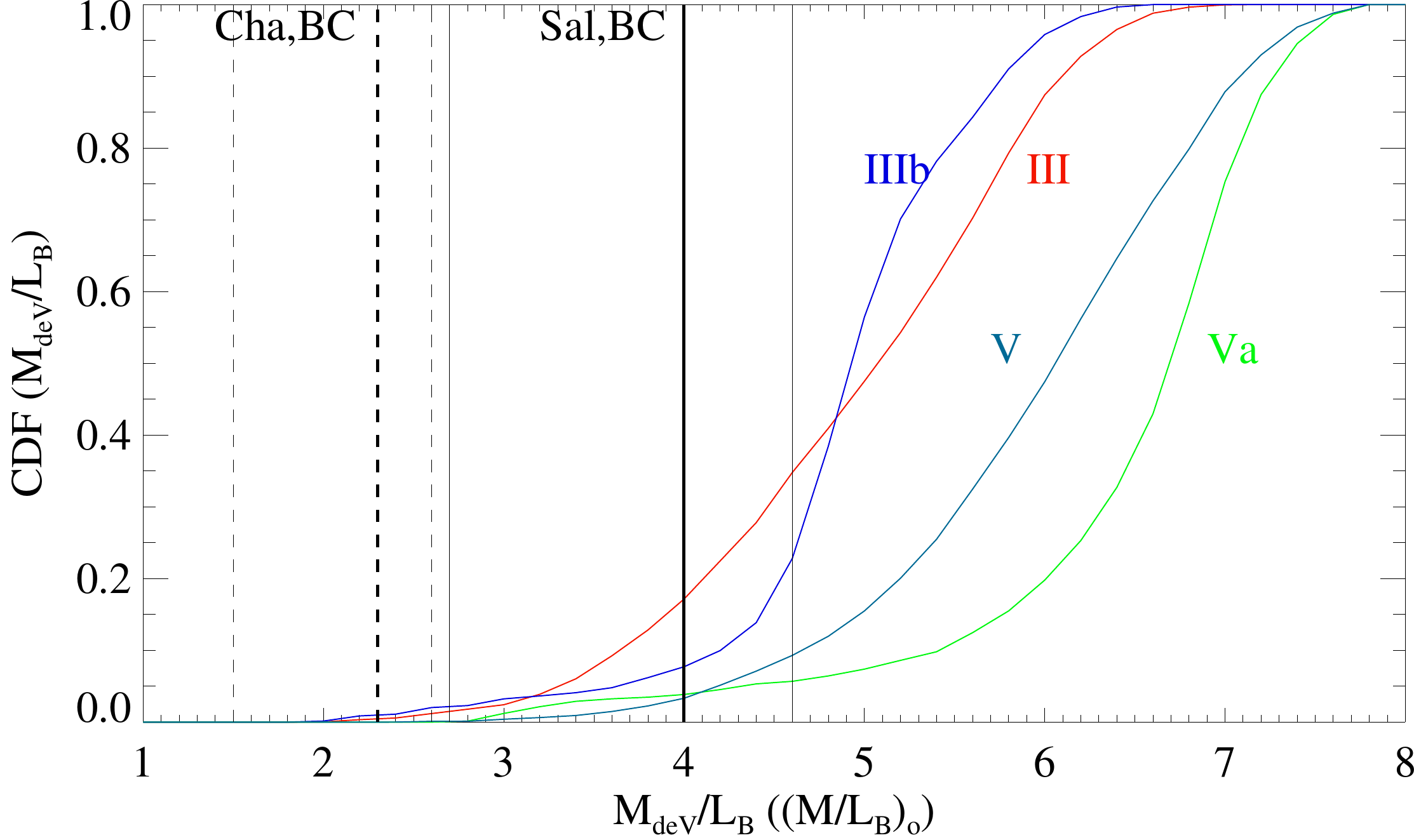}
\caption{Cumulative probability distribution functions of the mass to light ratios
  for the de Vaucouleurs components of the
  models III and V. Models III, IIIb, V and Va are marked by red, blue,
  turquoise and green. The vertical
  lines mark the derived stellar mass to light ratios with its 1 $\sigma$
  errors from
  \protect\cite{claudio_SED_fits} for this system fitting SFH
  to broad band SDSS photometry using a Salpeter
  IMF (solid line) and Chabrier IMF (dashed line). The
    mass to light ratios are as observed at $\rm z=0.285$ and not
  corrected for luminosity evolution to redshift zero.}
\label{mass:devauc:mtol}
\end{figure*}

\begin{figure}
\centering
\includegraphics[height=80mm]{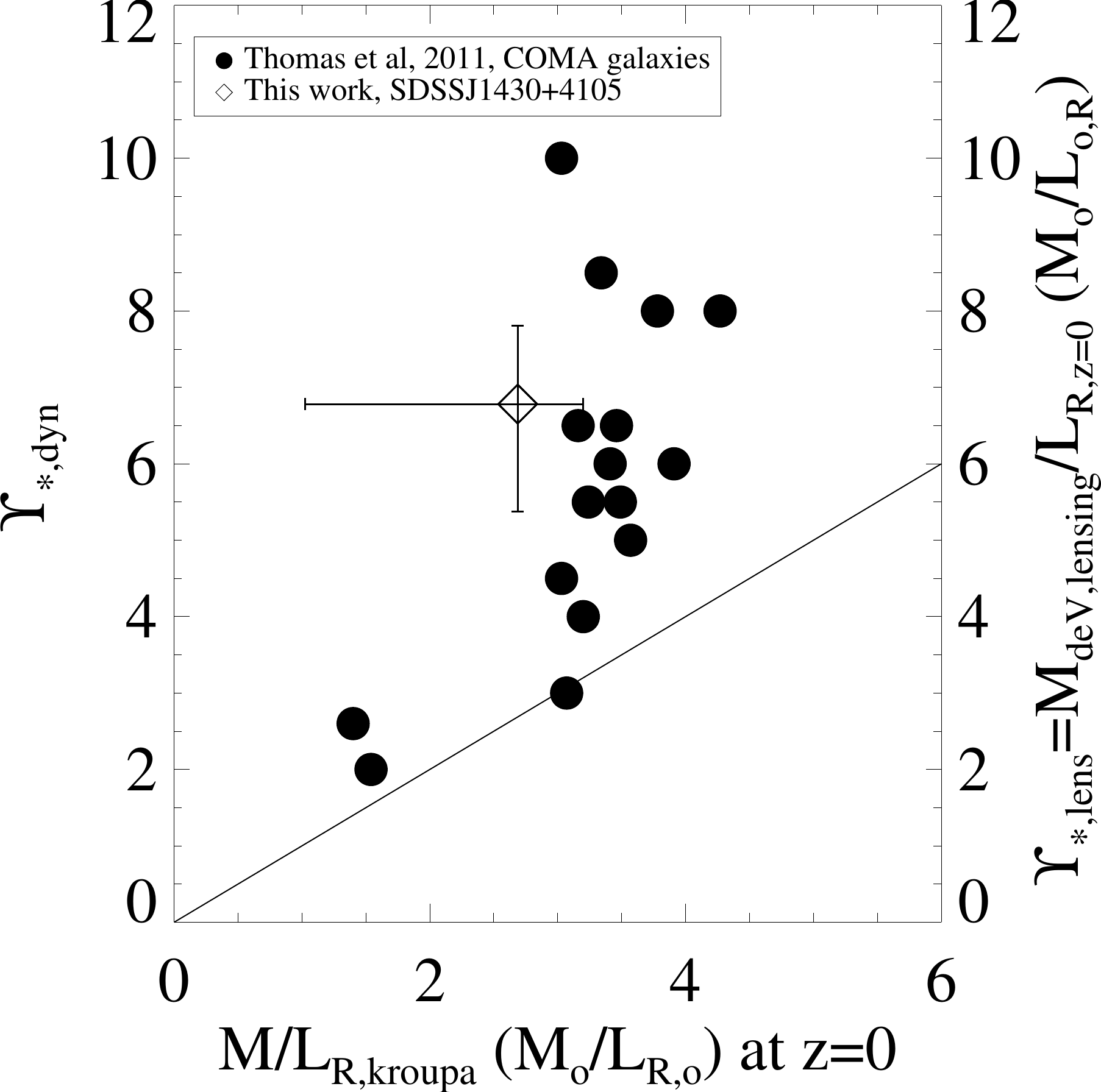}
\caption{MtoL in R from lensing versus stellar mass using a
  Kroupa-IMF, see \protect\cite{claudio_SED_fits}. Both values are evolved
  passively to the redshift 0. The lensing estimate is taken from
  Model III. The black points show the mass to light
  ratios derived from stellar dynamics analysis of the COMA cluster
  galaxies versus their stellar MtoL ratio derived from photometry for
  a Kroupa IMF, taken from \protect\cite{thomas_slacs_vs_coma_IMF}. The solid
  line marks the one to one correspondence.}
\label{mass_to_light_lensing_vs_kroupa}
\end{figure}

Since we calculated the de Vaucouleurs masses for this
galaxy, we now want to estimate the rest frame mass to light ratios of
this galaxy. Further, we evolve this mass to light ratios to present
day values in order to compare it with Coma galaxies. First, we
calculate the dark matter fractions within the Einstein radius.
In Fig. \ref{mass:devauc:darktototal}, we
plot the dark over total enclosed mass fraction within the Einstein radius. The error bars
  are estimated as before
from the central 68 \% and 90 \% entries of the random sample drawn from the
MCMC for Models III and V, respectively. The fractions for Model III and V
are: $\frac{\rm M_{\rm NFW}}{\rm M_{\rm tot}}=(0.40_{-0.09}^{+0.13})$
and $\frac{\rm M_{\rm NFW}}{\rm M_{\rm tot}}=(0.27_{-0.11}^{+0.12})$. These fractions indicate a
substantial amount of dark matter within the Einstein radius of this
lens. As can be seen, this picture is not significantly altered
by locating the group at GI in Model V. Although the actual
numbers change, we still need dark matter
associated with the lens in the centre of the
galaxy. Since we 
ignore the dark matter contribution associated with the group GI in
Model V for the total mass, we get a lower
dark matter fraction for Model V compared to Model III.\\
For the Models III and V, we calculate the mass to light
ratios for the de Vaucouleurs component at the
redshift of the lens. We use the masses from lensing and the light
(in rest frame B and R) as obtained from
photometric data. To compare the mass to light ratios
with present day galaxies, we also need the luminosity
evolution to redshift zero in these bands. We take the observed griz SDSS photometry for this
system which covers the rest frame B and R filters to calculate the B-R
rest frame colour.\footnote{All magnitudes are given
in AB.} We calculate a colour of $\rm (B-R)=(0.80\pm0.03)$ and luminosities of $\rm L_{\rm
    B,rf}=(1.66\pm0.03)\times10^{11}\rm L_{\odot}$ and $L_{\rm R,rf}=(1.92\pm 0.02) \times 10^{11}\rm L_{\rm
    R,\odot}$ from the absolute rest frame
  magnitudes. To estimate the luminosity evolution until today, we
fit 3 extinction-free BC \citep{bc03} composite stellar population (CSP) models to the observed griz SDSS photometry:
A Salpeter initial mass function (IMF) \citep{Salpeter_IMF} with solar
metallicity (Model A) and 2 models with Chabrier IMF \citep{Chabrier_IMF} and solar /
super--solar metallicity (Models B,C), respectively, see \cite{dro01,
  dro04}. The best-fitting results are stated in Table
  \ref{mtol:SEDfits} together with the luminosity evolutions in B and R bands and the best-fitting stellar
masses, which agree with \cite{claudio_SED_fits}. From the
best-fitting star formation histories (SFH)
  to the spectral energy distribution (SED) we obtain a stellar age of
  typically 8 Gyrs and a B-R colour of 0.8. Therefore, this galaxy has
  a formation redshift of approximately 2 to 3 which is a typical
  value for elliptical galaxies.
If we divide the de Vaucouleurs masses derived by lensing in Sec. \ref{sec:strong lensing} by the
luminosities derived from the SDSS photometry we obtain the
mass to light ratios (MtoL) for the de Vaucouleurs component of this galaxy.
For  Models III and  V, we find a MtoL of
 $\frac{\rm M_{\rm deV}}{\rm L_{\rm B}}=(5.3_{-1.1}^{+0.8})\frac{\rm
  M_{\odot}}{\rm L_{\odot,\rm B}}$ and $\frac{\rm M_{\rm deV}}{\rm
  L_{\rm B}}=(6.2_{-1.0}^{+0.9})\frac{\rm
  M_{\odot}}{\rm L_{\odot,\rm B}}$ in
the B-band rest frame at the redshift of the lens.
These two Models give the same MtoL within the errors,
  although including the group GI explicitly increases the most-likely MtoL.
We compare this with the total light of the galaxy and the stellar
mass derived in \cite{claudio_SED_fits}, who use composite stellar population models with a
Salpeter or Chabrier IMF, a delayed exponential star formation
history, and solar metallicity to model the SDSS multi band
photometry. First, we compare in the rest frame B band. In
Fig. \ref{mass:devauc:mtol} we plot the cumulative
  distribution function for the stellar mass to light
ratios derived from the respective de Vaucouleurs parts of the Models
III and V and Models IIIb and Va from Appendix \ref{sec:parametric
  models:additions}. We overplot the stellar
MtoL ratios derived in \cite{claudio_SED_fits}
for this system and get the best agreement for a NFW like halo and a
Salpeter IMF.\\
In the R-band, we get a MtoL for Models III and V of $\frac{\rm M_{\rm deV}}{\rm L_{\rm R}}=(4.6_{-1.1}^{+0.8})\frac{\rm
  M_{\odot}}{\rm L_{\odot,\rm R}}$ and $\frac{\rm M_{\rm deV}}{\rm L_{\rm R}}=(5.4_{-0.9}^{+0.7})\frac{\rm
  M_{\odot}}{\rm L_{\odot,\rm R}}$.
To compare the lensing galaxy SDSSJ 1430+4105 with present day Coma galaxies \citep{thomas_slacs_vs_coma_IMF}
we have to account for the luminosity evolution between redshift 0.285
and now. We use the average evolution factor from the SFH models stated in
Table \ref{mtol:SEDfits}, derived from the
  extrapolations of the fitted SFH models to redshift 0, which
increases the MtoL in the R band by a factor of
$1.48$ for z=0, 
 giving $(\frac{\rm M_{\rm deV}}{\rm L_{\rm R}})_{\rm pres}=(6.8_{-1.6}^{+1.2})\frac{\rm
  M_{\odot}}{\rm L_{\odot,\rm R}}$ and $(\frac{\rm M_{\rm deV}}{\rm L_{\rm R}})_{\rm pres}=(8.0_{-1.3}^{+1.1})\frac{\rm
  M_{\odot}}{\rm L_{\odot,\rm R}}$ for Models III and V, respectively.
In Fig. \ref{mass_to_light_lensing_vs_kroupa}, we plot this R-band de
Vaucouleurs mass to light ratio at redshift zero against the
present day R-band mass to light ratio for a
  Kroupa-IMF (\citealt{kro01}), obtained again from the SFH fit of
\cite{claudio_SED_fits}, translated to R-band and evolved to redshift zero. 
We added the results for a dynamical study of Coma
galaxies by \cite{thomas_slacs_vs_coma_IMF}. This allows us to conclude
that SDSSJ 1430+4105 evolves into a galaxy with mass to light ratio
similar as the Coma galaxies, and shows the same conflict with respect
to a Kroupa IMF as they do. This conflict to a Kroupa IMF
would be resolved if, for example, the de Vaucouleurs component is not made of stars
only but contains dark matter as well.


\begin{table}
\centering
\caption{Galaxy luminosity evolution factors for the different IMF
  and metallicity models}
\begin{tabular}{ccccccc}
\hline
 & $\rm (B-R)_{\rm rf}$ & T & $\tau$ & $\rm
\frac{L_{B,rf}}{L_{B,z=0}}$ & $\rm \frac{L_{R,rf}}{L_{R,z=0}}$ & $\rm
M_{\star}$ \\
 &ABmag & Gyrs & Gyrs & & & $\rm \left ( 10^{11} M_{\odot}\right )$ \\
\hline
A & 0.77 & 8 & 2 & 1.77 & 1.48 & 6.7 \\
B & 0.85 & 8 & 2 & 1.92 & 1.55 & 4.4 \\
C & 0.81 & 9 & 2 & 1.66 & 1.42 & 4.0 \\
\hline
\label{mtol:SEDfits}
\end{tabular}\\
\end{table}

\section{Discussion and conclusions}

In this paper, we studied the lensing properties of
SDSS J1430+4105.
From the complex source, we identified 5
double image systems, spanning a radial range from below $0.9 \arcsec$
to almost $2.1\arcsec$. The source is spectroscopically confirmed at a redshift of
$0.575$. Parametric models can match the observed image
positions well with an average scatter in the position
comparable to the pixel size of the ACS camera input
image, which is
$0.05\arcsec$. 
Our results are:
\begin{itemize}
\item[I)] 
The best-fitting reconstruction of the profile favours a
profile slope shallower than isothermal for the
best-fitting model. However, profiles with
free slope for the density steepness are consistent
  with an isothermal profile at 90\%
c.l.. This is also true when combining an explicit model for the
de Vaucouleurs-like light distribution with a NFW-like dark matter component.
\cite{slacs10} found a steepness for the 3d density profile for this system of
$\rho\sim r^{-(2.06 \pm 0.18)}$ by using the location of the Einstein radius only and
combining this with stellar dynamics, in agreement with our results
for the one component powerlaw total mass distribution of $\rho\sim r^{-(1.73_{-0.13}^{+0.21})}$
within the errors.
\item[II)]
The galaxy is part of a group of galaxies listed in
  the maxBCG cluster catalogue.  
Using a lens mass component following the stellar light, we can not model the strong
lensing signal for this galaxy if we use this component alone or
combine it with a dark matter halo not centred on SDSS
J1430+4105, called galaxy A. Therefore, this leads to two possibilities: Either A is indeed the
centre of the galaxy group, or the galaxy A is a satellite of this
group, residing in its own dark matter halo. Since we cannot
distinguish between these 2 cases, neither from the lensing signal nor
from external data, we model both scenarios and show that these lead
to similar results regarding the mass distribution of
  the galaxy. We show that the dark matter
halo of galaxy A must not be singular and isothermal at the same time, since this would
suppress the de Vaucouleurs component. Both a non-singular, isothermal halo
and a NFW-like halo for the dark matter halo of galaxy A fit the data
well. We find agreeing dark matter
fractions and distributions for both cases. From the
lensing data, we cannot distinguish whether the dark matter halo
follows a NFW or NSIE profile in the centre, since we cannot constrain the
concentration c and $r_{200}$ -- or $\Theta_{\rm c}$ for a NSIE dark matter
halo -- of the dark matter component well. From the models taking
explicitly the environment into account 
we conclude that the dark matter and the light of the galaxy have the
same axis ratio and are likely coaligned.
\item[III)]
We estimate the rest frame B-band mass to light ratios for the lensing
galaxy from the de Vaucouleurs lensing component. For the case of a
  deVauc+NFW mass model,  we  obtain a total mass of  $\rm M_{deV,tot}=(8.8_{-1.9}^{+1.3})\times
  10^{11} \rm M_{\odot}$ for the de
  Vaucouleurs component. \cite{claudio_SED_fits} have obtained stellar
  mass estimates for the luminous component using the ugriz broad band
  SDSS photometry and SFH fits. They assumed solar metallicity
  composite stellar populations with a delayed exponential SFH, and
  examined the Salpeter IMF (with BC and MAR
  \citep{mar05} single stellar populations (SSP)), and the Chabrier
  and Kroupa IMF  (based on BC-SSPs). They
  obtained stellar masses of $\rm M_{\star}=(5.6_{-1.8}^{+0.8})\times
  10^{11} \rm M_{\odot}$, $\rm M_{\star}=(3.9_{-2.2}^{+1.6})\times
  10^{11} \rm M_{\odot}$, $\rm M_{\star}=(3.2_{-1.1}^{+0.4})\times
  10^{11} \rm M_{\odot}$, and $\rm M_{\star}=(2.9_{-1.8}^{+0.6})\times
  10^{11} \rm M_{\odot}$ for these four cases. The stellar mass
  agrees best with the mass of the de Vaucouleurs
  component obtained from lensing if
  we assume a Salpeter IMF. In principle, the mass to light
  ratios also depend on the age of the galaxy and its metallicity. According to \cite{del06}, solar
  metallicity and a formation redshift of 3 to 5 as used in \cite{claudio_SED_fits} are good assumptions
  for a galaxy of the measured stellar mass. Thus, the IMF must not be Chabrier
  or Kroupa like unless a fraction of the de Vaucouleurs component is
  not of stellar origin, i.e. part of the dark matter follows the
  light distribution.  We measure a mass to light ratio for the stellar
  component of $\frac{\rm M_{deV}}{\rm L_{\rm B}}=(5.3_{-1.1}^{+0.8})\frac{\rm
    M_{\odot}}{\rm L_{\odot,\rm B}}$ using
    gravitational lensing and assuming a NFW-like dark
  matter halo. If we allow for a group
 halo at galaxy I, we obtain $\frac{\rm M_{deV}}{\rm L_{\rm B}}=(6.2_{-1.0}^{+0.8})\frac{\rm
    M_{\odot}}{\rm L_{\odot,\rm B}}$. These results again favor a
  Salpeter IMF, and are in agreement with the Fundamental
 Plane results $\frac{\rm M}{\rm L_{\rm B}}=(4.8 \pm 1.4)\frac{\rm
    M_{\odot}}{\rm L_{\odot,\rm B}}$ for this galaxy from
 \cite{claudio_SED_fits}. These results hold as long as the
 metallicity is approximately solar.
We compare the mass to light ratio, passively
  evolved to z=0, to Coma galaxies analysed in \cite{thomas_slacs_vs_coma_IMF}.
We confirm their trend towards
  a Salpeter IMF, again disfavouring a Kroupa-IMF. This trend also
  is seen in \cite{cap12} for the most massive galaxies. Their data
  indicate a more Salpeter-like IMF for high velocity dispersion galaxies.
The dark to total mass fraction of SDSSJ 1430+4105 rises from the centre outwards, giving
a value of $\rm \frac{M_{dark}}{M_{tot}}=(0.4_{-0.10}^{+0.14})$ at the Einstein radius. In this
galaxy, we need a significant amount of dark matter in its projected centre
 to explain the observations. 

\item[IV)]
We compare the 3d densities of total, dark and luminous dark matter
with Coma galaxies analysed by
\cite{thomas_modelling_coma_galaxies}, based on dynamical
modelling, especially their Figure 5.
Our galaxy has an effective radius of $10.96\rm kpc$ and a velocity dispersion of
$322 \pm \, 32\rm kms^{-1}$.
Concerning the effective radius it is most similar to the Coma galaxies
GMP 0144, GMP 4928 and GMP 2921, a cD galaxy, which have effective radii of 8.94
kpc, 14.31 kpc and 16.43 kpc (\citealt{thomas_modelling_coma_galaxies}) and effective
velocity dispersions of $211.8 \pm 0.4\,\rm kms^{-1}$ and $314.8 \pm 2.9
\,\rm kms^{-1}$ and $\approx 400 \,\rm kms^{-1}$
(\citealt{thomas_modelling_coma_galaxies}, \citealt{cor08}). 
Since our 3 dimensional matter densities are reliably known only between
3 kpc and about 6.5 kpc we
decide to compare the matter densities at 3 kpc. At this radius the
matter density values of \cite{thomas_modelling_coma_galaxies} are
reliable as well. At the same time this radius is within the core
radius  ($\rm r_c$ in Table 2 of \citealt{thomas_modelling_coma_galaxies}) for
all 3 GMP galaxies and thus the densities at this radius define the
central dark matter densities in these galaxies. We read off dark
matter and total densities of $\rm 6\times10^{-2} M_{\sun}pc^{-3}$ and
$\rm 3\times10^{-1} M_{\sun}pc^{-3}$
for GMP 0144, of $\rm 1.5\times10^{-2} M_{\sun}pc^{-3}$ and $\rm 2\times10^{-1} M_{\sun}pc^{-3}$ for GMP 4928
and $\rm 1\times10^{-1} M_{\sun}pc^{-3}$ and $\rm 3\times10^{-1}
M_{\sun}pc^{-3}$ for GMP 2921.
For SDSSJ 1430+4105 these numbers are $\rm 4\times10^{-2}
M_{\sun}pc^{-3}$ and $\rm 3.5\times10^{-1}
M_{\sun}pc^{-3}$ for the dark matter and the total density at 3 kpc with
fractional errors of about 25\%. This means that the dark matter
and total densities at 3 kpc for our galaxy and the 3 Coma galaxies are
comparable, and that the ratio of dark to total matter density of
about 1:10 is consistent within the error with the ratios of 1:5 and
1:13 for the non-central Coma galaxies.
\end{itemize}

\paragraph*{Acknowledgements}
We acknowledge the support of the European DUEL Research Training
Network, Transregional Collaborative Research Centre TRR 33, and
Cluster of Excellence for Fundamental Physics, further the use of data
from the Sloan Digital Sky Survey, and the Hubble Space
Telescope. Further we are grateful to the SLACS collaboration for
discovery and follow up observations of the galaxy scale lens sample
SDSSJ 1430+4105 is part of. Based on observations made with the
NASA/ESA Hubble Space Telescope, obtained from the data archive at the
Space Telescope Institute. STScI is operated by the association of
Universities for Research in Astronomy, Inc. under the NASA contract NAS 5-26555.
 We thank Natascha Greisel for making SFH
fits for us and Niv Drory for providing us with his SFH-fit code
SEDfits. We thank Claudio Grillo for many discussions and
  comments on this topic. We thank Jens Thomas for
    discussions about dynamical constraints on Coma galaxies. We thank
    Ralf Bender for discussions. We want to thank the
    anonymous referee for the numerous and elaborate comments, helping
    to improve the manuscript and making it more concise and understandable.


\appendix

\section{Additional strong lensing models}
\label{sec:parametric models:additions}

To check for the robustness of the previously derived lensing results,
we also examine some different strong lensing models to the ones
presented in Secs. \ref{sec:parametric models:point like}
and \ref{sec:lens modelling of the environment}. These
models confirm the previous results without adding new
implications for the results, therefore we add these models in
this appendix.  

\begin{figure}
\centering
\includegraphics[scale=0.4]{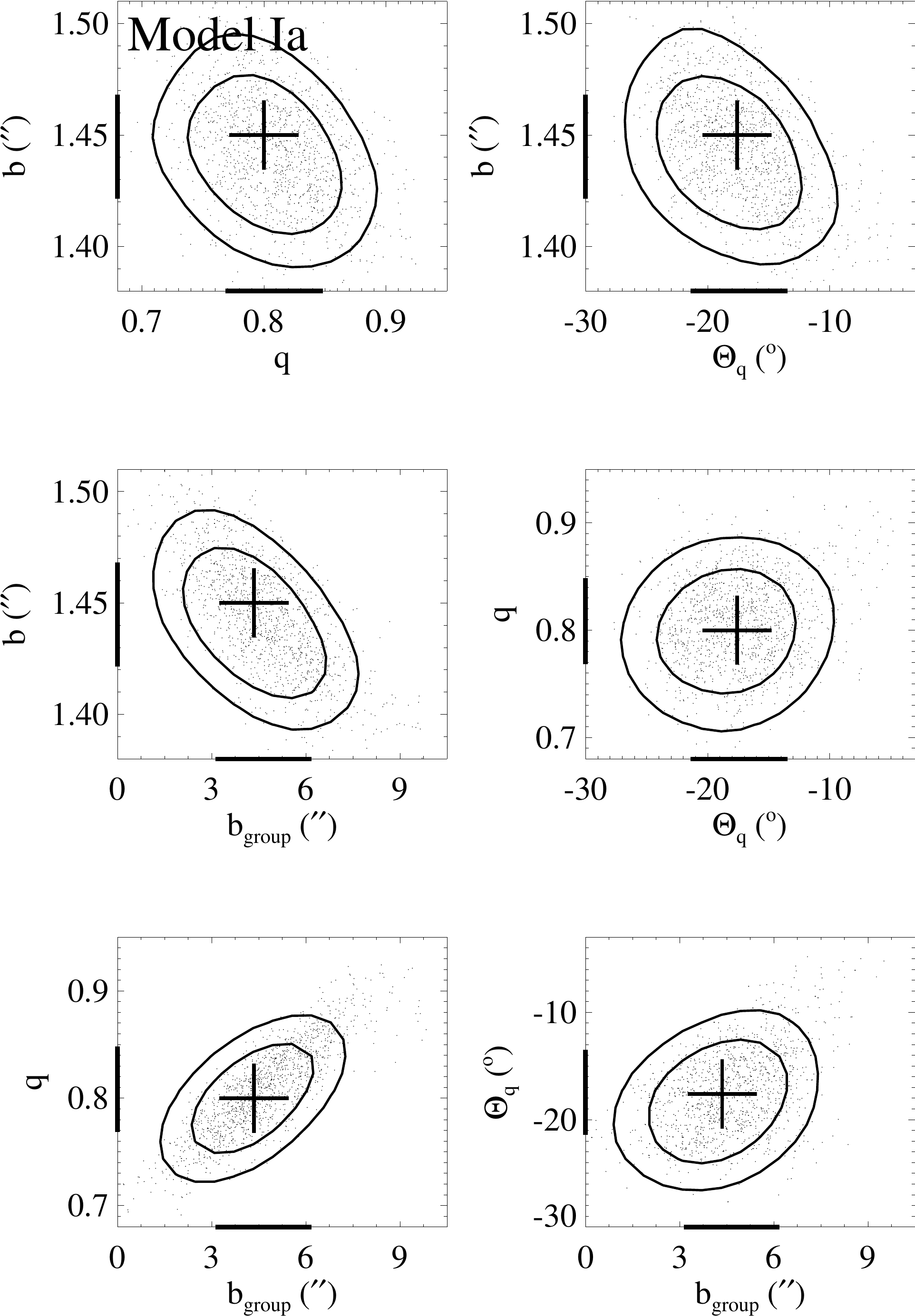}
\caption{Error estimates of the MCMC for the SIE case with GI
  (Model Ia),
  plotted are the individual points of the MCMC together with the 68
  and 90 \% confidence regions of the distribution. The crosses mark
  the minimum-$\chi^2$ value from Table
  \ref{lensing:parametric:table:best_fit_sie}. The bars on the axes mark
  the respective 68 \% marginalised error intervals. }
\label{lensing:parametric:SIE_wsdegeneracy}
\end{figure}

\paragraph*{Model Ia}:
To account for the environment, we include the galaxy group explicitly as a SIS
profile centred at galaxy I in
Tab. \ref{table:observations:environment:galaxies}.
We use a prior on the group Einstein radius of $\rm b_{\rm
  group,prior}=(3.6 \pm 1.5) \arcsec$, as derived in Sec. \ref{subsec:smooth group mass distribution centred at galaxy I}.
The results and 68 \% c.l. marginalised errors of this SIE+GI (Model Ia) case are: 
$\rm b=(1.45_{-0.02}^{+0.02})\arcsec$, $\rm q=(0.81_{-0.04}^{+0.04})$,
$\Theta_{\rm q}=(-17.4_{-4.0}^{+3.9})^\circ$ and $\rm b_{\rm
  group}=(4.6_{-1.4}^{+1.6})\arcsec$, see Fig.
\ref{lensing:parametric:SIE_wsdegeneracy} for the derived parameter errors.
This plot shows an anti-correlation of $\rm b$
and $\rm b_{\mbox{group}}$. This is expected since the total convergence
needed at the position of the main lens can either be provided
by the main lens or by the mass associated with GI.
We also include the environment as external shear, as
  calculated in Sec \ref{subsec:clumpy group}, see Model Ib in Appendix \ref{sec:parametric
    models:additions}. This has only small effects on the derived
  parameter values.

\paragraph*{Model Ib}:
Model Ib is for a SIE with external shear $\gamma$, hence it has 1 more free
parameter relative to Model Ia. The external shear priors are based on
the environment models derived in Sec. \ref{sec:observed environment}
and Appendix \ref{sec:environment:clumpy group nfw}:
We use $\gamma_{\rm prior}=(0.012 \pm 0.031)$ and $\Theta_{\rm
  \gamma,prior}=(-10 \pm 25)^{\circ}$.
The marginalised errors are: $\rm b=(1.50^{+0.02}_{-0.02})\arcsec$,
$\rm q=(0.81_{-0.07}^{+0.08})$, $\Theta_{\rm
  q}=(-13.4^{+12.5}_{-8.7})^\circ$,
$\gamma=(0.050^{+0.025}_{-0.025})$ and
$\Theta_{\gamma}=(-29.6_{-15.8}^{+7.9})^\circ$. 
There is a correlation present between the axis ratio q and the
external shear $\gamma$, reflecting the fact that the shear and the
ellipticity can compensate each other in its effects on the deflection
angle, since both are pointing in the same direction within $\approx
16^{\circ}$.

\paragraph*{Model IIa}:
If we add the group GI,
we obtain a marginalised steepness value of
$\beta=(1.71_{-0.13}^{+0.33})$ together with
$\rm b=(2.09_{-0.69}^{+0.80})\arcsec$, $\rm q=(0.91_{-0.11}^{+0.05})$,
$\Theta_{\rm q}=(-15.3_{-5.6}^{+10.2})^\circ$ and
$\rm b_{\rm group}=(4.5_{-1.5}^{+1.5})\arcsec$. Since there is no correlation
between $\beta$ and $\rm b_{\rm group}$, the details of the
environment implementation have no systematic influence on the derived
steepness of the lens mass profile. The shear and convergence provided
by the group are $\gamma_{\rm GI}=\kappa_{\rm GI}=0.037$ in agreement
with our expectations.

\paragraph*{Model IIb}:
A mass density profile which is flatter than isothermal at the Einstein
radius can also be achieved by an isothermal mass
distribution with a finite core radius. Therefore, Model IIb is for an
isothermal ellipsoid with a core radius (NSIE) with
$\beta=2$ and arbitrary value for the core radius $\Theta_{\rm
  c}$. For such a model one expects  to
also find a demagnified third image, which is not observed.  We assume
that the demagnified third image in the centre produced
by a non-singular mass profile could be
detected if its flux exceeds $3\sigma$ of the sky+object noise in the
image for the brightest source pixel. We exclude a region of
$0.2\arcsec$ in the centre due to residuals of the galaxy subtraction,
where we have no limits on the image fluxes at all.
We then get the following marginalised errors:
$\rm b=(1.63_{-0.10}^{+0.17})\arcsec$, $\rm q=(0.75_{-0.03}^{+0.03})$,
$\Theta_{\rm q}=(-21.5_{-2.3}^{+2.0})^\circ$ and a core
radius of $\Theta_{\rm c}=(0.11_{-0.075}^{+0.13}) \arcsec$. 
However, the best-fitting model is identical to Model I, i.e. purely
isothermal. There is a linear dependency between b and $\Theta_{\rm
  c}$ due to the definition of the non-singular profile: A larger core
radius needs to be compensated by a larger lensing strength $\rm b$ to
get the same enclosed mass within the Einstein radius. 

\paragraph*{Model IIIa}:
In Model IIIa we allow for a dark matter component that is
centred on the lensing galaxy. Here, we test if we can improve the modelling
by combining the de Vaucouleurs profile with a SIE halo. 
For the dark matter SIE halo, we impose a prior on the axis ratio based on the
\cite{slacsVII} work of $\rm q_{dark,prior}=(0.79\pm0.12)$. For the
errors, we get $\rm b=(1.43_{-0.08}^{+0.05})$, $\rm q_{\rm
  d}=(0.71_{-0.02}^{+0.02})$, $\Theta_{\rm
  q,d}=(-21.9_{-2.6}^{+2.1})^\circ$ and $\rm
M_{deV}=(0.6_{-0.4}^{+0.7})\times 10^{11} M_{\odot}$.
Since the best-fitting model has a $\chi^2=11.8$ and almost zero de Vaucouleurs mass, this implies that a de
Vaucouleurs like mass model for the light plus a purely isothermal
density profile for the dark matter are not compatible with the
data. The Einstein radius of the SIE component and the de Vaucouleurs
mass are anticorrelated, forcing the total projected mass within the
Einstein radius to be constant. 

\begin{figure*}
\centering
\includegraphics[scale=0.8]{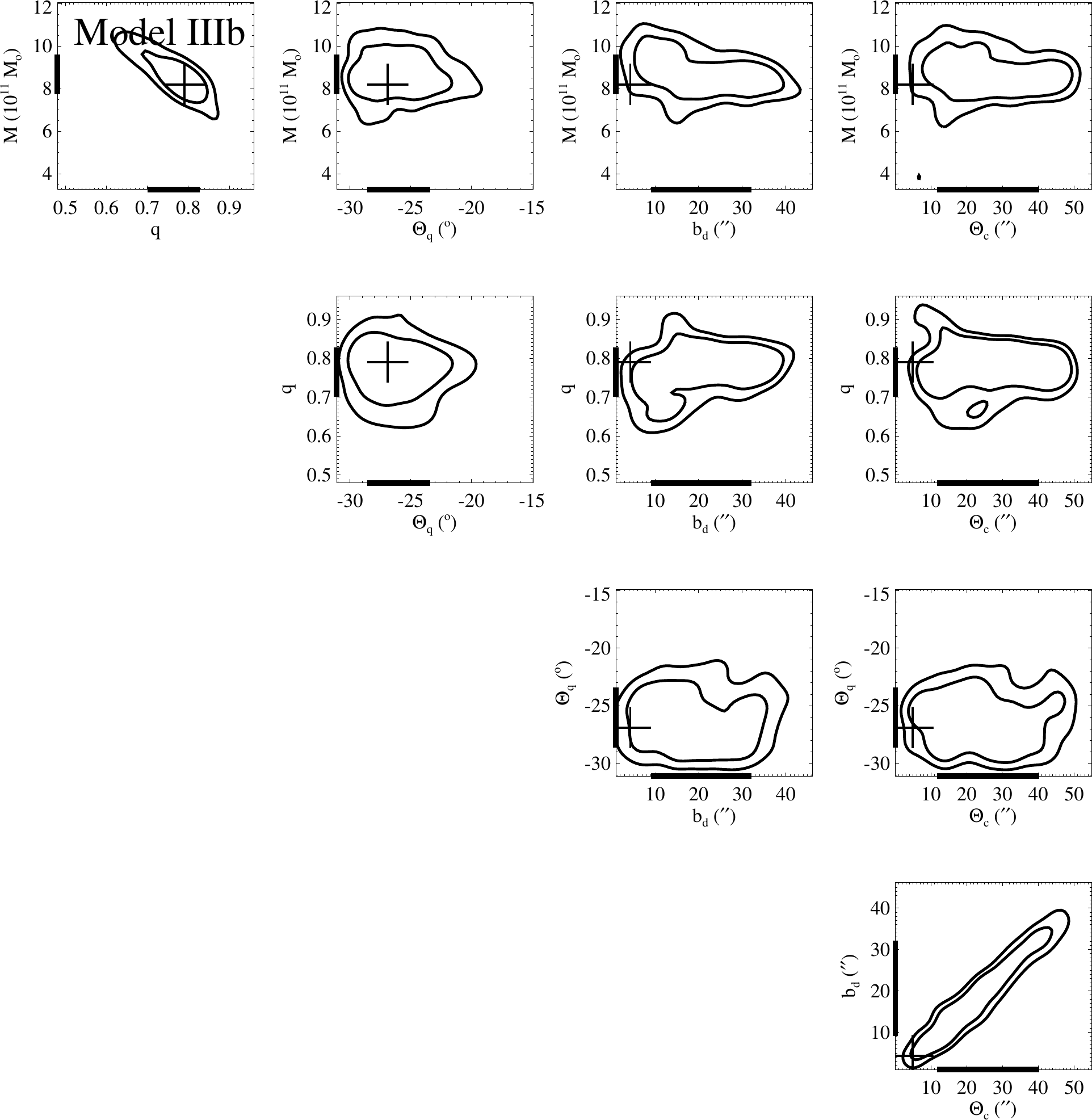}
\caption{Error estimates of the MCMC for the de Vaucouleurs+NSIE model
  (Model IIIb),
  plotted are the individual points of the MCMC together with the 68
  and 90 \% confidence regions of the distribution. The crosses mark
  the minimum-$\chi^2$ value from Table
  \ref{lensing:parametric:table:best_fit_devauc}. The bars on the axes mark
  the respective 68 \% marginalised error
  intervals. The individual points of the MCMC are
    omitted for clarity.}
\label{lensing:parametric:gravlens:NSIEdeVauc_ns}
\end{figure*}

\paragraph*{Model IIIb}:
Instead of the NFW component, we can also model the dark matter component
with a NSIE. We limit the core radius of this component to be between
$0$ and $50\arcsec$, and use the same prior on the axis ratio
as before.
For the mass of the de Vaucouleurs component, we get:
$\rm M_{deV}=(8.5_{-0.7}^{+1.1})\times 10^{11} M_{\odot}$. For the other parameters, we get (see
Fig. \ref{lensing:parametric:gravlens:NSIEdeVauc_ns}):
$\rm q_{\rm d}=(0.78_{-0.08}^{+0.05})$, $\Theta_{\rm
    q}=(-26.0_{-2.4}^{+2.6})^\circ$, $\rm b_{\rm
    d}=(19.2_{-10.1}^{+13.0})\arcsec$ and $\Theta_{\rm
  c}=(24.7_{-12.9}^{+15.5})\arcsec$. Again, we note that there is a
  degeneracy between $\rm b_{\rm d}$ and $\Theta_{\rm c}$, emerging from the
  profile definition. Since there are no observed
  images for radii larger than $2.32\arcsec$, this leaves the upper
  limit of the core radius $\Theta_{\rm c}$
  totally unconstrained. Large $\Theta_{\rm c}$ make this dark
  matter distribution flat at the Einstein radius, with $\rm b_{\rm
    d}$ giving its density value.

\paragraph*{Model IVa}:
We model the mass distribution by a de Vaucouleurs model with the shape
parameters following the light profile as stated in 
Table \ref{observations:table:slacs_values} Therefore, the mass of the
de Vaucouleurs component is the only free parameter in this model. The best-fitting light model (Model IVa) in Table
\ref{lensing:parametric:table:best_fit_devauc2} has a $\chi^2=118$,
meaning that a pure de Vaucouleurs profile is a bad fit to the observations. The
de Vaucouleurs mass in this case is
$\rm M_{\rm deV}=(15.0^{+0.2}_{-0.2})\times 10^{11} M_{\odot}$. This badness of the fit implies that there must be a
(dark) mass component not following a de Vaucouleurs profile.

\paragraph*{Model Va}:
Here we do the same as in the Model V before: We combine the 2 component
model (Model IIIb) with an explicit description for the galaxy group
GI. For the de Vaucouleurs component, we get:
$\rm M_{deV}=(11.5_{-1.5}^{+0.7})\times 10^{11} M_{\odot}$. For the other parameters, we get $\rm q_{\rm d}=0.74_{-0.10}^{+0.11}$, $\Theta_{\rm
    q}=(-21.8_{-5.3}^{+6.7})^\circ$, $\rm b_{\rm
    d}=(5.5_{-3.0}^{+4.1})\arcsec$, $\Theta_{\rm
  c}=(16.4_{-9.1}^{+12.1})\arcsec$, and for GI $\rm b_{\rm
    group}=(5.1_{-1.6}^{+1.3})\arcsec$. As one can see, again there is
no significant difference between this model's parameters and the one
of Model IIIb. As for the NFW-like dark matter halo, the $\rm M_{deV}$ is increased
relative to Model IIIb by introducing the group halo GI.

\paragraph*{Model Vb}:
Model Vb is motivated by the fact that for the preceding models
(Models III and V), the axis of the dark matter halo is always offset
from that of the light by about $-10^{\circ}$, which is statistically
significant on a more than $3 \sigma$ level for Models III and IIIb, see
Figs. \ref{lensing:parametric:gravlens:NFWdeVauc_ns} and
\ref{lensing:parametric:gravlens:NSIEdeVauc_ns}. At the same time, the
axis ratio of the dark matter haloes are consistent with the axis ratio of the
stellar component, see Table
\ref{observations:table:slacs_values}. This could be mimicked by a non
accounted external shear which is present if galaxy A is not the
centre of the group. From the results for Models Ia, IIa, V and Va, we
conclude that if we include the group explicitly as centred on galaxy I the matter
gets more aligned with the light. Looking at Model Ib we see that
using an external shear instead of GI changes the best-fitting orientation
of the total mass distribution more towards the observed light's
angle. So, in this Model, we combine Model IIIb with the external shear
of Model Ib. In numbers, we get here: $\rm
  M_{deV}=(9.9_{-2.7}^{+2.1})\times 10^{11} M_{\odot}$, $\rm q_{\rm d}=(0.77_{-0.10}^{+0.10})$, $\Theta_{\rm
    q}=(-14.2_{-11.2}^{+13.0})^\circ$, $\rm c_{\rm
    d}=(4.1_{-1.7}^{+4.1})$, $\rm
  r_{200}=(166_{-55}^{+61})\arcsec$, and for the external shear $\gamma=(0.038^{+0.023}_{-0.021})$ and
$\Theta_{\gamma}=(-35.2_{-15.2}^{+9.2})^\circ$. We also note
  that with this improvement, the dark matter profile becomes more
  concentrated, at a level expected for galaxies.

\begin{table*}
\centering
\begin{minipage}{290mm}
\caption{minimum-$\chi^2$ values and parameter estimates, derived with
  {\sc gravlens} for the one component
  isothermal and powerlaw models}
\begin{tabular}{cccccccccccc}
\hline
& & $b$& $q$ & $\Theta_{\rm q}$ & $\beta$ & $\Theta_{\rm c}$ & $\rm b_{\rm group}$
& $\chi^2$ &d.o.f.& $\frac{\chi^2}{\rm d.o.f}$\\
& &($\arcsec$)& & ($^{\circ}$)& &($\arcsec$) &($\arcsec$) & & & \\
\hline
Model Ia &SIE+GI & 1.45 & 0.80 & -17.6 &2.00\footnotemark[1]& & 4.4  & 8.7 & 6&1.5 \\
 & & 1.43 -- 1.47 & 0.77 -- 0.85 & -21.4 -- -13.5 & & & 3.2 -- 6.2 & \\
Model Ib & SIE+es &1.50 &0.82 &-7.2&2.00\footnotemark[1] & 
&  &9.1 &5&1.8\\
& & 1.48 -- 1.52 & 0.74 -- 0.89 & -22.1 -- -0.9 & & &  & &
 \\
Model IIa &PL+GI & 2.53 & 0.93 & -13.0 &1.60& & 4.3  &  7.6 & 5 &1.5 \\
& & 1.40 -- 2.89 & 0.80 -- 0.96 & -20.9 -- -5.1 & 1.58 -- 2.04 & & 3.0 --
6.0  & &\\
Model IIb & NSIE &1.49 &0.71 & -21.6 &2.00 \footnotemark[1]
&$3.8\times10^{-5}$   & &11.5 &6 & 1.9\\
 & & 1.53 -- 1.80 & 0.72 -- 0.78 & -23.8 -- -19.5 & & 0.035 -- 0.24  & &
& & \\
\hline
\label{lensing:parametric:table:best_fit_sie2}
\end{tabular}
\end{minipage}
\footnotemark[1]{fixed value}
\end{table*}
\begin{table*}
\centering
\begin{minipage}{290mm}
\caption{minimum-$\chi^2$ values and parameter estimates, derived with
  {\sc gravlens} for the two component
   de Vaucouleurs plus dark matter models}
\begin{tabular}{ccccccccccc}
\hline
& & $\rm M$ & $q_{\rm d}$ & $\Theta_{\rm q,d}$ &
 $\rm b_{\rm d}$ & $\Theta_{\rm c}$ & $\rm b_{\rm group}$
& $\chi^2$ & d.o.f. & $\frac{\chi^2}{\rm d.o.f}$\\
& &$\left (\rm 10^{11}M_{\odot}\right )$ & & ($^{\circ}$)
&($\arcsec$) &($\arcsec$) &($\arcsec$)  & & &  \\
\hline
Model IVa& deVauc & 15.0  & &  &  & & &
118 & 9 & 13.1  \\
 & & 14.8 -- 15.2 & & & & & & & & \\
Model IIIa & deVauc+SIE & 0.0004 & 0.72 & -21.5 & 1.50& & & 11.8& 6& 2.0\\
& & 0.2 -- 1.3 & 0.69 -- 0.73 & -24.5 -- -19.8 & 1.38 -- 1.48  & & & & &\\
Model IIIb & deVauc+NSIE & 8.2 & 0.79 & -26.9  & 4.3 & 4.9 & & 7.6 &
5 & 1.5 \\
& & 7.8 -- 9.6 & 0.70 -- 0.83 & -28.4 -- -23.4  & 9.1 -- 32.2 & 11.8 -
40.2 & & & & \\
Model Va & deVauc+NSIE+GI & 10.0 & 0.79 & -25.5  & 2.1 & 3.6 & 3.6 & 7.6 &
4 & 1.9 \\
& & 10.0 -- 12.2 & 0.64 -- 0.86 & -27.1 -- -15.1  & 2.5 -- 9.6 & 7.3 --
28.5 & 3.5--6.4 & & & \\
\hline
\label{lensing:parametric:table:best_fit_devauc2}
\end{tabular}
\end{minipage}
\end{table*}

\begin{figure*}
\centering
\subfigure[]{\includegraphics[scale=.55]{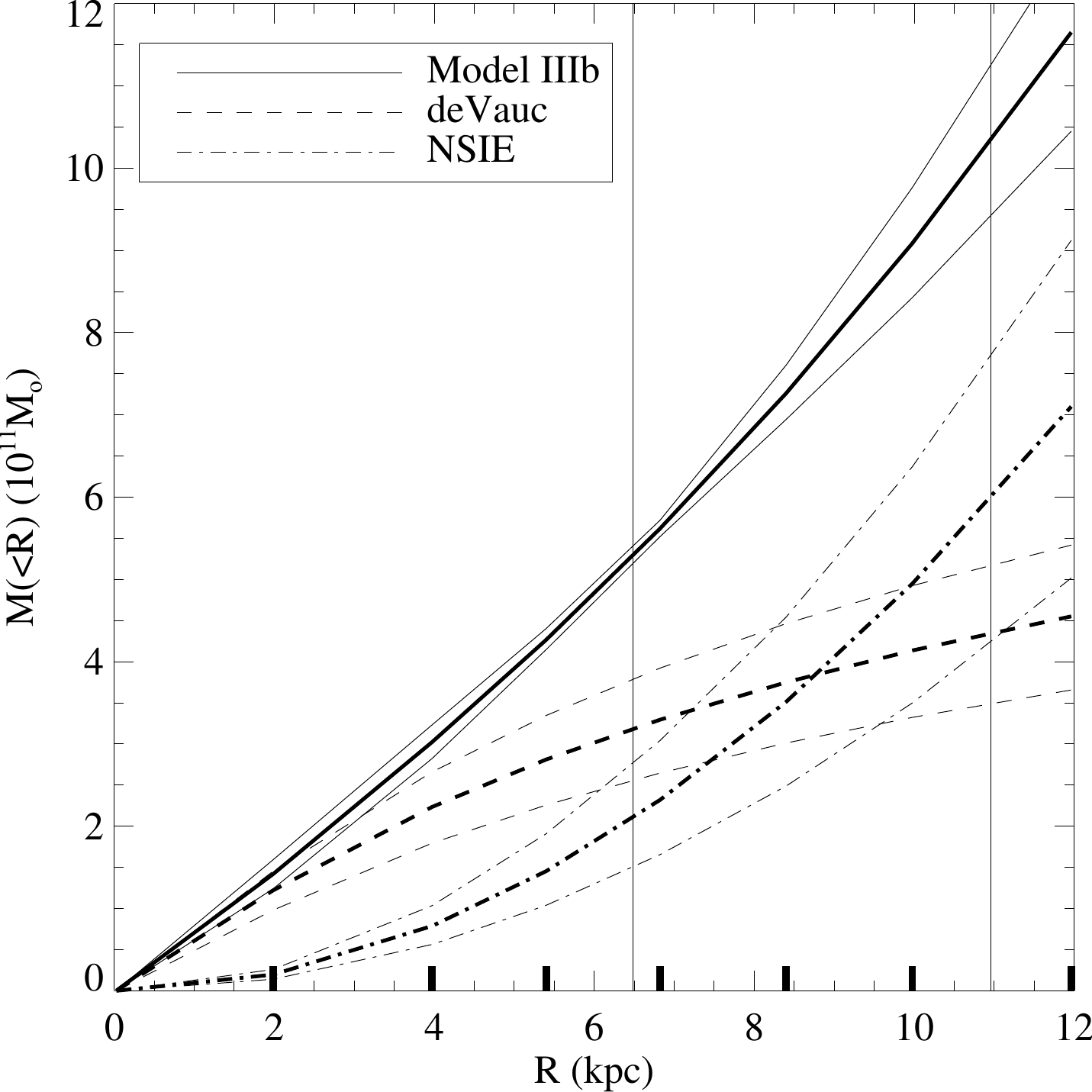}\label{lensing:parametric:devauc_NSIE_masses}}
\subfigure[]{\includegraphics[scale=.55]{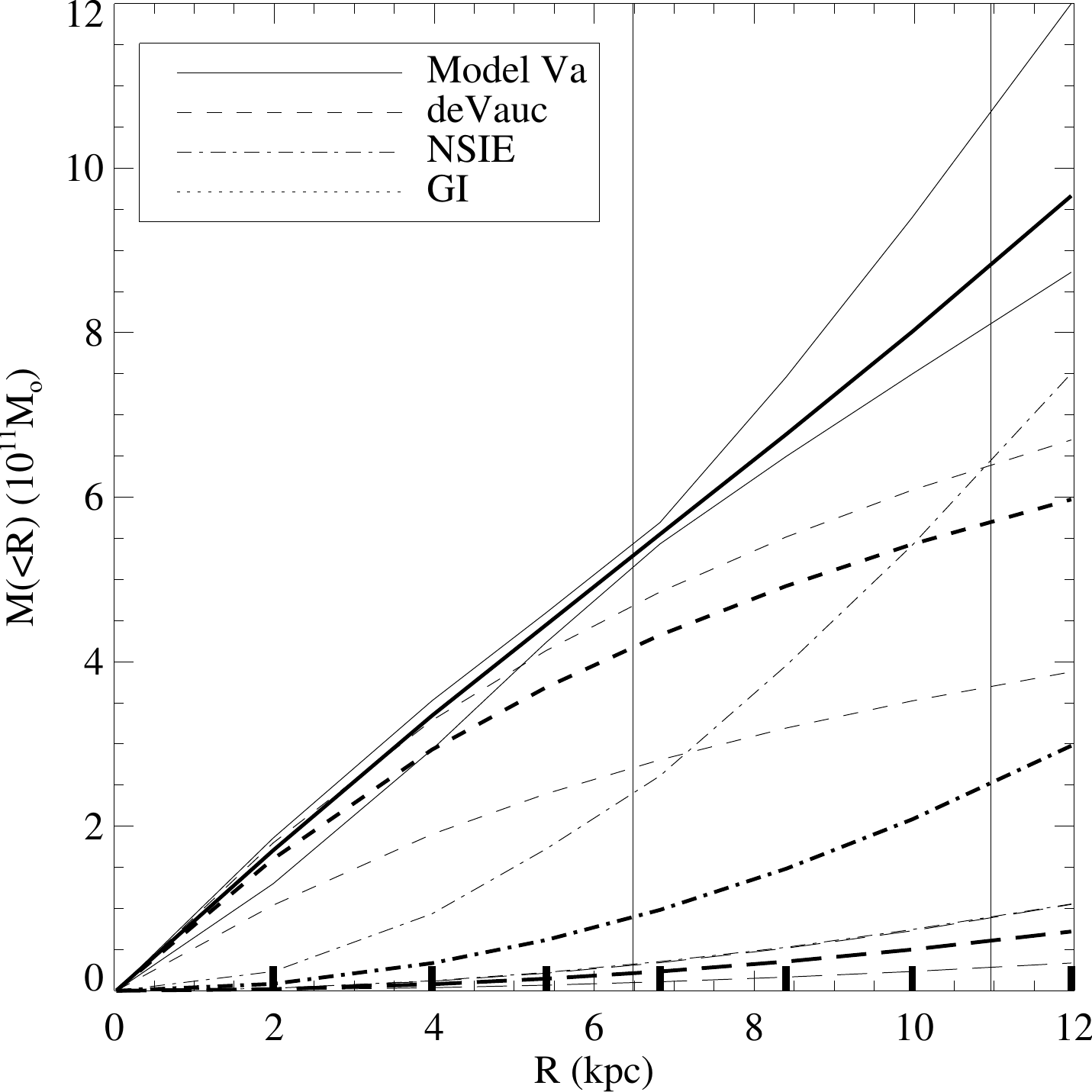}}
\subfigure[]{\includegraphics[scale=.55]{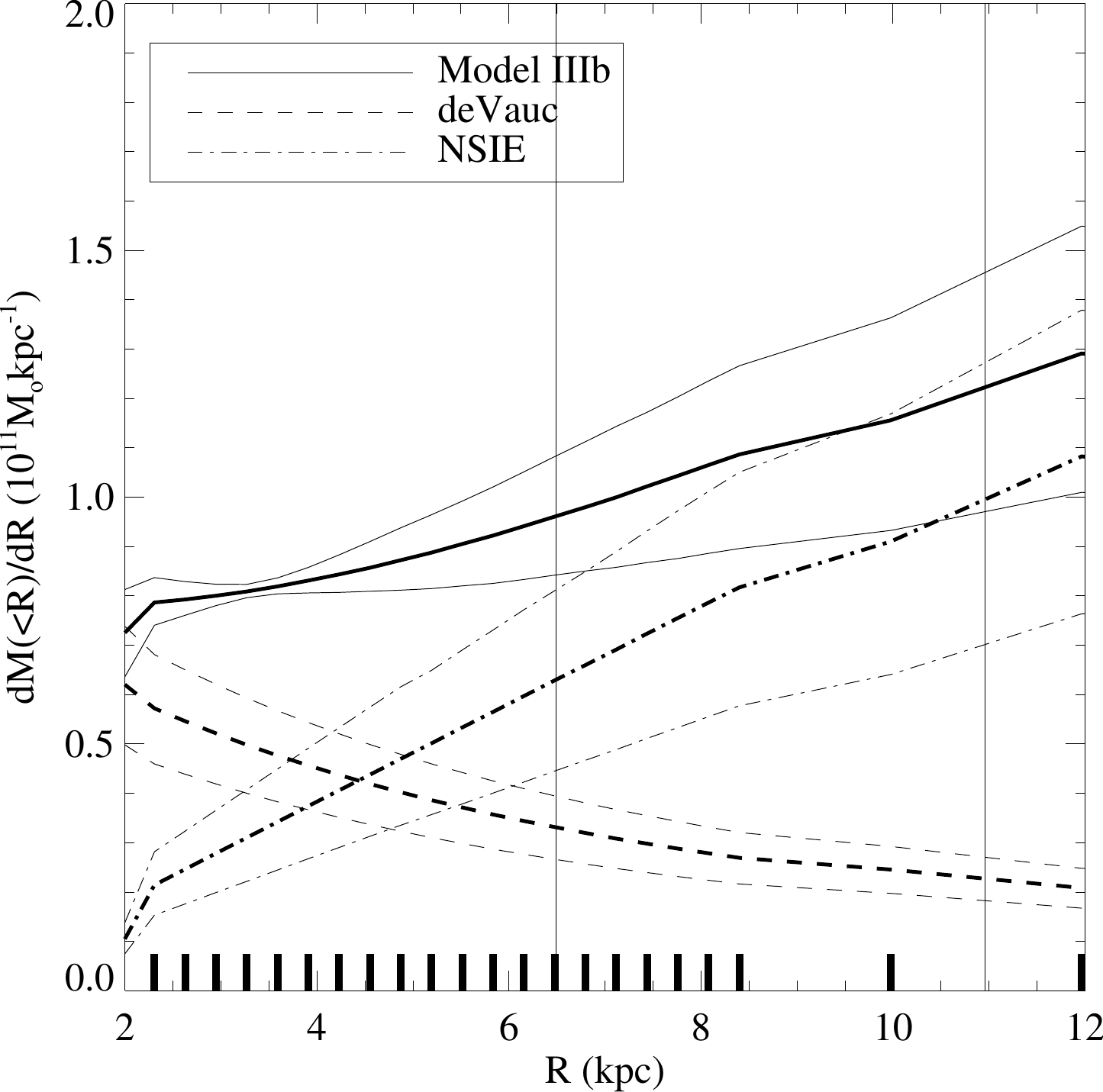}\label{lensing:parametric:devauc_NSIE_masses_deriv}}
\subfigure[]{\includegraphics[scale=.55]{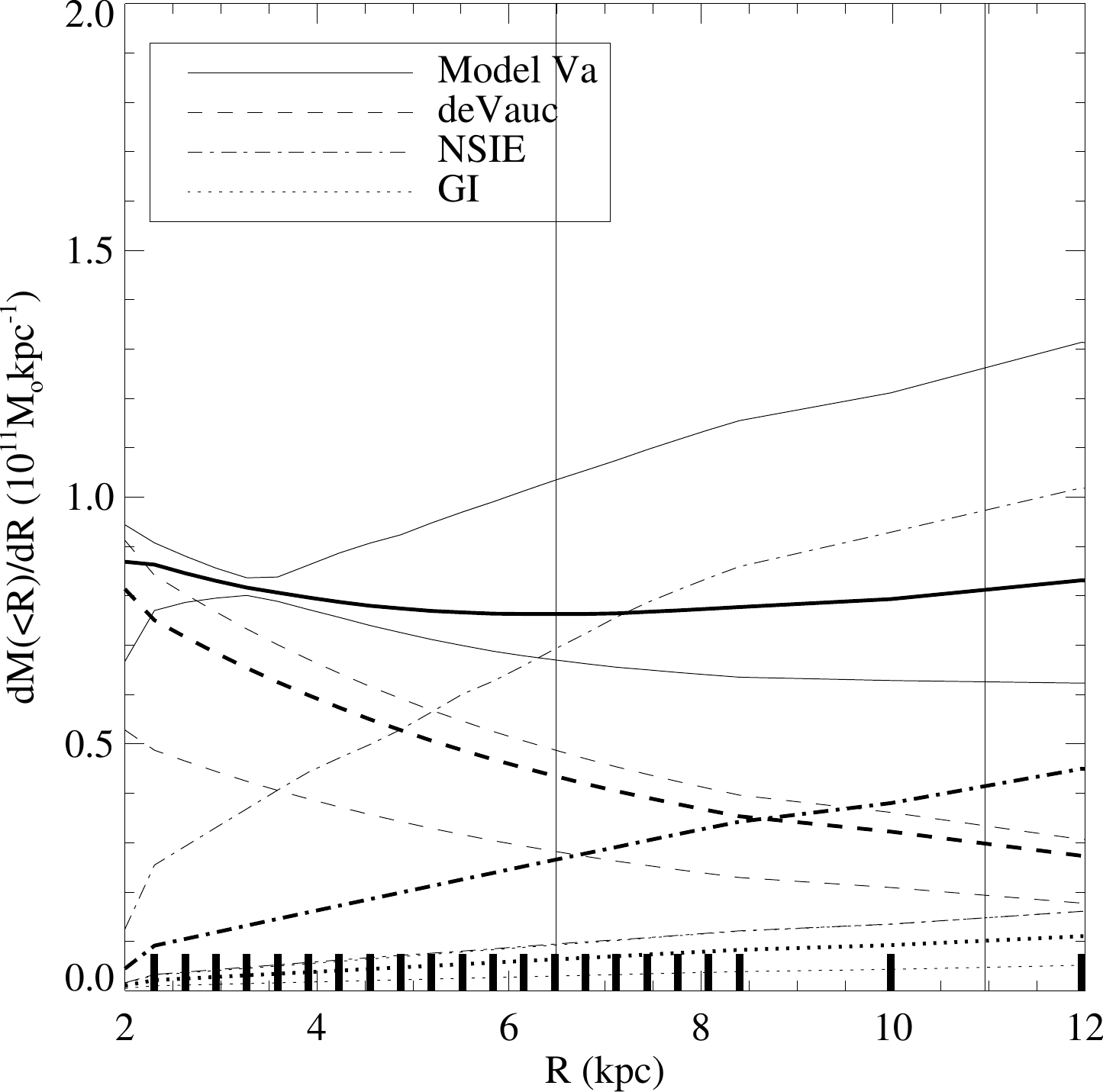}}
\caption{The same as Fig. \ref{lensing:parametric:gravlens:masses_devauc}, only for Models
IIIb and Va}
\label{lensing:parametric:gravlens:masses_devauc+GI}
\end{figure*}


\section{Alternative descriptions for the lens environment}
\label{sec:environment:clumpy group nfw}

In this Appendix, we discuss 2 alternative scenarios for the
environment, firstly a scenario in which the group is only consisting of its members
(``clumpy group'') without a reference to a group halo, secondly a
scenario where the group is a typical group with 12 members, but
centred on galaxy A instead of galaxy I.

\subsection{Clumpy group}
\label{subsec:clumpy group}
A clumpy group model is obtained if all group mass is considered to be
associated to the group galaxies. We describe the galaxies as singular
isothermal spheres (SIS) without truncation of their mass profiles and
obtain at the position of the lens:

\begin{equation}
\begin{aligned}
\kappa_{\rm A}^{\rm clumpy \,
    group}=\sum\limits_{\rm n}\kappa_{\rm SIS,n}\quad, \\
\gamma_{\rm A}^{\rm clumpy \, group}=\sum\limits_{\rm n}\gamma_{\rm SIS,n}\quad.
\label{clumpy_group_kappa_gamma}
\end{aligned}
\end{equation}

In this model, the shear and surface density at the location of the
lens depends on the 2-dimensional galaxy distribution and not at all on
the centre of mass of the group.
The galaxies themselves are 
parametrized only by their positions and velocity dispersions
$\sigma_{\rm n}$.  
The value of the velocity dispersion $\sigma_{\rm A}$ for SDSS
J1430+4105 is taken from the central
velocity dispersion measured by the SDSS. The estimates $\sigma_{\rm n}$ for the 
neighbouring galaxies are obtained from the Faber-Jackson relation (\citealt{faber-jackson-relation}),
\[\sigma_{\rm n} = \sigma_{\rm A} \left(\frac{i_{\rm n}}{i_{\rm A}}\right)^{0.25}\quad,\]
where $i$ is the SDSS $i$ band flux of the neighbours and $i_{\rm A}$
the flux of the lens galaxy A.  
The shear $\gamma_{\rm n}(d_{\rm n})$ and convergence $\kappa(d_{\rm n})$  for a SIS at a projected
angular distance $d_{\rm n}$ from its centre are
\[\kappa(d_{\rm n})=\gamma_{\rm n}(d_{\rm n}) = \frac{2 \pi
  \sigma_{\rm n}^{2}}{c^{2}d_{\rm n}}\left(\frac{D_{\rm ds}}{D_{\rm s}}\right)\quad,\]
with $c$ denoting the speed of light and $D_{\rm ds}$
  and $D_{\rm s}$
mark the angular diameter distances from the lens to the source and
from the observer to the source, respectively. The 
proper (vector) addition of these convergence and shear values yields
a prediction of
\begin{equation}
\begin{aligned} 
\gamma_{\rm A}^{\rm clumpy\, group} = 0.012\quad ,\\
\kappa_{\rm A}^{\rm clumpy\, group}=0.023\quad.
\label{environment: clumpy group values}
\end{aligned}
\end{equation}
The angle of the shear is $-10^{\circ}$
in the local coordinate system.
The fact that we model the galaxies as SISs, ignoring the
finite halo sizes which
would keep the mass associated to galaxies limited, is not relevant,
since finite halo sizes can only lead to lower estimates
  for the convergence and shear at the position of galaxy
  A. Therefore, we get an upper limit of the clumpy group estimates
  using this assumption.
As we  see in
Sec. \ref{sec:lens modelling of the environment}, the parameters of the
lensing galaxy only mildly depend on the assumptions about the
group as long as it is centred on galaxy I.
To calculate the mass of this clumpy group, we first need the $r_{200}$.
We adopt the definition of \cite{maxBCG} of
$r_{200}$ as a function of number of group members. We sum up all mass
contributions of member galaxies in Table
\ref{table:observations:environment:galaxies} within this $r_{200}=3.8\arcmin$ centred
on A or I, respectively and calculate the total projected mass of the group within its $r_{200}$:
 $\rm M_{200}=5.5 \times 10^{14}\,\rm M_{\odot}$. This value gives an
upper limit of the mass associated with the group,
since SIS profiles for its members overestimate the densities of each member at large radii. 


\subsection{Smooth group mass distribution centred at galaxy A}
\label{subsec:smooth group mass distribution centred at galaxy A}

\begin{figure}
\centering
\includegraphics[height=80mm]{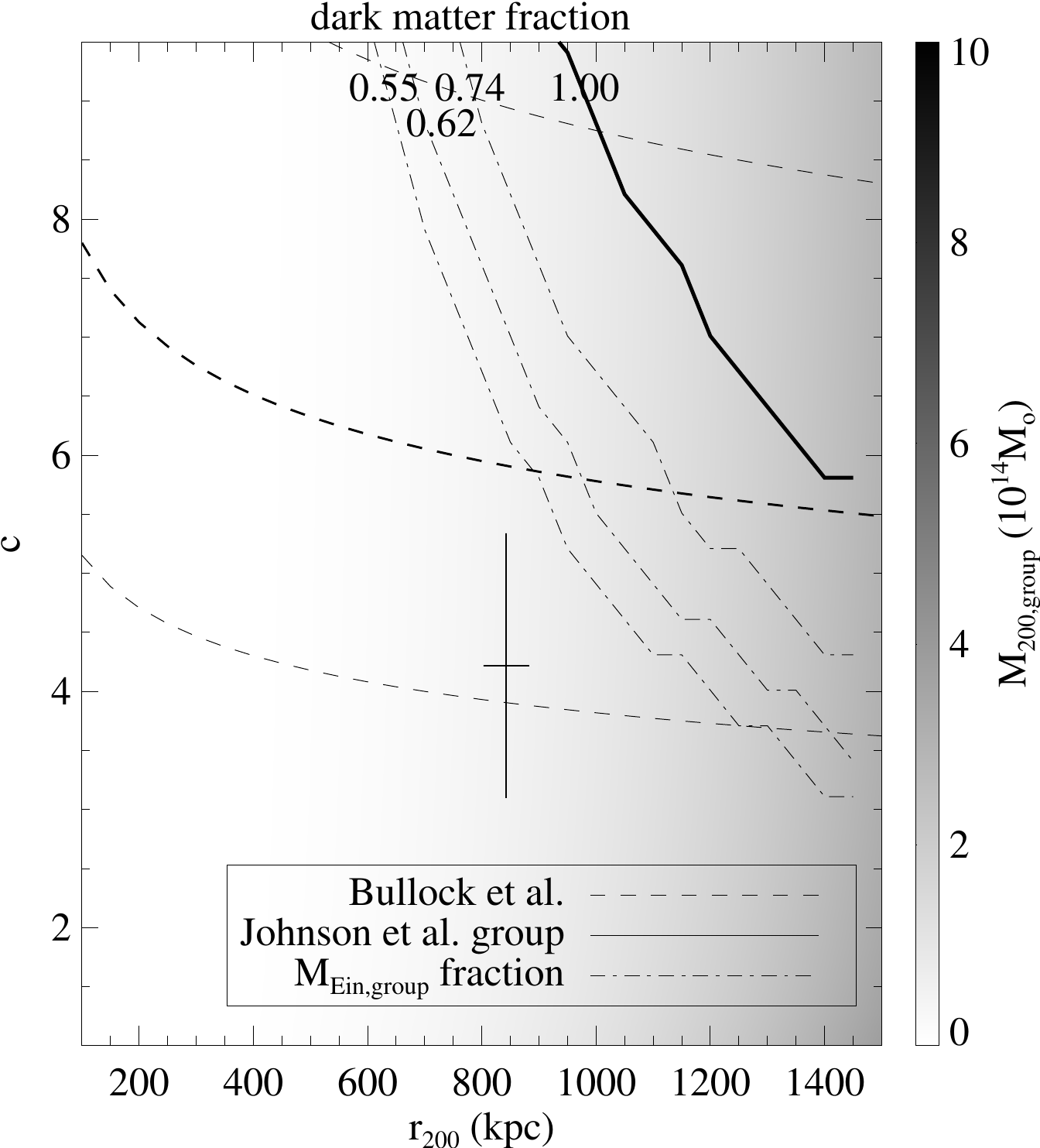}
\caption{This figure shows the concentration c - $\rm r_{200}$ properties for a
  NFW halo profile. The levels of grey show the
    virial masses of the dark matter haloes. Overplotted are several
    different lines: The dashed lines are the
  \protect\cite{bullock_c_r_200_relation} c-$\rm r_{200}$ relation with its
  $1\sigma$ error. This marks the area typically populated by galaxy
  groups. Further we overplot the c-$\rm r_{200}$ values for
    a typical richness N=12 group halo as found by
    \protect\cite{Johnston_2007} with its errorbars. This shows where we
  expect the group halo to lie approximately in this plane.  
The dash-dotted lines mark the transition above which more than 55,
62 and 74 \% of the observed critical mass within the observed Einstein radius
would be made up by the dark matter halo of the
group. All group haloes above this dash-dotted line
  in this c-$\rm r_{200}$ plane overpredict the observed total mass
  within the Einstein radius, therefore this lines mark regions with
  excluded group haloes. Since the typical \protect\cite{Johnston_2007} group
  halo lies below this lines, the observed critical mass within the
  Einstein radius does not exclude A as the group centre. The thick,
solid 1.00 line marks the transition where the dark matter group halo alone
would provide the observed critical mass within the Einstein
radius. Hence along this line no baryons (or dark matter) in the lensing galaxy A would be
required at all.}
\label{SDSSJ1430_group_mass_limits}
\end{figure}

In principle, the assumption of the group being located at A could
already be in conflict with the lensing observables. The most secure
strong lensing estimate is the observed
critical mass $\pi\rm R_{\rm Ein}^2\Sigma_{\rm crit}=5.43^{+0.15}_{-0.16}\times
10^{11}\rm M_{\odot}$ within the Einstein radius, obtained from all
models in Sec. \ref{Sec:mass} consistently. We now can model the group -- located at A
as an NFW or SIS (see Sec. \ref{sec:parametric models:point like}
for details) profile and estimate its projected mass within the observed
Einstein radius. If this halo mass estimate exceeds the observed critical mass, the assumption of this
group being a typical group with 12 members and with galaxy A as its
centre is already in conflict with the lensing observables.\\
In Fig. \ref{SDSSJ1430_group_mass_limits} we show the c-$\rm r_{200}$ diagram
for a NFW profile. The levels of grey indicate the virial
$\rm M_{200}$ mass of a group with parameter values c and $\rm r_{200}$.
The thick solid line marks the transition where the NFW group halo mass
within the observed Einstein radius alone (without baryons and dark
matter of the galaxy A) exceeds the critical mass, predicting a bigger
than the observed Einstein radius. Therefore all
  groups that lie above this line would -- from its group halo mass
  alone -- overpredict the observed total projected mass within the
  Einstein radius and cannot be centred at galaxy A.\\
In reality, some of the observed mass within the
  Einstein radius has to be contributed by the stars, giving an even
  smaller upper limit for the dark matter mass within the Einstein radius.
Hence, we plot the analogous curves for the case where the dark matter makes
up only a fraction of the total critical mass within the Einstein
radius. The dark to total matter fractions shown also in
Fig. \ref{SDSSJ1430_group_mass_limits} as dash-dotted lines are
$\rm f_{\rm dark}=0.55,\, 0.62,\, 0.74.$ To obtain these numbers, we
subtract the stellar mass measurements within the
  Einstein radius done in
\cite{claudio_SED_fits} from the derived lensing mass within the Einstein
radius in this work. If we attribute the missing mass
to the group dark matter halo, we get again upper limits for the possible group
halo mass contribution within the Einstein radius, allowing us to
exclude all groups that would exceed this upper mass limits. \cite{claudio_SED_fits} fit
  composite stellar population models to the SDSS photometry of this galaxy to derive its
  stellar mass within the observed Einstein
  radius. We use the Salpeter IMF stellar
masses of \cite{claudio_SED_fits}, since these give the highest
mass in stars. Now we plot the model group with
    richness 12 in Fig. \ref{SDSSJ1430_group_mass_limits} to see
    where it resides. From \cite{Johnston_2007}, we obtain c-$\rm
    r_{200}$ values of 4.22 and 848 kpc for a richness 12 group. Since
  this group therefore does not fall into the excluded regions of
  Fig. \ref{SDSSJ1430_group_mass_limits}, we cannot exclude A as the
  group centre from the lensing observables.
This conclusion also holds in the
picture where the group is modelled as SIS. If the group follows a SIS
matter profile it has an Einstein radius of $\Theta_{\rm E}=3.6 \pm
1.5\arcsec$, see Sec. \ref{subsec:smooth group mass distribution
  centred at galaxy I}.
This is consistent within the errors with the value derived from the
strong lensing models in Sec. \ref{Sec:mass}. 
Therefore a typical group with richness 12, as seen
  in the vicinity of SDSSJ 1430+4105, does not violate the observed
  critical mass within the Einstein radius, nor the Einstein radius
  itself. Hence, galaxy A could
also be the group centre without violating the lensing observables for
a typical group of this richness.

\clearpage
\end{document}